\documentclass[a4paper,11pt]{article}
\usepackage[utf8]{inputenc}
\usepackage[left=2.5cm,right=2.5cm,top=2.5cm,bottom=2.5cm]{geometry}
\usepackage{amsmath,amssymb,amsfonts,amsthm,caption}
\usepackage{amscd,fancyhdr,graphicx,latexsym,mathrsfs,multicol,setspace,wrapfig}
\usepackage{psfrag}
\usepackage{color,epsfig}
\usepackage[
  bookmarksopen,
  colorlinks,
  linkcolor = blue,
  urlcolor  = blue,
  citecolor = blue,
  menucolor = black
]{hyperref}
\usepackage{natbib}

\def\argmin{\text{argmin}}

\newcommand\R{\mathbb{R}}

\newcommand\N{\mathbb{N}}

\def\argmin{\text{argmin}}

\newtheorem{problem}{Problem}
\newtheorem{df}{Definition}

\newcommand{\vdd}{{\mathcal{V}^\delta}}

\newcommand{\tvf}{{\widetilde{\mathcal{V}}}}

\newcommand{\af}{{\mathcal{A}}}
\newcommand{\vf}{{\mathcal{V}}}
\newcommand{\Q}{{\mathfrak{Q}}}
\newcommand\X{H^1(0,\overline{s};H^{1+\varepsilon}(D))}

\newcommand{\s}{{\overline{s}}}
\newcommand\he{H^{1+\varepsilon}(D)}
\newcommand\Y{L^2(0,\s;L^{2}(D))}

\title{\bf Local Volatility Models in Commodity Markets and Online Calibration}

\author{Vinicius Albani\thanks{Computational Science Center, University of Vienna, 
1090 Vienna, Austria, \href{mailto:vvla@impa.br}{\tt vvla@impa.br}}, \,
	Uri M. Ascher\thanks{Dept. of Computer Science, University of British Columbia, Canada, \href{mailto:ascher@cs.ubc.ca}{\tt ascher@cs.ubc.ca}} \,
and Jorge P. Zubelli\thanks{IMPA, 
Rio de Janeiro, Brazil, \href{mailto:zubelli@impa.br}{\tt zubelli@impa.br}}}
\date{\today}
\begin{document}

\maketitle

\begin{abstract}
We introduce a local volatility model for the valuation of  options on commodity futures by using 
European vanilla option prices.
The corresponding calibration problem is addressed within  
an online framework, allowing the use of multiple price surfaces.
Since uncertainty in the observation of the underlying future prices translates to uncertainty in data locations, we propose a model-based adjustment of such prices that improves
reconstructions and smile adherence.
In order to tackle the ill-posedness of the calibration problem we
incorporate a priori information through a judiciously designed
Tikhonov-type regularization. 
Extensive empirical tests with market as well as synthetic data are used to demonstrate
the effectiveness of the methodology and algorithms.

\end{abstract}

\noindent {\bf keywords:} commodity future options, local volatility calibration, online approach, inverse problem, Tikhonov-type regularization.

\section{Introduction}
\label{sec:introduction}

Commodity futures and their derivatives have become key players in the portfolios of many corporations, 
especially those in the energy sector. As a consequence, problems arise in the pricing of more complex derivatives, 
since plain vanilla options are generally not sufficient to address all the exposure 
to which such companies are subjected. 
Moreover, well-known models that had been developed to accommodate the futures term-structure did not 
necessarily fit the market implied volatilities (or market ``smile'').

We propose a local volatility model to price European vanilla options on commodity futures. 
Specifically, we consider a special class of local volatility surfaces and assume that 
future prices are zero-drift diffusions driven by such volatility surfaces. 
After some simple technical adaptations, i.e., re-parametrization, normalization and change of variables, 
we obtain an initial-boundary value problem that uniquely determines
the option prices on futures with different maturities. 
This leads to a model that fits the market ``smile'' and the futures term-structure.

The corresponding inverse problem is addressed by adapting the {\em online} calibration technique introduced in
\citep{vvla2} to the case of scarce data~\citep{AlbAscYanZub2015}. 
This allows us to use option prices for different dates in the calibration, without making any data interpolation, 
and to consider a regular dependence of sets of local volatilities on an index, 
i.e, we assume that local volatility evolves in a well-behaved way. 
The inverse problem under this online setting is then solved using a Tikhonov-type regularization, 
with a penalization functional including the squared $L_2$-norm of the derivatives of the local volatility 
with respect to the index from the online setting, the time to maturity and the log-moneyness. 
Such components have different weights which must be selected carefully.
This selection is much more involved than the usual one-parameter discrepancy-like principles 
considered in~\citep{vvla2}. It is carried out heuristically, using numerical experiments. 
We also take into account the uncertainty in the observation of the underlying asset prices, 
which in turn translates into uncertainty in the log-moneyness. 
In our methodology these asset prices are included in the set of unknowns,
and their values are adjusted simultaneously with the calibration of the local volatility surface. 

Note that, in general, market vanilla options are American.
In commodity markets, convenience yield implies that American call options are more expensive than the European ones. 
This suggests that a calibration procedure must take into account an American pricing procedure,
and it was studied in~\citep{achdou1}, Chapter~9 in~\citep{achpi}, and Section~8 in~\citep{crepey2}. 
However, because of the intrinsic nonlinearity of the American option pricing problem, 
an approach similar to 
the one proposed in~\citep{dupire} is not available. 
As a consequence, the problem of American option pricing must be solved for each strike and maturity, 
leading to a computationally intensive undertaking. 
Therefore, we convert American into European prices, by extracting implied volatilities and using 
Black's formula (see~\citep{black}). 
We have performed some tests to illustrate that the noise introduced by this operation does not 
compromise the reconstructed local volatility surface.

\paragraph*{Main Contributions}
A local volatility model is introduced in the context of commodity markets in such a way that 
the existing techniques for equity markets also hold in the present context. 
The online calibration technique is adapted to consider scarce data as well as uncertainty in undelying asset price. 
The results of several numerical tests with market and synthetic data are reported, 
illustrating and validating the techniques presented here. 
More precisely, we test the reliability of the transformation of American call option prices 
into European ones by using implied volatilities and the trinomial-tree model presented in~\citep{crepey2} 
when calibrating local volatility. We also test the robustness of 
the online calibration method with scarce data, and the introduction of the asset prices in the set of unknowns.

The plan for this article is as follows.
In Section~\ref{sec:forward2} we introduce the direct (or forward) problem. 
Section~\ref{sec:discrete} is devoted to the inverse problem under a discrete setting. 
Some theoretical results required for our model calibration are collected here as well. 
Section~\ref{sec:calibration} describes details of our numerical algorithms, including
the discretization and optimization schemes, the choice of regularization functional,
the treatment of American options through European ones, and the introduction of specific
a priori information. 
Numerical experiments are performed in Section~\ref{sec:numerics}, using both synthetic
and real data, and addressing the questions of adjusting the underlying asset price and
assessing the online approach. In addition, we evaluate exotic options with a calibrated local volatility surface.
Conclusions are drawn in Section~\ref{sec:future}.

\section{Pricing Problem}\label{sec:forward2}
We start by defining the dynamics followed by futures prices. 
The corresponding version of the Dupire equation for pricing European call options on such futures 
is then presented. After some technical adaptations we show that, for a fixed time, 
option prices on futures with different maturities satisfy the same partial differential equation. 

\subsection{The Term-Structure Model} 

We consider the risk-neutral filtered probability space $(\Omega,\mathscr{U},\mathbb{F},\mathbb{Q})$, 
with $\mathbb{F} = \{\mathcal{F}_t\}_{t \geq 0}$ a filtration.
Assume that commodity futures contracts are positive-valued $\mathbb{Q}$-martingales, 
and let $F_{t,T}$ denote the commodity future price at time $t \geq 0$, maturing at $T \geq t$. 
Its corresponding spot price at time $t$, $F_{t,t}$, is denoted by $S_t$. 
The relation between future and spot prices is given by the well-known expression \citep{black,geman}
\begin{equation}
F_{t,T} = \mathbb{E}^{\mathbb{Q}}[S_T~|~\mathcal{F}_t].
\label{eq:1}
\end{equation}
Let us assume that, for each fixed maturity $T$, the dynamics of the log-future price $y_{t,T} = \log(F_{t,T}/F_{0,T})$, is independent of $T$, i.e., $y_{t,T} = y_t$, and it is given by
\begin{equation}
dy_t = -a(S_0;t,y_t)dt + \sqrt{2a(S_0;t,y_t)}dW^{\mathbb{Q}}_t
\label{com:eq1}
\end{equation}
for $t>0$, with $S_0 = F_{0,0}$, and $a$ a bounded and positive function. 
The process $W^{\mathbb{Q}}$ is a Brownian motion under the risk-neutral measure $\mathbb{Q}$. 

Since
\begin{equation}
 F_{t,T} = F_{0,T} \text{e}^{y_t},
\end{equation}
by It\^o's formula,
\begin{equation}
\frac{dF_{t,T}}{F_{t,T}} = \sqrt{2a(S_0;t,\log(F_{t,T}/F_{0,T}))}dW^{\mathbb{Q}}_t,
\label{com:eq2}
\end{equation}
for $0 < t \leq T <\infty$, with $F_{t,t} = S_t$, $\forall ~t \leq T$, and $F_{0,T}$ known. 

Under this framework, the term structure of the futures is completely determined by the 
initial curve of prices $T\mapsto F_{0,T}$ and the family of local volatility surfaces, defined by: 
$$T\mapsto 
\sqrt{2a(S_0;t,\log(F_{t,T}/F_{0,T}))}.$$

\subsection{Pricing European Call Options}

Let $C(t,F_{t,T},T^\prime,K)$ denote the price of a European call option at time $t \geq 0$, 
not discounted by the interest rate $r$, on the future $F_{t,T}$, 
with maturity $T^\prime$ such that $ t\leq T^\prime \leq T$, and strike $K > 0$. 
Fixing the current time by setting $t = 0$, the current future price is $F_{0,T}$. 

Following~\citep{dupire,volguide}, we must apply the Fokker-Planck equation to the pseudo-probability density 
of the future price at the option's maturity and consider call option prices as functions of $T^\prime$ and $K$. 
Then $C(T^\prime,K)$ satisfies the following initial-boundary value problem:

\begin{equation}
\left\{
\begin{array}{rcl}
\displaystyle\frac{\partial C}{\partial T^\prime}(T^\prime,K) &=&a\left(S_0;T^\prime,\log\left(\displaystyle\frac{K}{F_{0,T}}\right)\right)K^2\displaystyle\frac{\partial^2 C}{\partial K^2}(T^\prime,K), \quad 0 < T^\prime \leq T, ~ K>0,\\
\\
C(T^\prime = 0,K) &=& \max\{0,F_{0,T} - K\}, \quad K>0,\\
\\
\displaystyle\lim_{K\rightarrow 0} C(T^\prime,K) &=& F_{0,T}, \quad 0 < T^\prime \leq T,\\
\\
\displaystyle\lim_{K\rightarrow +\infty} C(T^\prime,K) &=& 0, \quad 0 < T^\prime \leq T.
\end{array}
\right.
\label{dup1}
\end{equation}

In practice, on each commodity future there is only one vanilla option maturity. 
For example, Light Sweet Crude Oil (WTI), Heating Oil (HO) and Gasoline (RBOB) vanilla options generally 
expire three business days before the maturity of the corresponding underlying futures, 
whereas Henry Hub natural gas vanilla options expire one business day before maturity. 
For more examples and the contractual details see the CME website (\url{www.cmegroup.com}). 
Hence, it is not possible to find a surface of option prices related to a fixed future,
unlike in the case of equity markets. 

Let us normalize and re-parameterize option prices in order to eliminate the explicit dependence on the underlying futures. So, the resulting option prices on futures with different maturities satisfy the same PDE problem. 

Define $\widetilde{C}(T^\prime,K) := C(T^\prime,K)/F_{0,T} \text{  for every } T > 0 \text{  and  } K >0, \text{   with   } T^\prime = T^\prime(T).$ 
The coefficients in the PDE problem~\eqref{dup1} are
not dimension-less, and its diffusion coefficient is unbounded. 
Hence, we make the standard change of variables
$$
\tau := T^\prime ~(\text{time to maturity}), \quad y := \log(K/F_{0,T}) ~(\text{log-moneyness}),
$$
and define
$$
v(\tau,y) = \widetilde{C}(\tau,F_{0,T}\exp(y)).
$$

\noindent Then $v = v(\tau,y)$ satisfies the differential problem
\begin{equation}
\left\{
\begin{array}{rcl}
\displaystyle\frac{\partial v}{\partial \tau}(\tau,y) &=&\displaystyle a(S_0;\tau,y)\left(\displaystyle\frac{\partial^2 v}{\partial y^2}(\tau,y) - \frac{\partial v}{\partial y}
(\tau,y)\right), \quad 0 < \tau \leq T, ~ y \in \mathbb{R},\\
\\
v(\tau = 0,y) &=& \max\{0,1 - \exp(y)\}, \quad y \in \mathbb{R},\\
\\
\displaystyle\lim_{y\rightarrow -\infty} v(\tau,y) &=& 1,\quad \tau>0,\\
\\
\displaystyle\lim_{y\rightarrow +\infty} v(\tau,y) &=& 0, \quad \tau>0.
\end{array}
\right.
\label{dup2}
\end{equation}

The resulting initial-boundary value problem~\eqref{dup2} no longer depends explicitly on $F_{0,T}$ 
and is defined for any $\tau >0$. Further, it is more amenable to a uniform discretization than~\eqref{dup1}. 
However, note that in cases where there is uncertainty in $F_{0,T}$, 
there is also corresponding uncertainty in the new independent variable $y$;
see~\citep{AlbAscYanZub2015}.

The problem~\eqref{dup2} is defined on the domain $D = \R_+\times \R$. 
The inverse problem consists of finding its diffusion coefficient $a$ in such a manner that explains data
observed on $v$ at some points in $D$.

Let $a_1,a_2 \in \R$ be scalar constants such that $0 < a_1 \leq a_2 < +\infty$. 
We also consider a fixed, continuous and bounded function $a_0 = a_0(\tau,y)$, 
such that $\partial_y a_0, \partial_\tau a_0 \in L^2(D)$ and $a_1 \leq a_0(\tau,y) \leq a_2$ 
for almost every $(\tau,y) \in D$ and every $S_0$.

\noindent Define the set
\begin{equation}
Q:=  \{a \in a_0 + \he : a_1\leq a \leq a_2\},
\label{domopdi}
\end{equation}
where $\varepsilon > 0$. 
Existence and uniqueness results for the solution of the differential
problem~\eqref{dup2} with $a \in Q$ can be found in \citep{crepey,acpaper,eggeng,isakov}.


In order to allow the inclusion of more information in the calibration, we adapt the {\em online} 
model introduced in~\citep{vvla2} to the case of scarce data. 
So, instead of considering the solution of~\eqref{dup2} at a unique time instant, 
we index it by the commodity spot price or the closest to maturity future, after sorting in ascending order. 
Assuming that the local volatility surface has a well-behaved dependency on such index, 
we have the so-called {\em online} setting, which is defined by the map
$$
\begin{array}{rcl}
 F: \Q \subset \X & \longrightarrow & \Y \\
         \af\in \Q & \rightarrow & \vf(\af) - \vf(\af_0) \in \Y,
\end{array}
$$
where $\vf(\af): s \mapsto v(a(s))$ denotes the family of solutions of problem~\eqref{dup2}, 
with the corresponding diffusion coefficients $\af: s\mapsto a(s)$, and the index $s \in [0,\overline s]$. 
For further technical details concerning this online approach,  see~\citep{vvla2}.

\section{The Discrete Calibration Problem}\label{sec:discrete}
\subsection{Basic Setup}\label{sec:nounderlying}
Given a set of market European call option prices, we
want to identify the corresponding local volatility surface, 
assuming these prices were generated by the
initial-boundary value problem~\eqref{dup2}. 
Thus, within
the {\em online} setting, we consider a set of prices $\tvf$, such that $\tvf - \vf(\af)$ is in the range of the forward operator $\mathcal{R}(F)$, 
and search for the family of local volatility surfaces $\af^\dagger$ in $\Q$, satisfying
\begin{equation}
\tvf = \vf(\af^\dagger).
\label{ip1a}
\end{equation}
Since the operator $F$ is injective, there exists only one family of local volatility surfaces 
satisfying~\eqref{ip1a}. However, assuming that $\tvf - \vf(\af)\in \mathcal{R}(F)$ is unrealistic, 
since market prices are scarce and subject to noise. 
Moreover, the compactness of $F$ implies that the local volatility calibration problem cannot be directly solved. 
To be specific, we assume that the normalized price data, denoted  by $\vdd$, is related to $\tvf$ by
\begin{equation*}
\|\vdd - P\vf(\af^\dagger)\| \leq \delta ,
\end{equation*}
where $\delta$ is called noise level and $P$ is the projection of $\vf(\af^\dagger)$ onto some observation mesh.

Since the data is finite-dimensional, and the differential problem requires discretization to enable
a computer solution, we define the calibration problem under a discrete setting.
\begin{df}
Let $\{X_m\}_{m\in\N}$ and $\{Y_n\}_{n\in\N}$ be sequences of finite dimensional subspaces of $\X$ and $\Y$, respectively, satisfying
$$
\begin{array}{c}
X_m \subset X_{m+1}\subset ... \subset \X,\quad Y_n \subset Y_{n+1}\subset ... \subset \Y,\\
\\
\overline{\cup_{m\in\N}X_m} = \X, ~\text{  and  }~ \overline{\cup_{n\in\N}Y_n} = \Y.
\end{array}
$$
Define the finite-dimensional domains $\Q_m := \Q \cap X_m$, and assume that $\Q_m \not= \emptyset$ for every $m\in \N$.
\label{def3}
\end{df}

Let us consider some discrete approximation of the operator $F$ in the subspace $Y_n$:
$$
F_n : \Q \subset \X \longrightarrow Y_n.
$$
Let $P_n$ be the projection of $\vf(\af)$ onto some set of observation meshes, so $F_n(\af) = P_n\vf(\af) - P_n\vf(\af_0) \in Y_n$. 

The problem of minimizing $\|P_n\vf(\af) - \vdd\|$ in the discrete $2$-norm is typically underdetermined
and has infinitely many solutions. To obtain a locally unique solution we must add a priori information 
(or, a prior in a Bayesian framework),
and we do that using a Tikhonov regularization setting. Thus, we add to the objective function a penalty functional $\psi_{\af_0}(\af)$
scaled by a positive parameter $\alpha$.
So, the calibration problem considered is:
\begin{problem}
 Find a minimizer in $\Q_m$ of the functional
\begin{equation}
\mathcal{F}(\af) = \|P_n\vf(\af) - \vdd\|^2 + \alpha \psi_{\af_0}(\af).
 \label{tik2}
\end{equation}
\label{prob:1}
\end{problem}
Assume that $\psi_{\af_0}$ in \eqref{tik2} is convex, coercive and weakly lower semi-continuous. 
The existence and stability of minimizers, denoted $\af^{\delta,\alpha}_{m,n}$, 
follows from Theorems~3.22 and~3.23 in \citep{schervar}, respectively. 

If the data is not scarce, the noise is Gaussian with the covariance matrix being a scaled identity, and assuming that $\psi_{\af_0}(\af) = 0$ if and only if $\af = \af_0$, the choice of the discretization level $m$ and the regularization parameter $\alpha$ in \eqref{tik2}, 
for fixed $n$ and $\delta$, can be made with Morozov-like discrepancy principles. See \citep{acz2013b,acz2013a},
which contain further results regarding
convergence and convergence-rate with respect to $\delta$ and $n$.

We emphasize that in practice the challenge is often in choosing an appropriate model for the noise distribution
and an appropriate representation of available a priori information through the Tikhonov-type regularization term.
Numerical experiments with this scarce version of the online setting can be found in 
Sections~\ref{sec:synthetic}, \ref{sec:hh} and \ref{sec:wti}.


\subsection{Underlying Asset Prices as Additional Unknowns}\label{sec:underlying}
The underlying future prices form a random vector, which means that the observation of its entries is uncertain. 
Therefore, since they are used to normalize option prices and to define the log-moneyness meshes, 
including them  in the set of unknowns would improve calibration. Denoting the future prices vector by $\mathscr F$, it follows that $P_n = P_n(\mathscr F)$. So, Problem~\ref{prob:1} is replaced by:
\begin{subequations}
\begin{equation}
 \text{Find } (\af^{\delta,\alpha}_{m,n};\mathscr F) \in \text{argmin}\left\{\|P_n(\mathscr F) \vf(\af)-\vdd\|^2 + \psi_{\af_0}(\af;\mathscr F)\right\},
 \label{eq:tikscar}
\end{equation}
where
\begin{multline}
 \psi_{\af_0}(\af;\mathscr F) = \alpha_1\sum_{l=0}^L\|a(s_l)-a_0(s_l)\|^2 + \alpha_2\sum_{l=0}^L\|\partial_{y,m} a(s_l)\|^2 + \alpha_3\sum_{l=0}^L\|\partial_{\tau,m} a(s_l)\|^2 +\\ \alpha_4\sum_{l=0}^L\|q(\mathscr F(s_l),s_l) - q(\hat{\mathscr F}(s_l),s_l)\|^2 +\alpha_5\|\mathscr F - \hat{\mathscr F}\|^2 +\frac{\alpha_6}{\Delta s^2}\sum^L_{l=1}\|a(s_l) - a(s_{l-1})\|^2.
 \label{eq:psi}
 \end{multline}
Here $q(\mathscr F(s_l),s_l)$ represents the boundary and initial conditions for each $s_l$ (the discrete version of the index $s$), taking $\mathscr F(s_l)$ into account in the definition of the log-moneyness mesh, and $\hat{\mathscr F}$ is the observed future price vector. 
In addition, $\partial_{y,m}$ and $\partial_{\tau,m}$ denote the first order forward difference matrices approximating the derivative with respect to $y$ and $\tau$, respectively. When the dependence on $\mathscr F$ is ignored, as it is in Section~\ref{sec:nounderlying}, we set $\alpha_4 = \alpha_5 = 0$.
\label{9}
\end{subequations}%

Since $a$ and $\mathscr F$ are independent, we split the minimization as follows:
\begin{enumerate}
\item For a fixed set of underlying assets $\mathscr F$, we find a minimizer for~\eqref{eq:tikscar}. 
Since $\mathscr F$ is fixed, the term in \eqref{eq:psi} with $\alpha_4$ does not affect the 
minimization and is ignored.

\item Given a local volatility surface, we minimize~\eqref{eq:tikscar} with respect to the underlying asset 
(or assets). Since now $a$ is constant, the terms in~\eqref{eq:psi} with $\alpha_1$, $\alpha_2$ and $\alpha_3$ 
do not affect the minimization and are ignored.
\end{enumerate}  
We repeat steps 1 and 2 a few times until a tolerance for the fixed point process is reached. 
For more details, see~\citep{AlbAscYanZub2015}.

Section~\ref{sec:adjust} presents a numerical example of this calibration method with synthetic data.

\section{Model Calibration: Algorithms and Extensions}
\label{sec:calibration}

{In this section we first describe the numerical discretization of the Cauchy problem~\eqref{dup2},
followed by our optimization strategy for the problem~\eqref{9}, 
written more compactly as~\eqref{tik2}. 
We then describe and demonstrate} a technique to transform American call option prices into European ones, 
based on the calibration of Black's implied volatilities.
{The last subsection is concerned with improving the quality of the a priori information
inserted into the optimization formulation~\eqref{9}}.  

\subsection{Numerical {Discretization} of the Calibration Problem}
\label{sec:onlinenum}

Let us first approximate the initial-boundary value problem~\eqref{dup2} by restricting the range in
log-moneyness $y$ to be $[-5,5]$, thus defining the domain $D = [0,\tau_{\max}]\times[-5,5]$. 
The boundary conditions at $y \rightarrow \pm \infty$ are then imposed at $y = \pm 5$, respectively.
 
Let $I,J,L \in \N$ be fixed. We consider the discretization of $[0,\overline{s}]\times D$ with the notation $s_l = l\Delta s$, with $l = 0,1,2,...,L$, $\tau_i = i \Delta \tau$, with $i = 0,1,2,...,I$, 
and $y_j = j \Delta y$, with $j = -J,-J+1,...,0,1,...,J$. 
Let us denote $v^{i}_{j}(l):= v(s_l;\tau_i,y_j)$, $a^i_j(l) := a(s_l;\tau_i,y_j)$,  
$\beta:= \Delta \tau/\Delta y$ and $\eta = \Delta \tau/\Delta y^2$. 
The differential equation in~\eqref{dup2} is discretized by the Crank-Nicolson scheme 
\begin{eqnarray}
& &v^{i}_{j} - \displaystyle\frac{1}{2}\eta a^{i}_{j}(v^{i}_{j+1} - 2v^{i}_{j} +v^{i}_{j-1}) + \frac{1}{4}\beta a^{i}_{j}(v^{i}_{j+1} - v^{i}_{j-1}) = \nonumber \\ 
& &v^{i-1}_{j} + \displaystyle\frac{1}{2}\eta a^{i-1}_{j}(v^{i-1}_{j+1} - 2v^{i-1}_{j} +v^{i-1}_{j-1}) - \frac{1}{4}\beta a^{i-1}_{j}(v^{i-1}_{j+1} - v^{i-1}_{j-1}),
\label{cns}
\end{eqnarray}
where the dependence on $l$ is omitted to ease the notation. 
Note that, for this specific finite-difference scheme, the continuity of the forward operator implies continuity of its discrete version.

For each $l$ the scheme \eqref{cns} can be written in {matrix form as
\begin{equation}
 [Id + M(a^{i})]v^{i} + b(a^{i}) = [Id - M(a^{i-1})]v^{i-1} - b(a^{i-1}), \qquad i=1,...,I,
 \label{mcns}
\end{equation}
where
\begin{eqnarray*}
 a^{i} &=& (a^i_{-J+1},...,a^i_0,...,a^i_{J-1})^T, \\
 v^{i} &=& (v^i_{-J+1},...,v^i_0,...,v^i_{J-1})^T,\\
 \left[M(a^{i})\right]_{j,j} &=& \eta a^{i}_j,\qquad j = -J+1,...,0,J-1,\\
 \left[M(a^{i})\right]_{j,j+1} &=& \frac{1}{2}\left(\frac{1}{2}\beta-\eta\right) a^{i}_j,\qquad j = -J+1,...,0,J-1,\\
 \left[M(a^{i})\right]_{j,j-1} &=& -\frac{1}{2}\left(\frac{1}{2}\beta+\eta\right) a^{i}_j\qquad j = -J+1,...,0,J-1,\\
 b(a^{i}) &=& \left(-\frac{1}{2}\left(\frac{1}{2}\beta+\eta\right)a^{i}_{-J},0,...,0\right)^T,
\end{eqnarray*}
and denoting the identity matrix by $Id$.}

Replacing $\vf(\af)$ in \eqref{tik2} by the family indexed by $l$ of solutions of \eqref{mcns}, we solve the corresponding minimization problem by a gradient descent method. Thus, denoting the $k$th
iterate of $\af$ by $\af^{k}$ we evaluate
\begin{equation}
 \af^{k+1} = \af^{k} - \lambda_k \nabla \mathcal{F}(\af^{k}),
 \label{eq:gradient}
\end{equation}
until a stopping criterion, 
defined in Section~\ref{sec:choice}, is satisfied. The resulting family of volatility surfaces is taken as the regularized solution. 

Note that
$$\nabla \mathcal{F}(\af) = \nabla J^\delta(\af) + \nabla \psi_{\af_0}(\af),$$
where
\begin{equation}
J^\delta(\af) :=  \|P_n\vf(\af) - \vdd\|^2_{Y_n} \approx \sum_{l=0}^L\sum_{i=1}^{I-1} \|P_{n,l}v^i(l;a(l)) - v^{\delta,i}(l) \|^2.
\label{j}
\end{equation}
We set the step size $\lambda_k$ in \eqref{eq:gradient} as
\begin{equation}
\lambda_k = \min\left(2.5,~\displaystyle\frac{J^\delta(\af^k)}{2\|\nabla J^\delta(\af^k)\|^2}\right).
 \label{eq:step}
\end{equation}
This corresponds to 
a line search algorithm based on the Wolfe conditions 
(see, e.g.,~\citep{nocedal}). The formula 
controls the step size taking into account the fact that, 
if $\|\nabla J^\delta(\af^k)\|$ is small, the method may become unstable. 

In what follows, we omit dependence on $a$ and $l$ as much as possible to ease notation. 
For each $l=0,1,...,L$, let us consider the adjoint state $u$, solution of the adjoint equation
\begin{equation}
 [Id + M(a^{i+1})^T]u^{i}  = [Id - M(a^{i+1})^T]u^{i+1} + 2\left( P_{n,l}^*P_{n,l}\hat v^{i} - P_{n,l}^*\hat v^{\delta,i} \right),~ i=I-1,...,0,
 \label{eq:adj}
\end{equation}
with $u^I = 0$.

Let $a:b$ denote the product of two vectors $a $ and $b$ giving a vector with entries $a_jb_j$, $j=-J+1,...,0,...,J-1$, and 
$\partial_{y,m}$ and $\partial_{yy,m}$ the first order and second order centered difference matrices. 
Thus, for each $i=1,...,I-1$, and $l = 0,...,L$,
\begin{equation}
 \nabla J^\delta(l;\af)^i \approx -\Delta \tau [\partial_{yy,m} - \partial_{y,m}] v^{i}: u^{i} + [\partial_{yy,m} - \partial_{y,m}] v^{i}: u^{i+1}.
\end{equation}

\subsection{The Choice of the Parameters in the Regularizing Functional}
\label{sec:choice}

One of the main challenges in Tikhonov-type regularization is the appropriate choice 
of the weights in the regularizing functional. In the specific case of \eqref{eq:psi}, 
we choose the parameters heuristically, taking into account the data and our belief 
as to how the local volatility surface must look like. So, we have the following:

\begin{enumerate}
 \item The weight $\alpha_1$ states how much the {\em a priori} surface $a_0$ {should} be taken into account, 
 i.e., how accurate we think the first approximation $a_0$ is. 
 For example, if the {\em a priori} surface $a_0$ is constant, 
 it is away from the expected solution, so $\alpha_1$ should be small. 
In the following examples we take it as $0.1\alpha_2$ or $0.01\alpha_2$.
 
 \item The weights $\alpha_2$ and $\alpha_3$ allow us to state the relative importance of 
 the smoothness of the local volatility surface with respect to $\tau$ and $y$. 
 Since the data is much scarcer in $\tau$ than in $y$, 
we take $\alpha_2 > \alpha_3$. 
In the following examples we assume that $\alpha_3 = 0.5\alpha_2$, $0.1\alpha_2$, $0.02\alpha_2$ or $0.01\alpha_2$, 
and $\alpha_2 = 10^{-2}$, $10^{-3}$, or $10^{-4}$.

 \item The weight $\alpha_6$ allows us to penalize lack of sufficient smoothness of a family of 
 local volatility surfaces with respect to the index $s$.
 We set $\alpha_6 = c\Delta s^2/\Delta y^2\alpha_2$, with $c = 0.01$ or $0.1$. 
This enforces {the restriction that
$\| \frac{\partial \af}{\partial s} \|$ 
should not have more weight 
than $\| \frac{\partial \af}{\partial y} \|$ 
in the regularizing functional.}

 \item The weights $\alpha_4$ and $\alpha_5$ are related to the adjustment of the underlying asset prices and
should have the same weight as the data misfit $\|P\vf(\af) - \vdd\|^2$.
\end{enumerate}

The parameter $\alpha_2$ is a scaling factor that states how much the inverse problem must be regularized. 
So, we choose $\alpha_2$ in terms of a given tolerance;
specifically, we choose the largest $\alpha_2$ such that the normalized data misfit 
 \begin{equation}
R(\mathcal A) = \|P\vf(\af)-\vdd\|/\|\vdd \|
\label{eq:residual2}
\end{equation}
is below a tolerance, say $0.02$. To find an appropriate tolerance, we set $\alpha_2 = 10^{-4}$ and first set the tolerance at $0.01$. 
If the produced reconstructions are too noisy,
or the relative changes in the normalized data misfit become small, 
for example below $10^{-6}$, we increase the tolerance. 
However, the tolerance must always be 
approximately $1\%$ of 
$\| \vdd \|$.

The same tolerance on the normalized data misfit used to determine $\alpha_2$ is then set in the 
algorithm's stopping criterion.

\subsection{From American into European Prices}
\label{sec:transformation}

Under the risk-neutral measure, commodity futures have zero drift, although the interest rate is positive. 
Hence, future prices can be seen as a dividend-paying asset, implying that its American call options are more expensive than the European ones. 
Therefore, we must either use an American pricing technique in the calibration of local volatility 
or transform American into European prices. We choose the latter option since it is computationally cheaper, 
and the calibration using European pricing is better understood both computationally and theoretically.

In \citep{achdou1,achpi}, the authors introduced a local volatility technique to price American put options and its corresponding calibration problem, using a free-boundary value problem approach. 
In \citep{crepey2}, this problem is addressed under a discrete setting by trinomial trees. In both cases, to solve the inverse problem, the direct problem must be solved for each strike and maturity. In contrast, in the European pricing setting, the direct problem must be solved only once.

So, to evaluate the European prices, we extract the implied volatilities $\sigma_{\text{I}}$ of the American options 
using a nonlinear least square technique, i.e., find
$$
\sigma_{\text{I}} \in \argmin\{|C(\sigma) - C_{\text{M}}|^2 ~:~\sigma > 0\},
$$
where $C_{\text{M}}$ denotes the market call price and $C(\sigma)$ the model price. 
In this case the evaluation of $C(\sigma)$, based on the Black model, is solved by the trinomial tree method presented in Section~8 of~\citep{crepey2}. 
Instead of using trinomial trees, we could evaluate American prices by some approximation technique, such as~\citep{baw} or~\citep{bs93,bs02}. 		

We test the accuracy of this transformation technique as follows. 
Evaluating American call option prices under the local volatility model by the trinomial tree method, we extract the corresponding implied volatilities and evaluate European prices with the Black-Scholes formula. Then, we calibrate the local volatility surface using this European data. 
Note that this transformation is a data processing stage which must be performed before the 
local volatility calibration stage.

In order to evaluate the American call prices, we take $S_0 = 1$, the interest-rate as $r = 0.03$, the dividend yield as $q = 0.03$, and the local volatility surface
\begin{equation}
\sigma(\tau,y) = \left\{
\begin{array}{ll}
\displaystyle\frac{2}{5}-\frac{4}{25}\text{e}^{-\tau/2}\cos\left(\displaystyle\frac{4\pi y}{5}\right),& \text{ if } -2/5 \leq y \leq 2/5\\
\\
2/5,& \text{ otherwise.}
\end{array} \right.
\label{vol}
\end{equation}
The call prices are evaluated at the maturities $\tau_i = i\cdot 0.1$, with $i=1,2,...,5$, and strikes  $y_j=j\cdot0.05$, with $j = -10,-9,...,0,1,...,10$.
 
We hold the future prices $\mathscr F$ and the index $s$ fixed. So, $\alpha_3=\alpha_4=\alpha_5 = 0$ in the regularizing functional $\psi_{\af_0}$ in \eqref{eq:psi}. 
The operator $P$ in the regularized problem~\eqref{tik2}, or~\eqref{9}, 
projects the numerical solution given by the 
scheme~\eqref{cns}, with discretization level $n$, onto the strikes and maturities where the data is given. In the regularizing functional~\eqref{eq:psi}, 
we have taken the parameters as $a_0\equiv 0.08$, $\alpha_1=10^{-3}\alpha_2$, and $\alpha_3 = 0.2\alpha_2$.

We set $\alpha_2 = 10^{-4}$ and use the step sizes $\Delta y = 0.05$ (for log-moneyness), and $\Delta \tau = 0.01$ to solve the 
discretized direct problem. 
The initial guess for the calibration was the constant local volatility surface $a \equiv 0.08$.  

In all the following examples the mesh widths used to evaluate the local volatility 
were the same as those used to generate the data.
In addition, to evaluate the local volatility surface in the whole domain, we use
bilinear interpolation, 
and set 
\[ a(\tau,y) = \begin{cases} a(\tau,-0.5) & {\rm if~} \tau > 0.1, \ y \leq -0.5, (\rm{deep~in~the~money}) \\ 
a(\tau,0.5) & {\rm if~} \tau > 0.1, \ y\geq 0.5, (\rm{deep~out~of~the~money}) \\
a(0.1,y) & {\rm if~} \tau \leq 0.1 . \end{cases} \]

The tolerance on the normalized data misfit is taken 
as $tol = 2.55\times 10^{-3}$. 
\begin{figure}[!ht]
\centering
\includegraphics[width=0.2\textwidth]{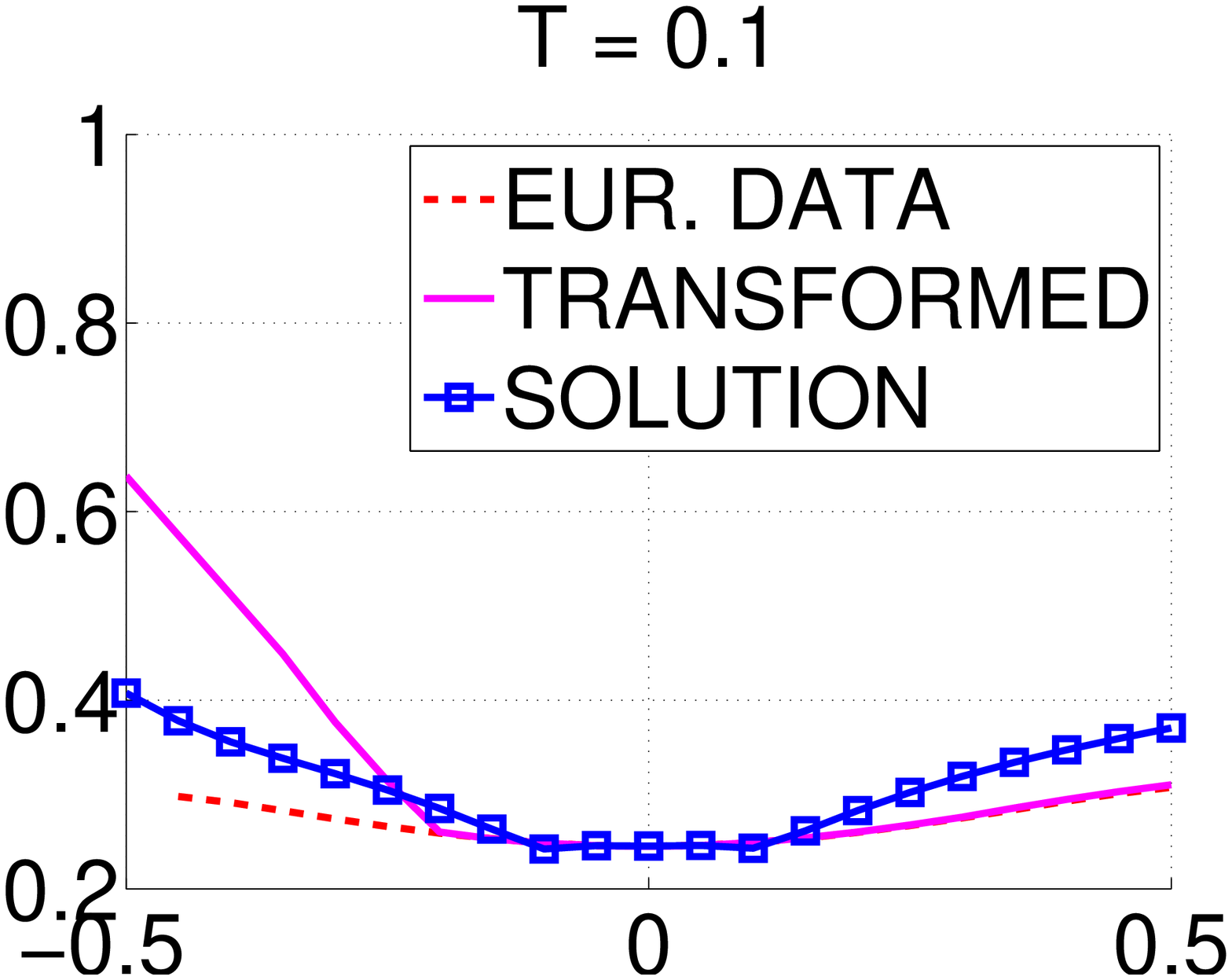}\hfill 
\includegraphics[width=0.2\textwidth]{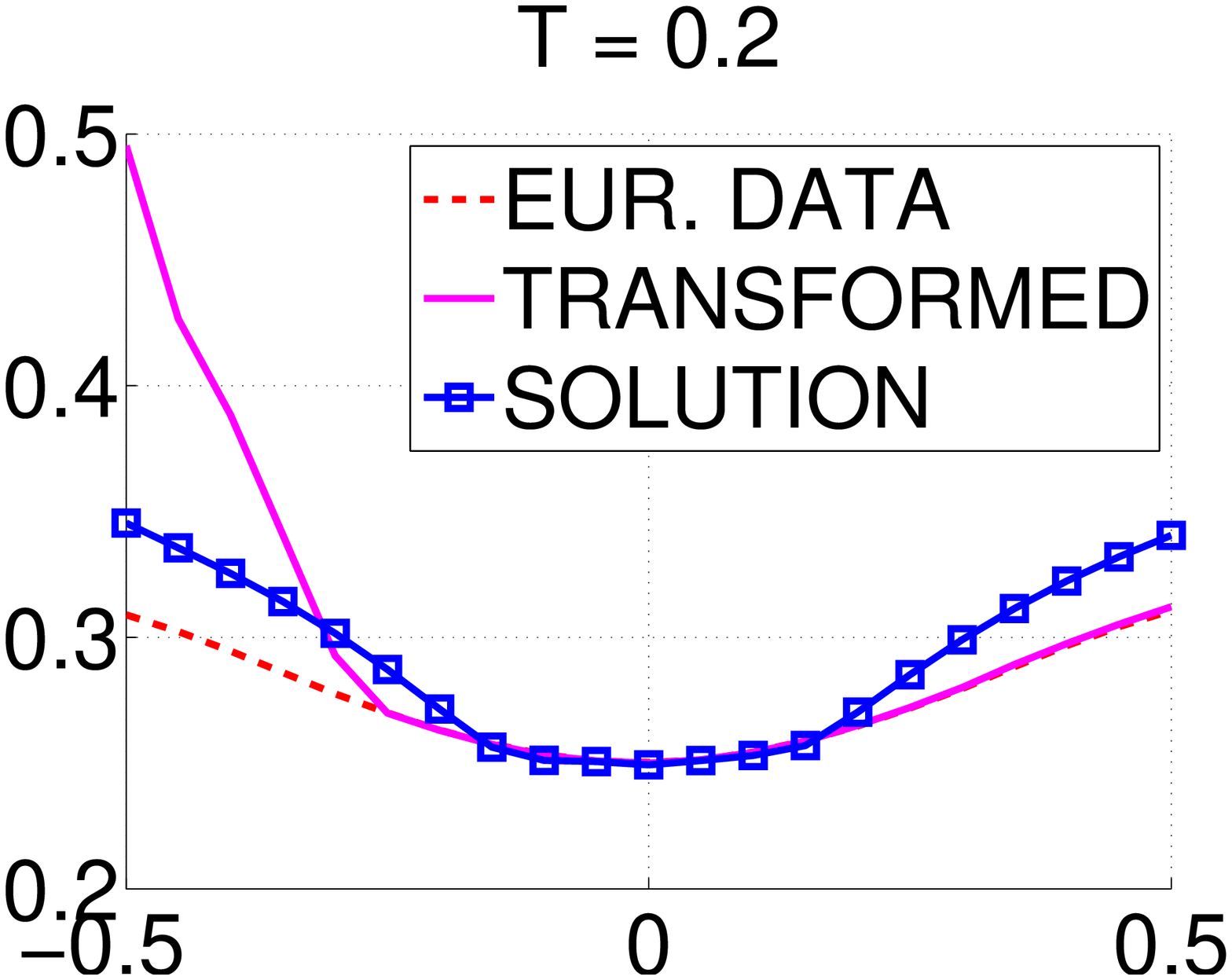}\hfill 
\includegraphics[width=0.2\textwidth]{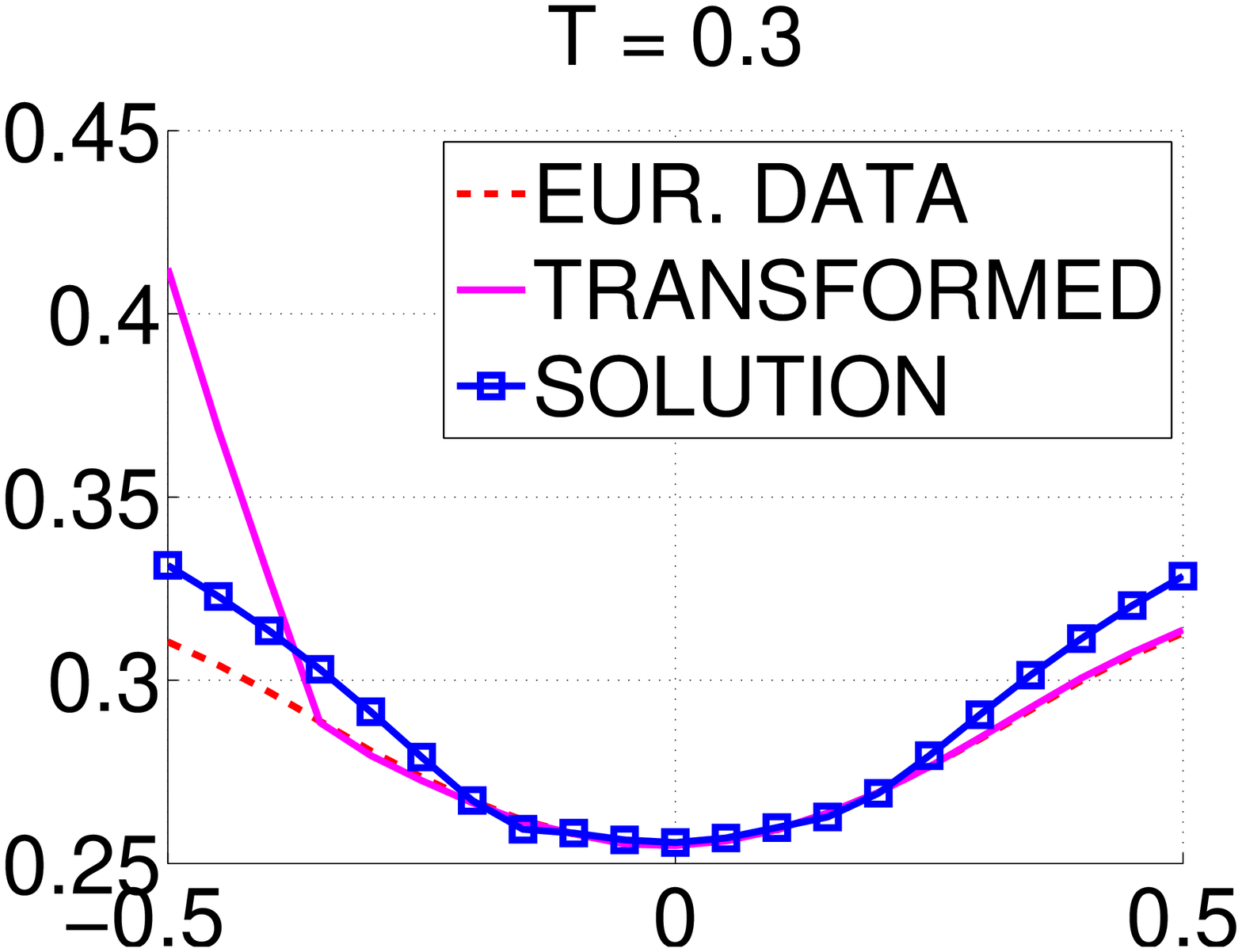}\hfill 
\includegraphics[width=0.2\textwidth]{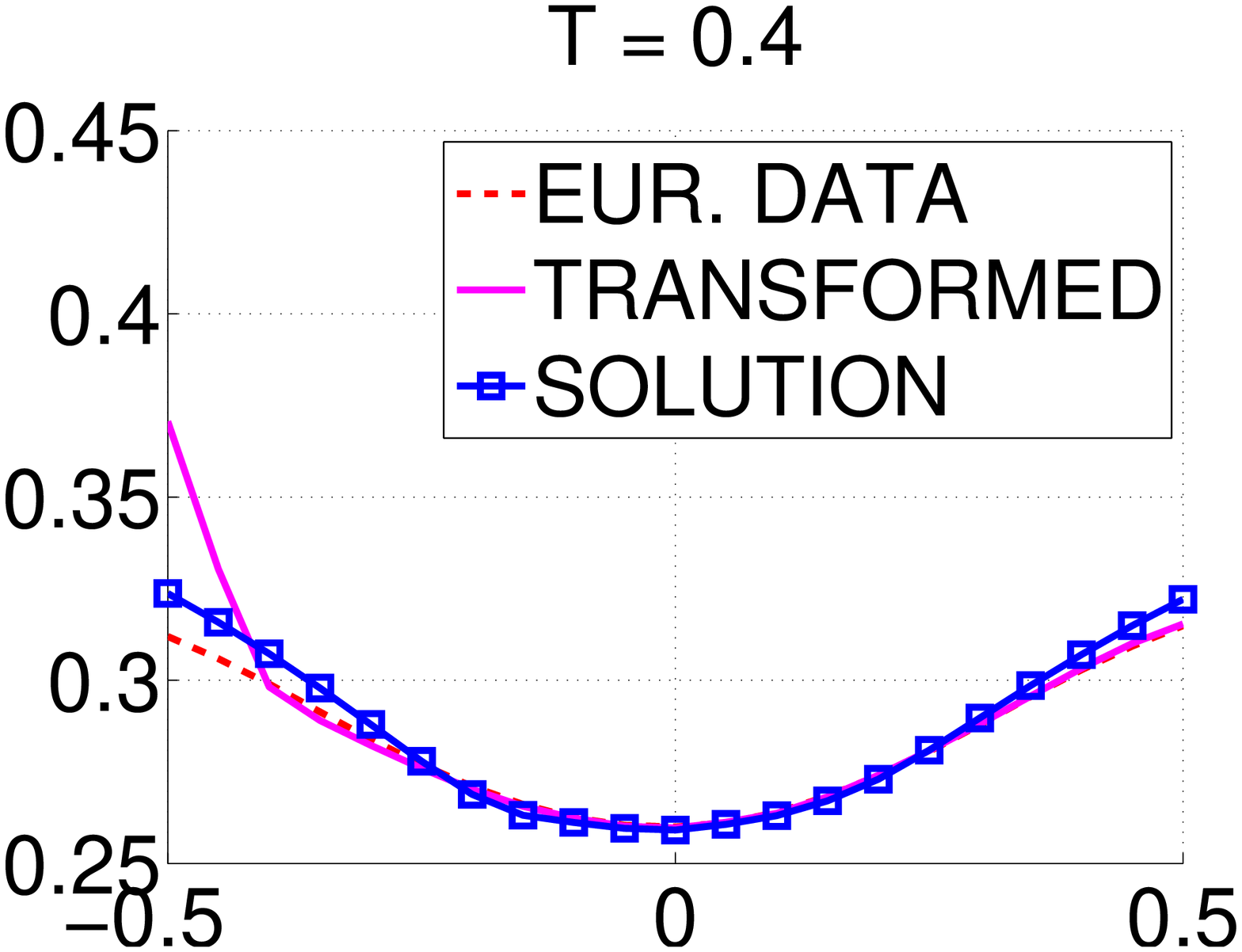}\hfill
\includegraphics[width=0.2\textwidth]{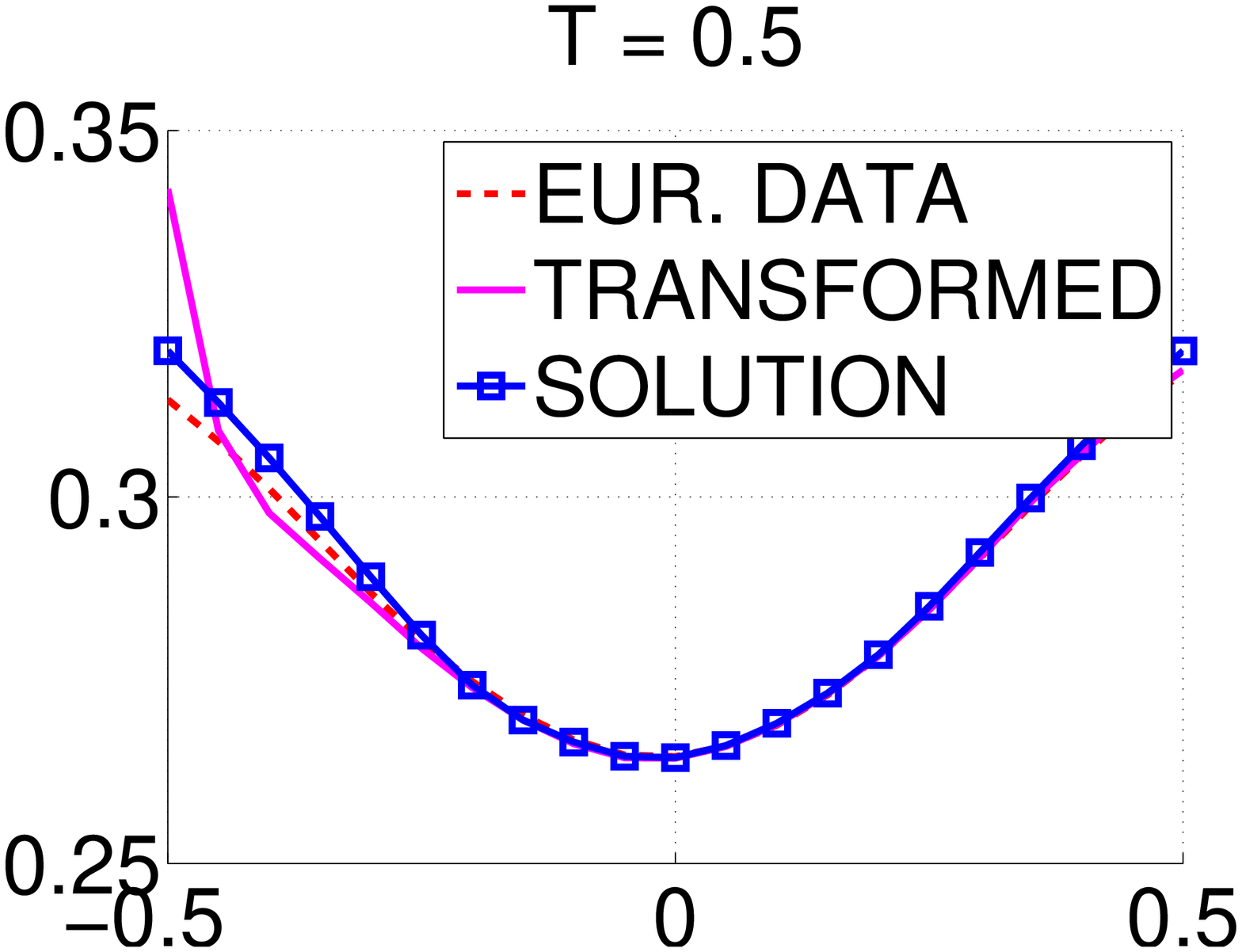}\hfill
\caption{Implied volatility of the European and American prices at each maturity.}
\label{fig1}
\end{figure}
 
\begin{figure}[!ht]
\centering
\begin{minipage}{.49\textwidth}
  \centering
\includegraphics[width=0.49\textwidth]{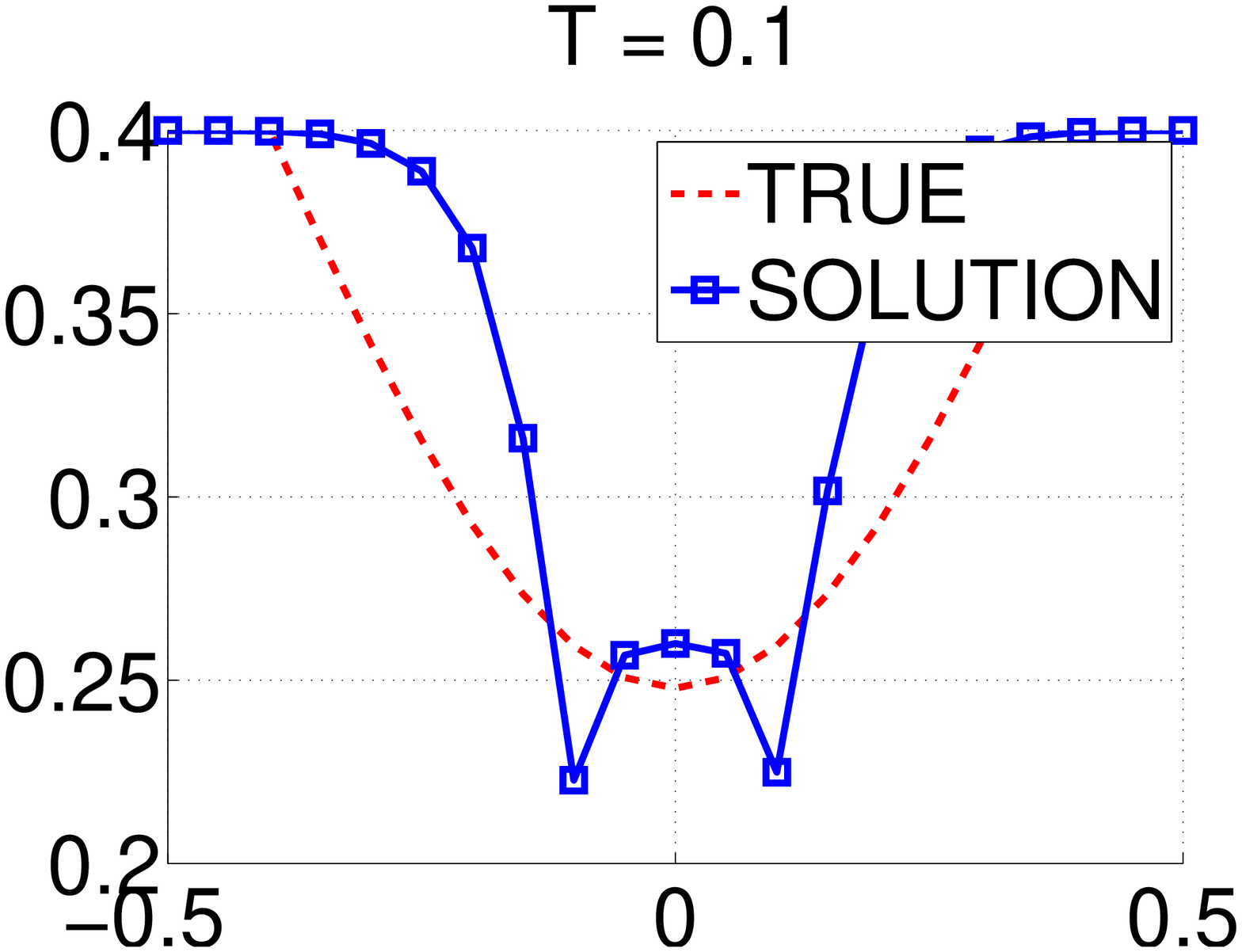}\hfill
\includegraphics[width=0.49\textwidth]{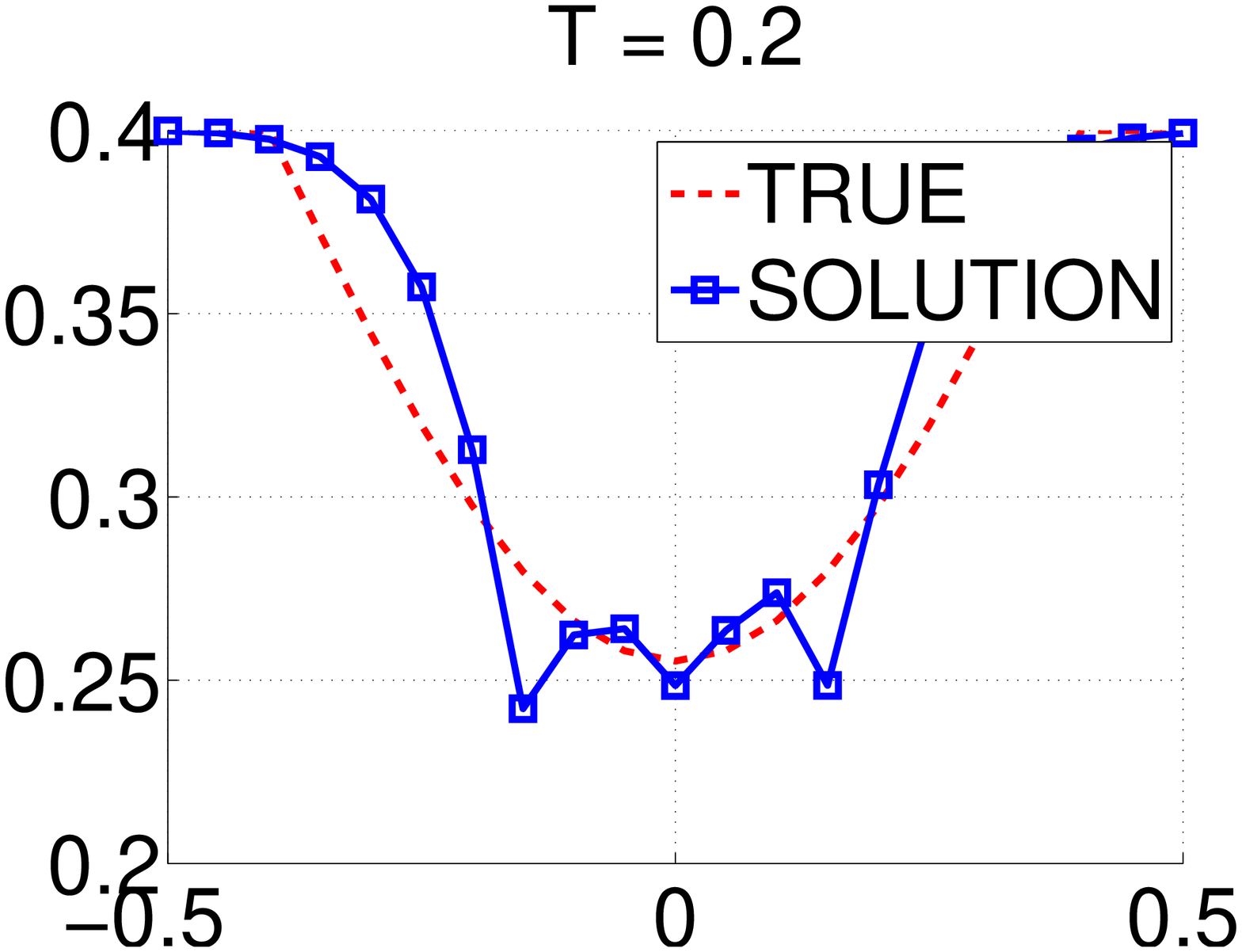}\hfill
\includegraphics[width=0.49\textwidth]{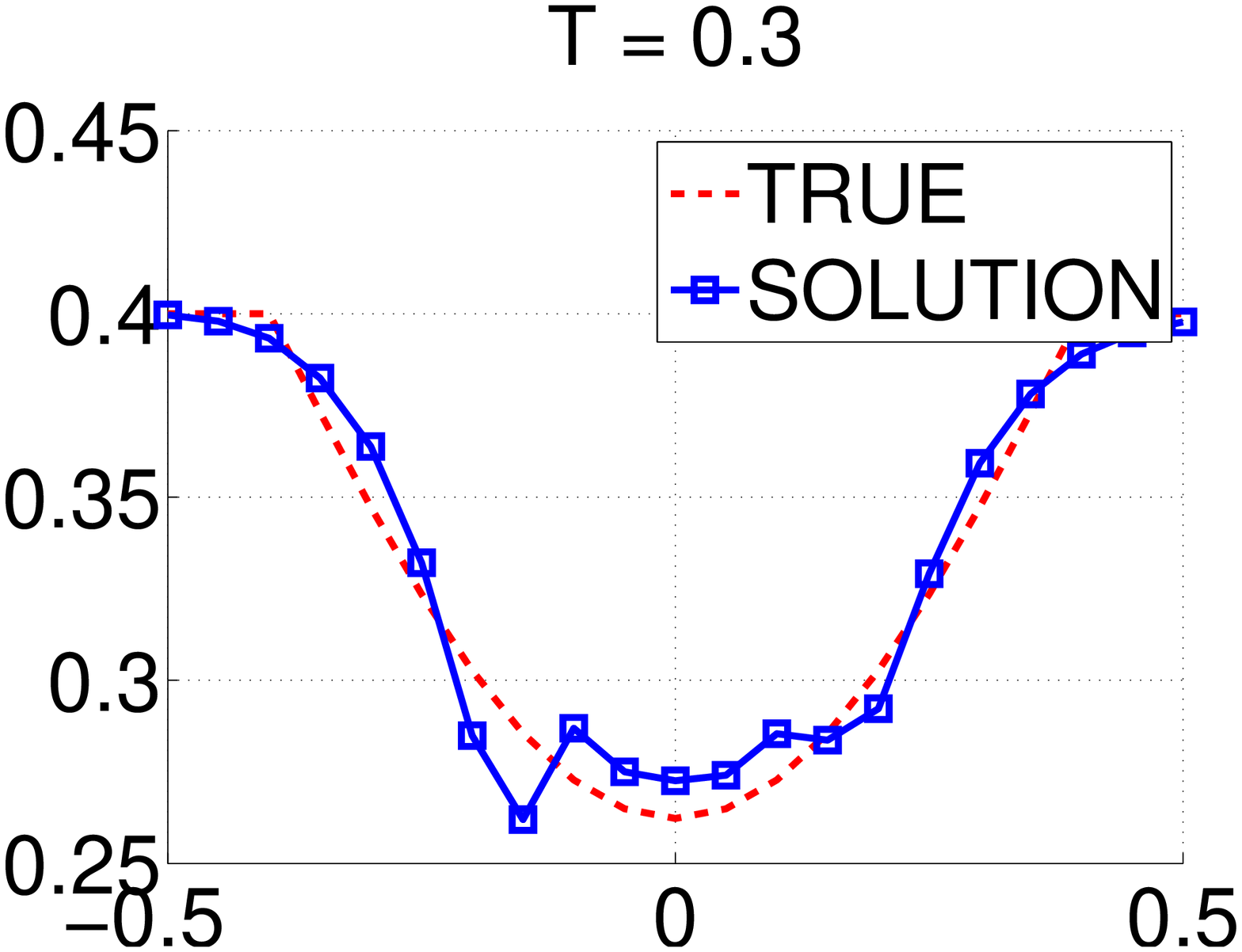}\hfill
\includegraphics[width=0.49\textwidth]{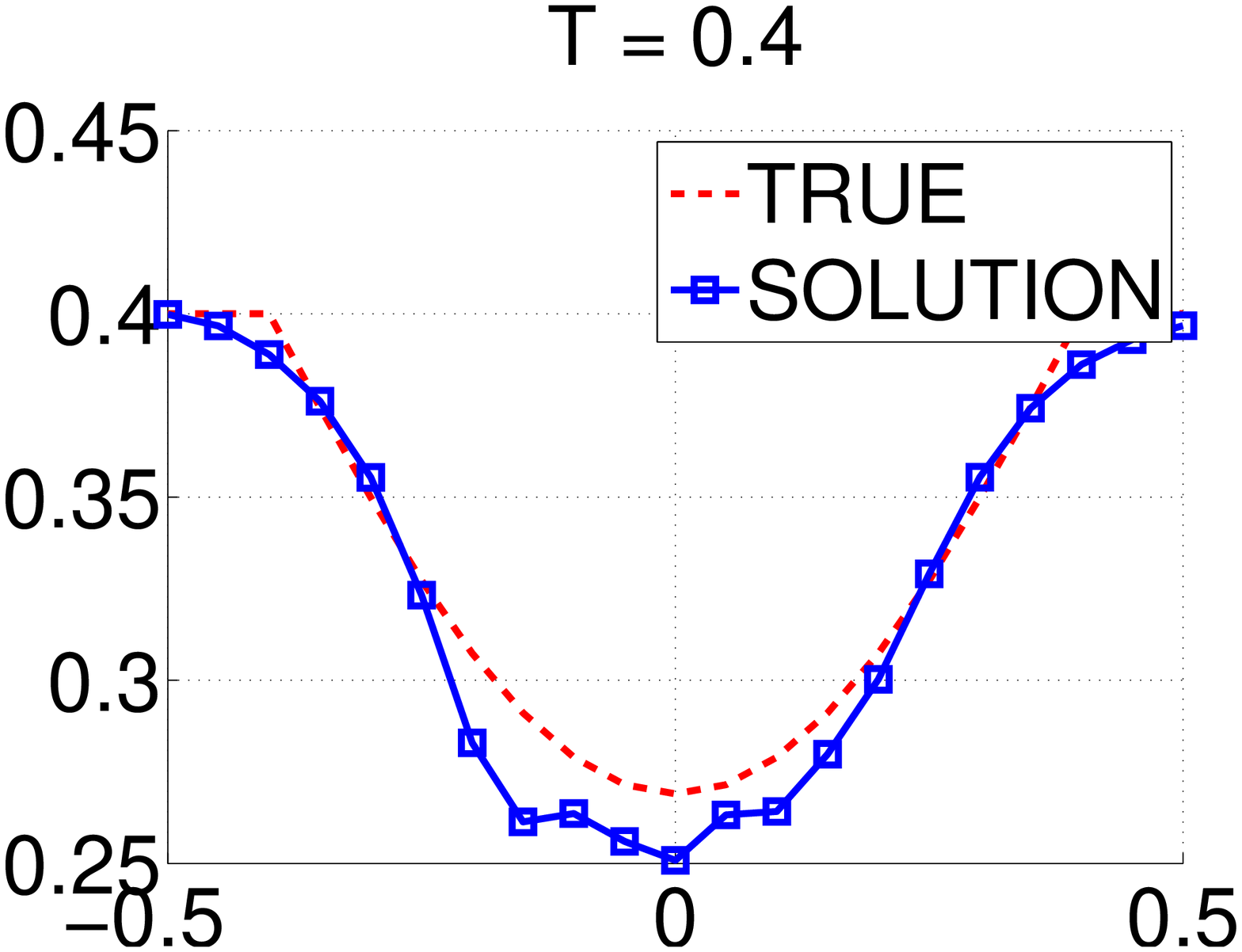}\hfill
\includegraphics[width=0.49\textwidth]{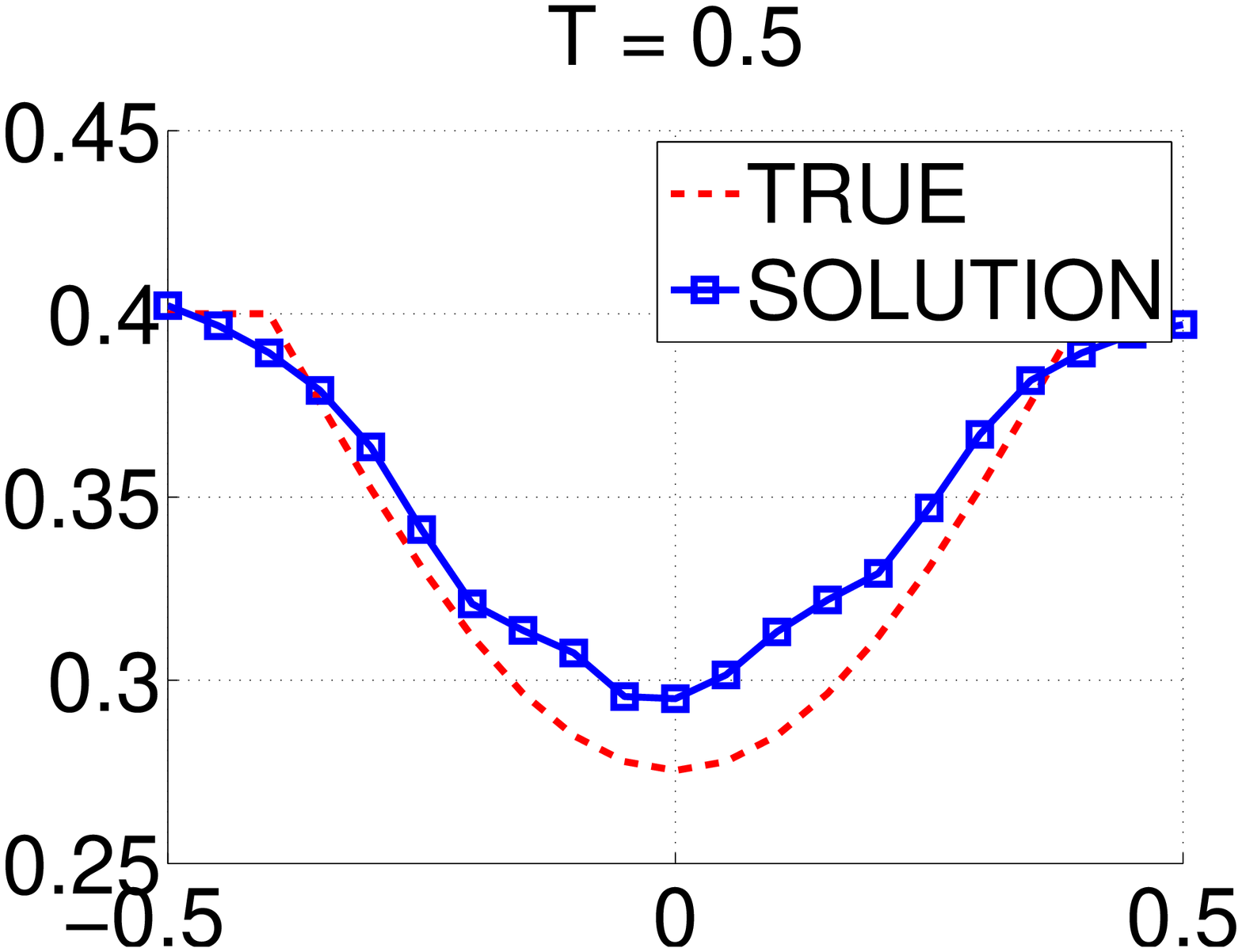}
\caption{True and calibrated local volatility surfaces at each maturity.}
\label{fig2}
\end{minipage}\hfill
\begin{minipage}{.49\textwidth}
  \centering
\includegraphics[width=0.66\textwidth]{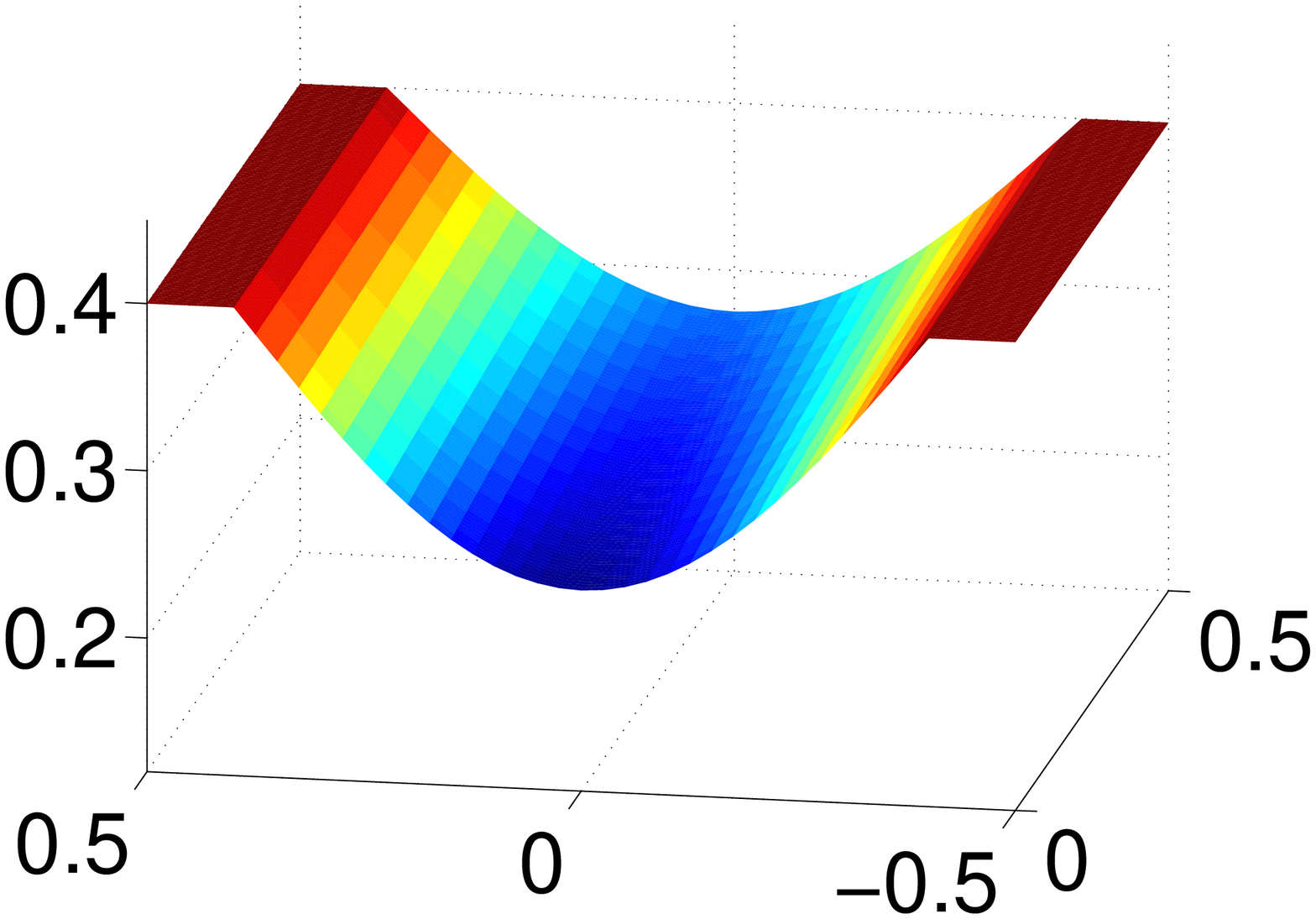}\hfill
\includegraphics[width=0.66\textwidth]{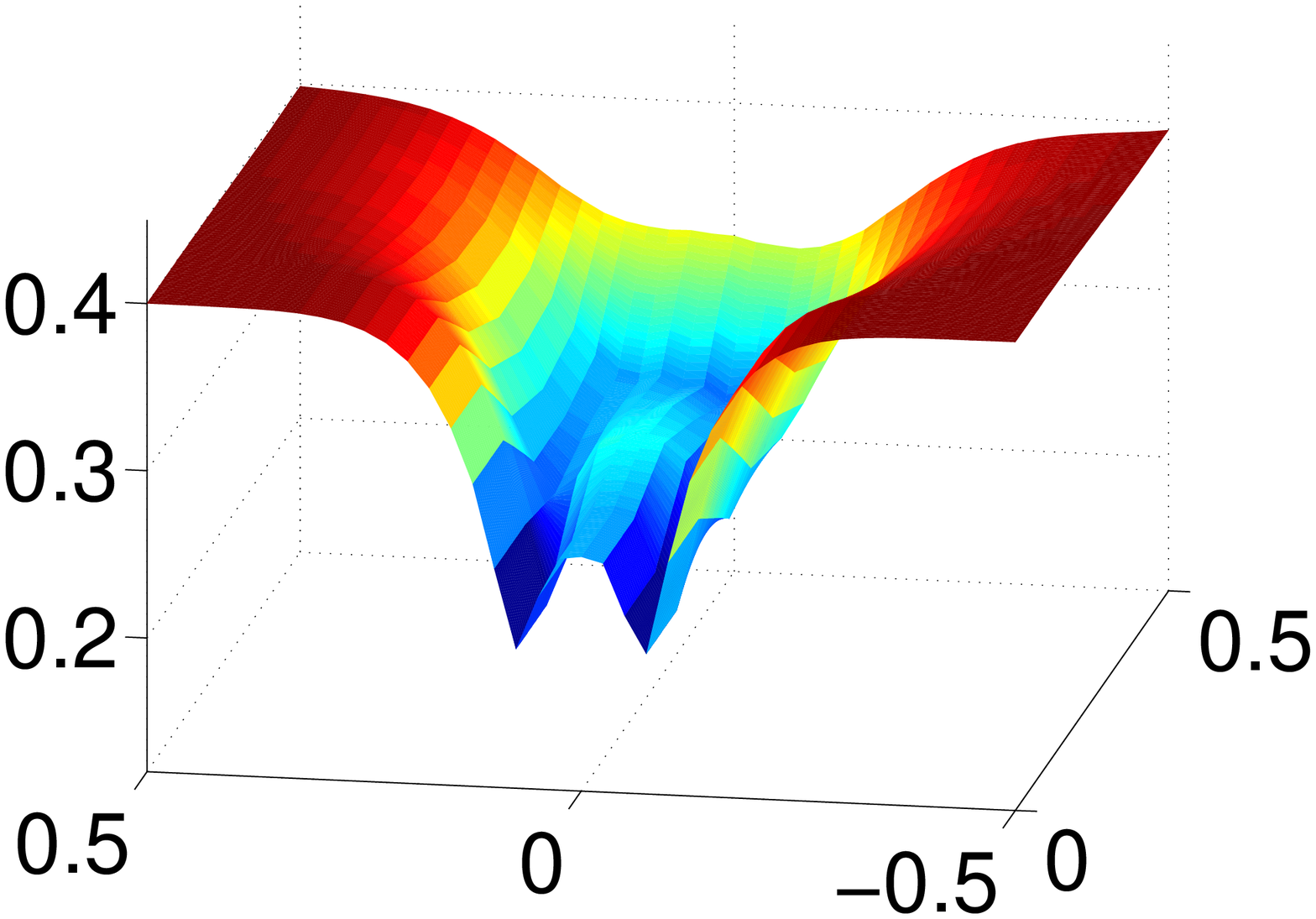}\hfill
\caption{Original (left) and calibrated (right) local volatility surfaces.}
\label{fig_3}
\end{minipage}%
\end{figure}

 In Figure~\ref{fig1}, at each maturity, we compare implied volatility of the data (the ones given by the reconstructed local volatility surface) and the implied volatility surface given by the European prices, evaluated with the true local volatility surface~\eqref{vol}, using \eqref{cns}.
 The comparison between the calibrated and the true local volatility surfaces at each maturity
 is presented in Figure~\ref{fig2}.

In Figure~\ref{fig_3}, we compare the true and the reconstructed local volatility surfaces. The normalized error
\begin{equation}
 E(a) = \|a-a^\dagger\|/\|a^\dagger\|,
 \label{eq:error}
\end{equation}
where $a^\dagger$ denotes the true local volatility surface given in~\eqref{vol}, is $0.1578$.

As can be gleaned from Figures~\ref{fig1}--\ref{fig_3}, 
the resulting local volatility surface was similar to the true one, although  
the calibration was not entirely qualitatively correct.
That is so 
since the transformation from American into European prices inevitably introduces noise. 
In addition, the set of locations where the data is defined, i.e., the mesh where the data is given, which is described in the first paragraph after Equation~\eqref{vol}, is sparse in comparison to the one 
on which the direct problem is solved. 

\subsection{A Parametric Local Volatility Surface}
\label{sec:parametric}

A crucial ingredient of the regularization approach is the introduction of {\em a priori} information. 
In our tests this is achieved by first considering a parameterized surface, which can be viewed as one step beyond the plain constant volatility model. This can be used also for testing the accuracy of the reconstructions under noise by producing synthetic data.

We chose the special family of parameterized surfaces 
\begin{equation}
 \sigma(\tau,y,\Theta) = \left\{
\begin{array}{ll}
a\tau + b - c \text{e}^{-d \tau}\cos\left(\displaystyle\frac{\pi}{2}\frac{y}{e}\right),& \text{if } y \in [-e,e]\\
a\tau + b, & \text{otherwise,}
\end{array}
\right.
\label{paramvol}
\end{equation}
where $\Theta = (a,b,c,d,e)$ is the vector of parameters that satisfy $0 < a \leq (1-b)/\tau_{\max}$, $0 < b \leq 1$, $0 < c \leq
b$, $-1 \leq d \leq 1$ and $0 < e \leq 2$.

\begin{figure}[!ht]
\centering
\begin{minipage}{.49\textwidth}
  \centering
      \includegraphics[width=0.49\textwidth]{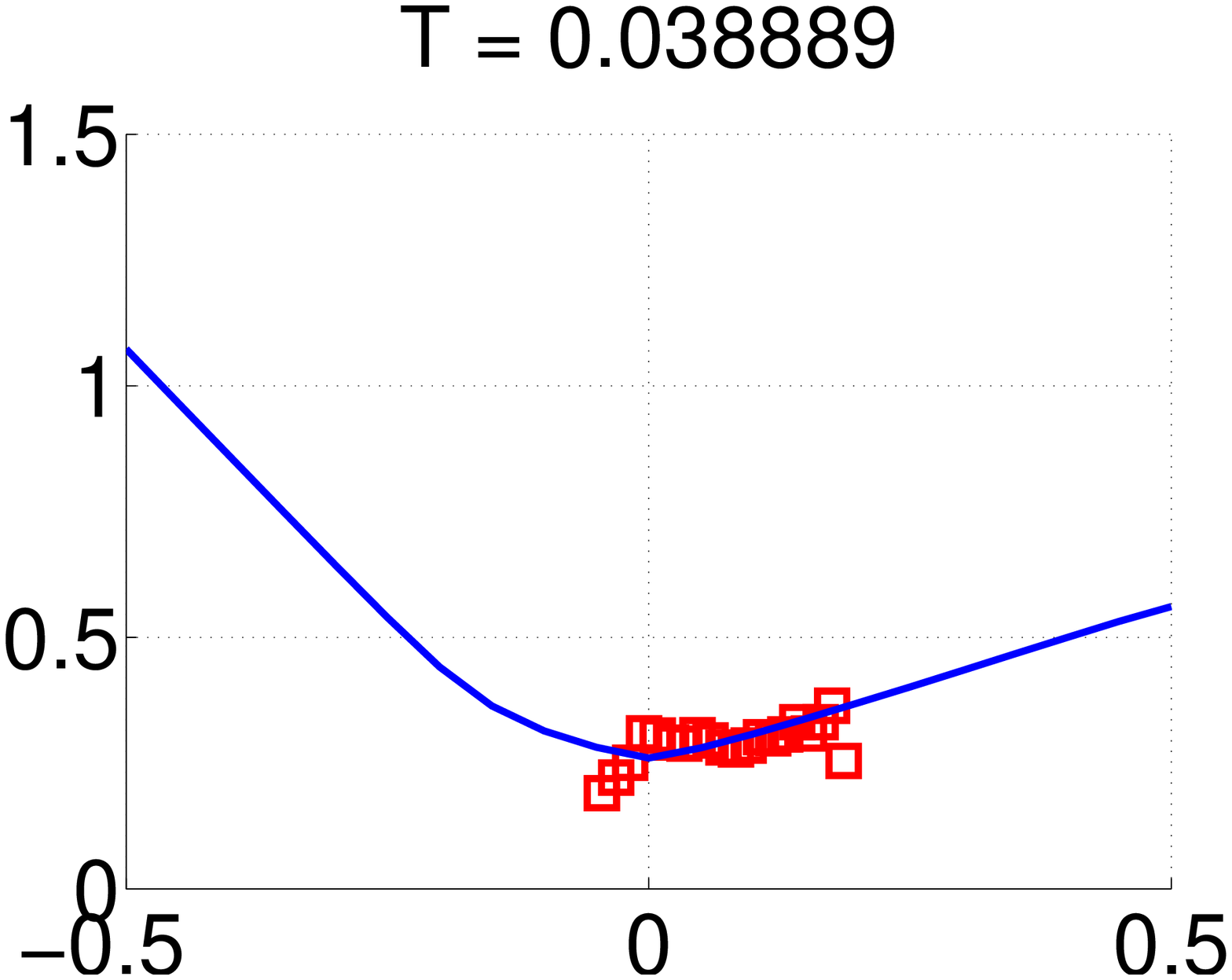}\hfill
      \includegraphics[width=0.49\textwidth]{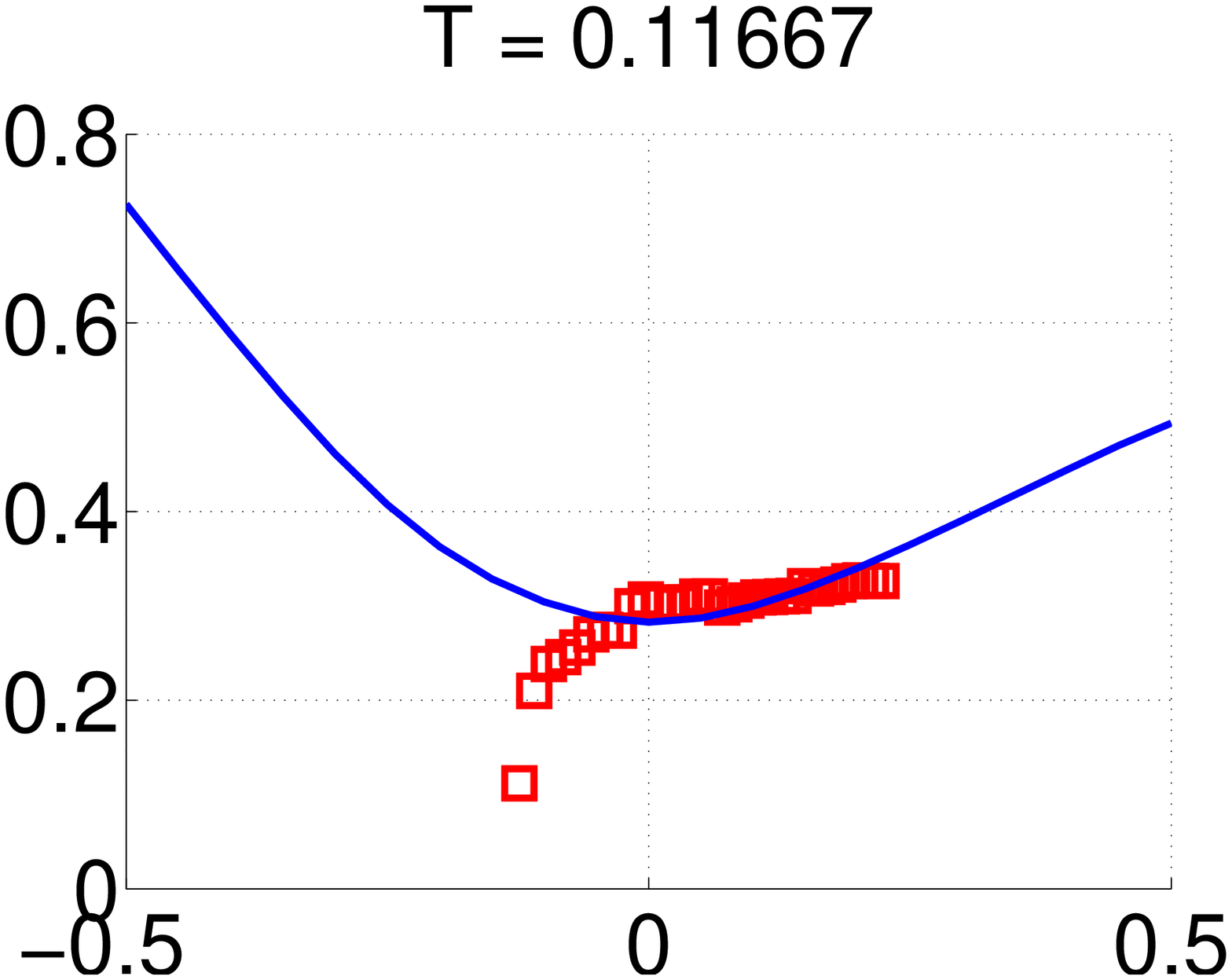}\hfill
      \includegraphics[width=0.49\textwidth]{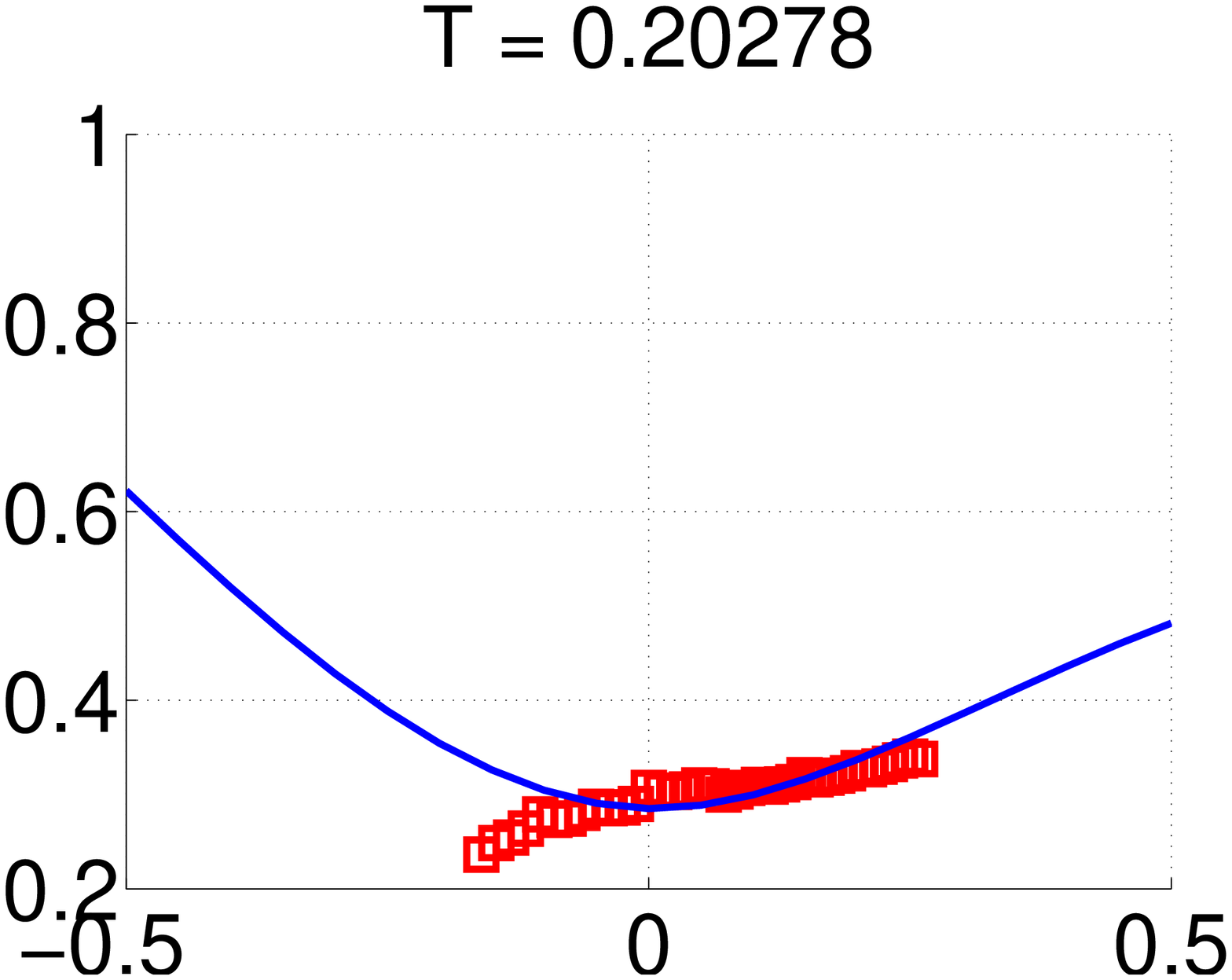}\hfill
      \includegraphics[width=0.49\textwidth]{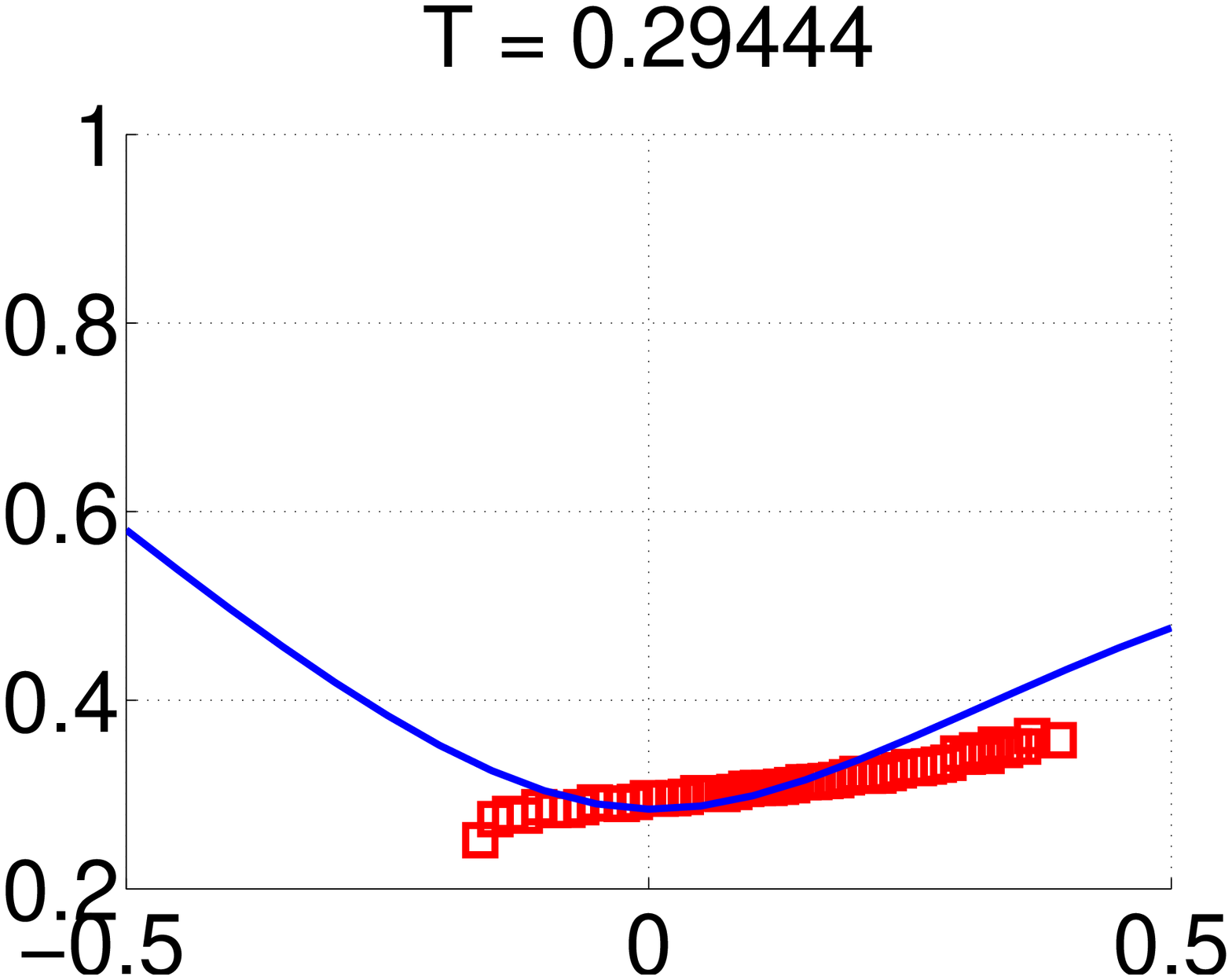}\hfill
      \includegraphics[width=0.49\textwidth]{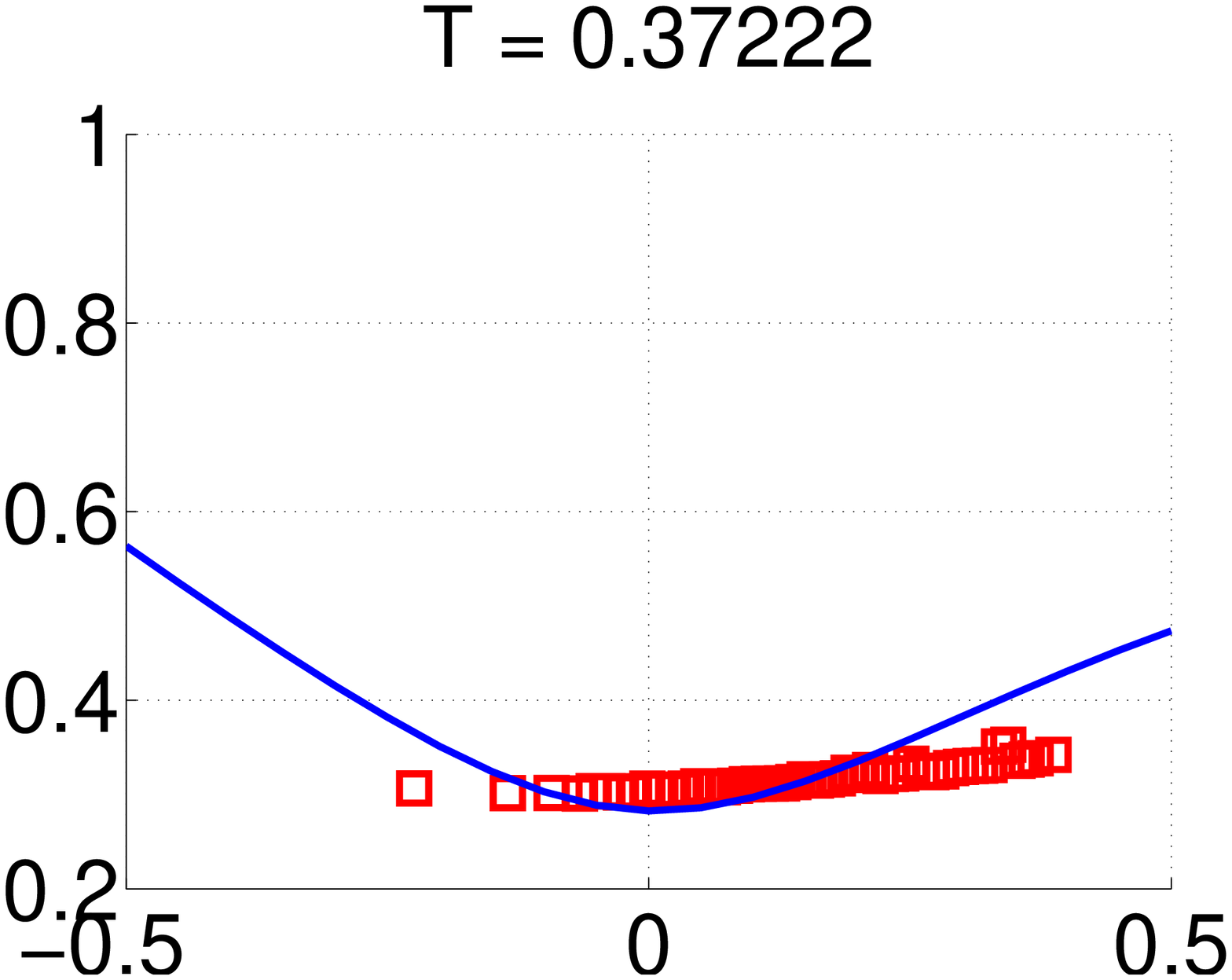}\hfill
      \includegraphics[width=0.49\textwidth]{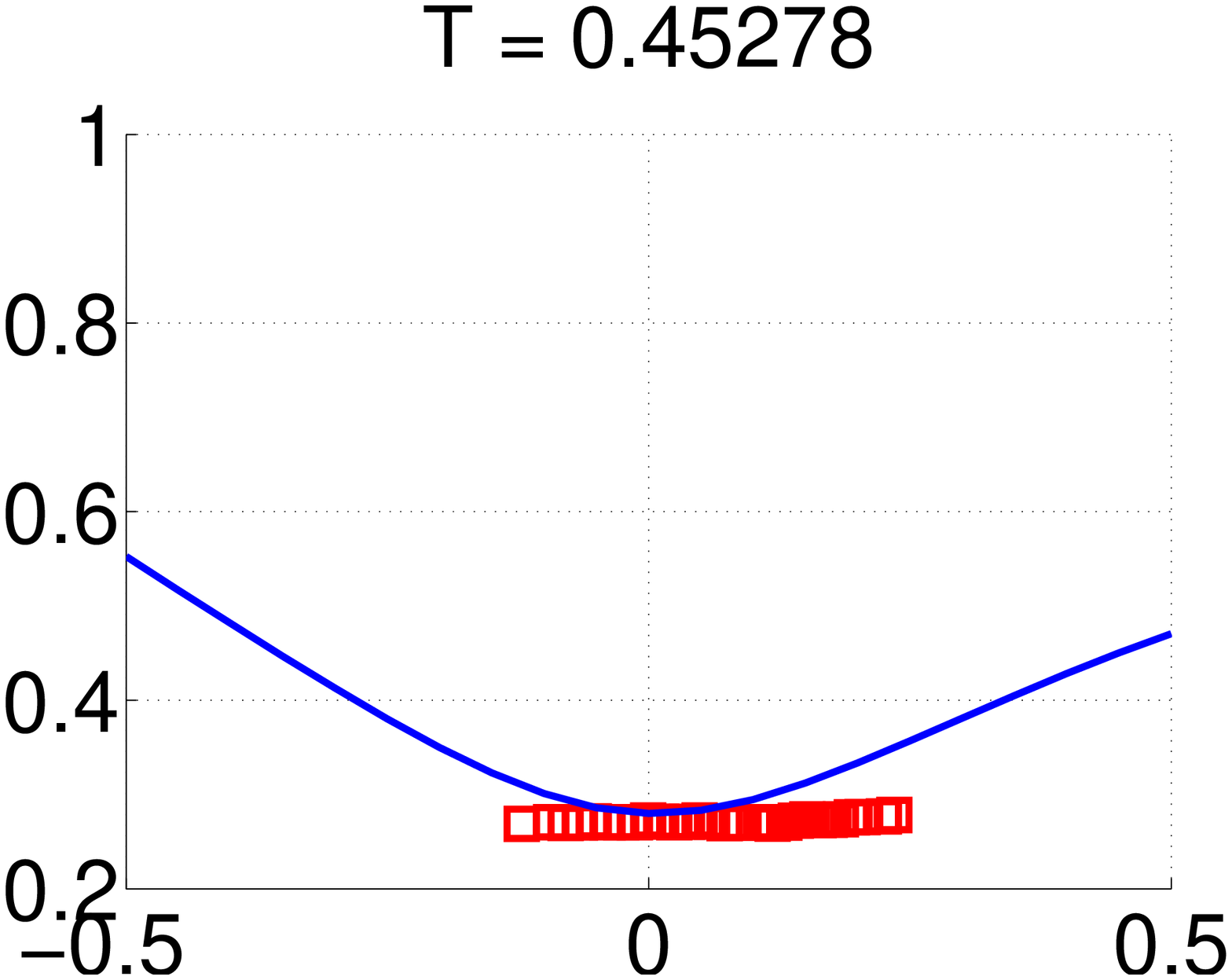}
\caption{Implied volatilities corresponding to the parametric model (continuous line) and market data (squares).}
\label{testh}
\end{minipage}\hfill
\begin{minipage}{.49\textwidth}
  \centering
\includegraphics[width=0.66\textwidth]{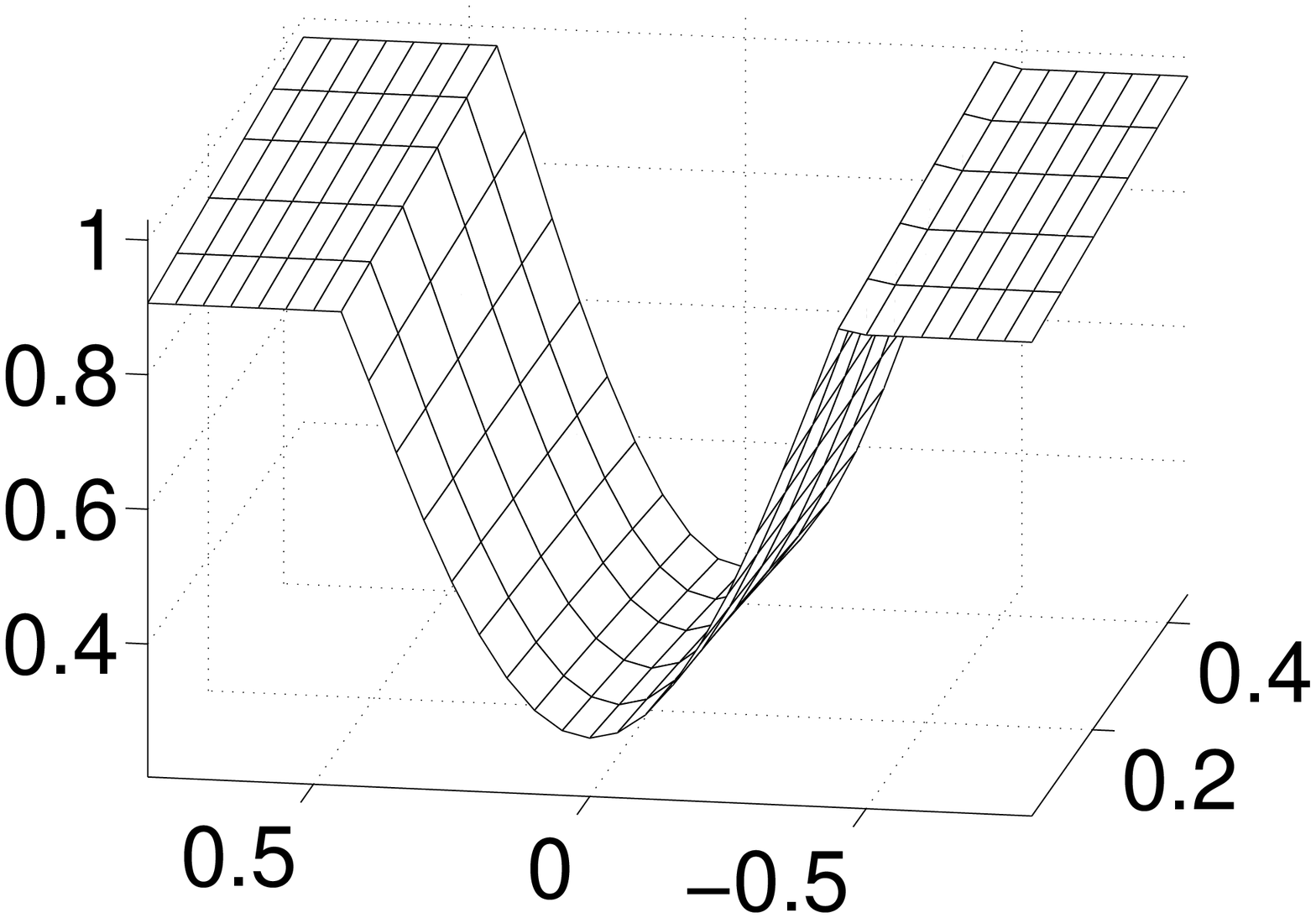}\hfill
\includegraphics[width=0.66\textwidth]{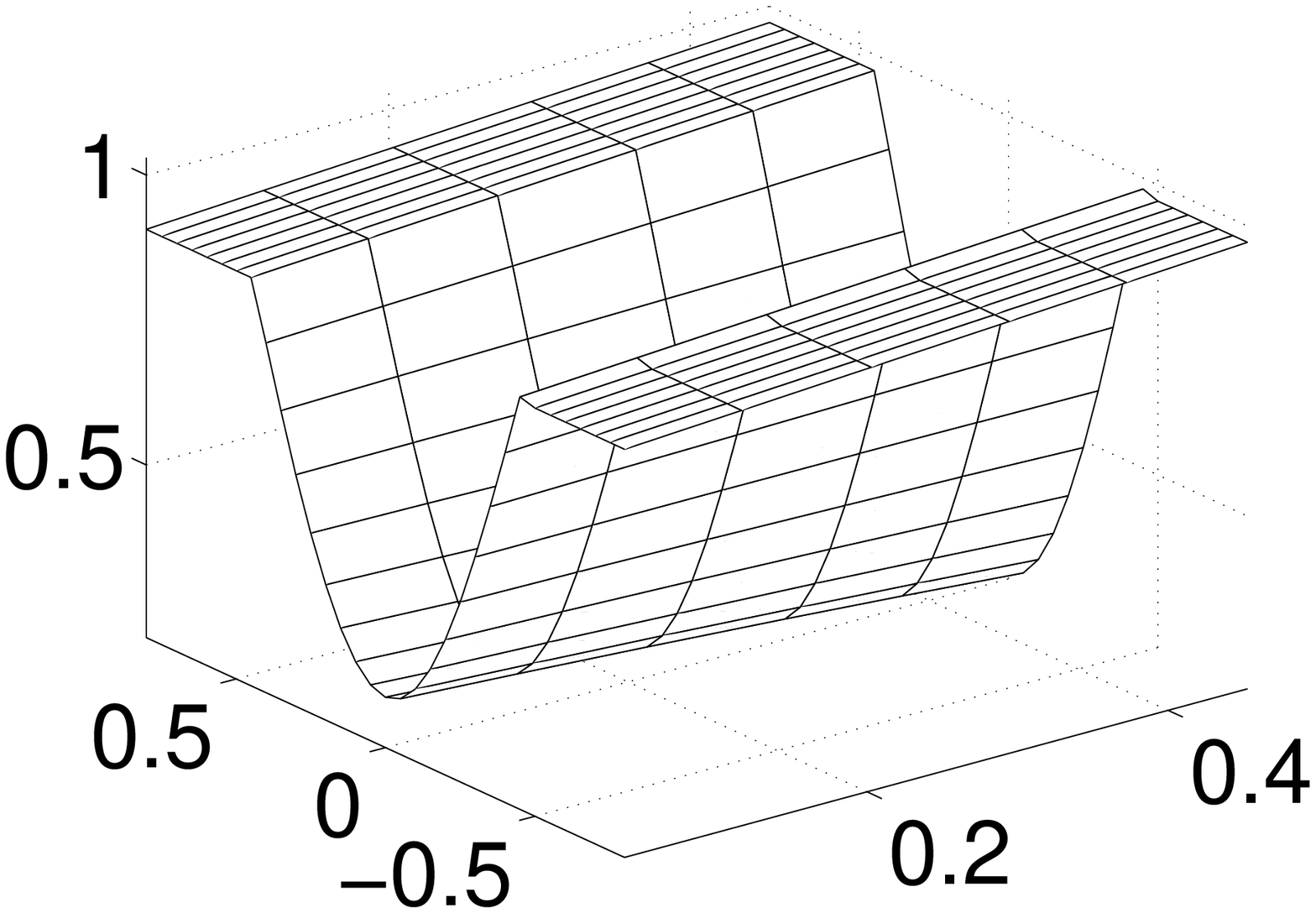}
\caption{Parametric local volatility surface with the parameters of Table~\ref{table:param}.}
\label{fig:paramvol2}
\end{minipage}%
\end{figure}

The quantity $b + a\tau$ is the level of local volatility for deep in-the-money and deep out-of-the-money prices, where $b$ is the initial level and $a$ gives its (linear) evolution. The parameters $c$, $d$ and $e$ define the evolution of the smile, where $c$ is the initial smile depth, $d$ determines its (exponential) evolution and $2\cdot e$ is the width of the smile centered at $0$.

Other parameters could be added to include other properties of the local volatility, such as the evolution of the center 
of the smile and asymmetry.

We calibrate the parametric volatility from transformed and normalized Henry Hub call option prices. The resulting parameters can be found on Table~\ref{table:param}. 
The Henry Hub call option prices used in this example were traded at 04-Sep-2013, and maturing at 29-Oct-2013, 27-Nov-2013, 27-Dec-2013, 29-Jan-2014, 26-Feb-2014, and 27-Mar-2014, respectively. We used the strikes between $-0.4$ and $0.4$, in log-moneyness.
\begin{table}[!ht]
\centering
\begin{tabular}{|c|c|c|c|c|c|}
\hline
&$a$ & $b$ & $c$ & $d$ & $e$\\
\hline\hline
Initial & $0.100$ & $0.400$ & $0.200$ & $0$ & $0.500$\\
Calibrated & $2.232\times 10^{-14}$ & $0.897$ & $0.562$ & $-0.500$ & $0.500$\\
\hline
\end{tabular}
\caption{Initial and calibrated parameters of the parametric local volatility surface, using Henry Hub prices.}
\label{table:param}
\end{table}

The parameters in~\eqref{paramvol} were calibrated by solving the nonlinear least-squares problem
$$
\min \left\{\|P_n\vf(a,b,c,d,e) - \vdd\|^2 ~: ~0 < a \leq \frac{1-b}{\tau_{\max}}, 0 < b \leq 1, 0 < c \leq b, -1 \leq d \leq 1, 0 < e \leq 2\right\}.
$$
For this test with Henry Hub call prices, the initial parameters can be found in Table~\ref{table:param}. The step sizes used for the PDE discretization 
were $\Delta \tau = 0.001$ and $\Delta y = 0.05$, 
and the interest rate as well as the dividend yield were assumed to be equal to $0.0325$.

The calibrated local volatility surface matched the market implied volatilities close to the at-the-money strikes. In addition, differently form Heston (see \citep{volguide}) and other parametric models, we have the same set of parameters for every maturity, which simplifies the model and speeds up the calibration. These features justify the use of our parametric surface as an initial guess or a {\em prior} 
to feed the online method or other local volatility calibration techniques.

\section{Numerical Experiments}
\label{sec:numerics}

The numerical experiments in the previous section were in the specific context of
transforming from American into European prices (Section~\ref{sec:transformation})
and constructing vital a priori information (Section~\ref{sec:parametric}).
Now we are ready for general numerical experiments, demonstrating the utility in adjusting
the asset price and our online approach. Moreover, we evaluate path-dependent exotic options to illustrate the accuracy of the local volatility model.

\subsection{Underlying Asset Prices as Additional Unknowns}
\label{sec:adjust}

In this subsection we first synthesize data from a known solution in order to study the advantage that may
be had in treating the underlying asset price as an uncertain quantity. Thus, rather than
fixing the price, we carefully penalize its
distance from a measured value (say, the average price over a day's trade).

To generate synthetic prices, 
we use the local volatility surface given in Equation~\eqref{vol}. 
The data is generated on a fine mesh with step sizes $\Delta \tau = 0.005$ and $\Delta y = 0.025$, and the maximum time to maturity is $\tau_{\max} = 0.5$. 
We use only the five future prices maturing at $\tau_i = i \cdot 0.1$, with $i=1,...,5$. They are given by
$$
F_{0,\tau_i} = 1+0.1\sin(3\tau_i \pi/2).
$$

We take the risk-free interest rate as $r=0$, and denote the noiseless call prices surface by $v = v(a)$. Then, at each $(\tau,y)$, we add a relative noise of $1\%$, defining noisy data using
\begin{equation}
v^{\delta}(\tau,y) = v(\tau,y) \left( 1 + \delta \eta\right),
 \label{eq:noise}
\end{equation}
where $\delta = 0.01$ and $\eta$ is a standard normal pseudo-random variable. 
We collect the price data at the strikes $K^i_j = F_{0,\tau_i}\exp(y_j)$, where $y_j = j\cdot 0.05$, 
with $j= -10,-9,...,0,1,...,10$, and at the the same maturities of its underlying futures. 

As mentioned in Section~\ref{sec:underlying}, an optimization step for the functional \eqref{eq:tikscar} uses a splitting
approach consisting of two stages, 
the minimization with respect to $\mathscr F$ and the minimization with respect to $\af$. 
In the first of these, 
we hold $s$ and the local volatility $a$ fixed, and set the parameters in the functional \eqref{eq:psi}
to $\alpha_1=\alpha_2=\alpha_3 = \alpha_6 =0$, $\alpha_4 = \alpha_5 = 1$ and $\hat{\mathscr F} \equiv 1$.
In the second stage, 
we hold the index $s$ and $\mathscr F$ fixed, and set the parameters in the functional \eqref{eq:psi} 
to $a_0 \equiv 0.101$, $\alpha_4 = \alpha_5 = \alpha_6 = 0$, $\alpha_2 = 10^{-4}$, $\alpha_1= \alpha_3 = 0.01\alpha_2$.

We started the minimization with a constant local volatility $a^0 \equiv 0.101$ and the observed underlying asset prices $\mathscr{F}^0 = 0.95 \mathscr{F}_{\text{true}}$, where $\mathscr{F}_{\text{true}}$ is the set of true underlying assets.  At each step $k$, we normalize the call prices by the current underlying asset price, i.e, $v^\delta(\tau_i,\cdot)/F^k_{0,\tau_i}$, with $i=1,...,5$, and assume that the data values are given at the nodes of the mesh defined by $(\tau_i,y_j)$, with $y_j = \log(K^i_j/F^k_{0,\tau_i})$, with $i=1,...,5$.

\begin{figure}[!ht]
\centering
\begin{minipage}{.44\textwidth}
   \centering
       \includegraphics[width=0.9\textwidth]{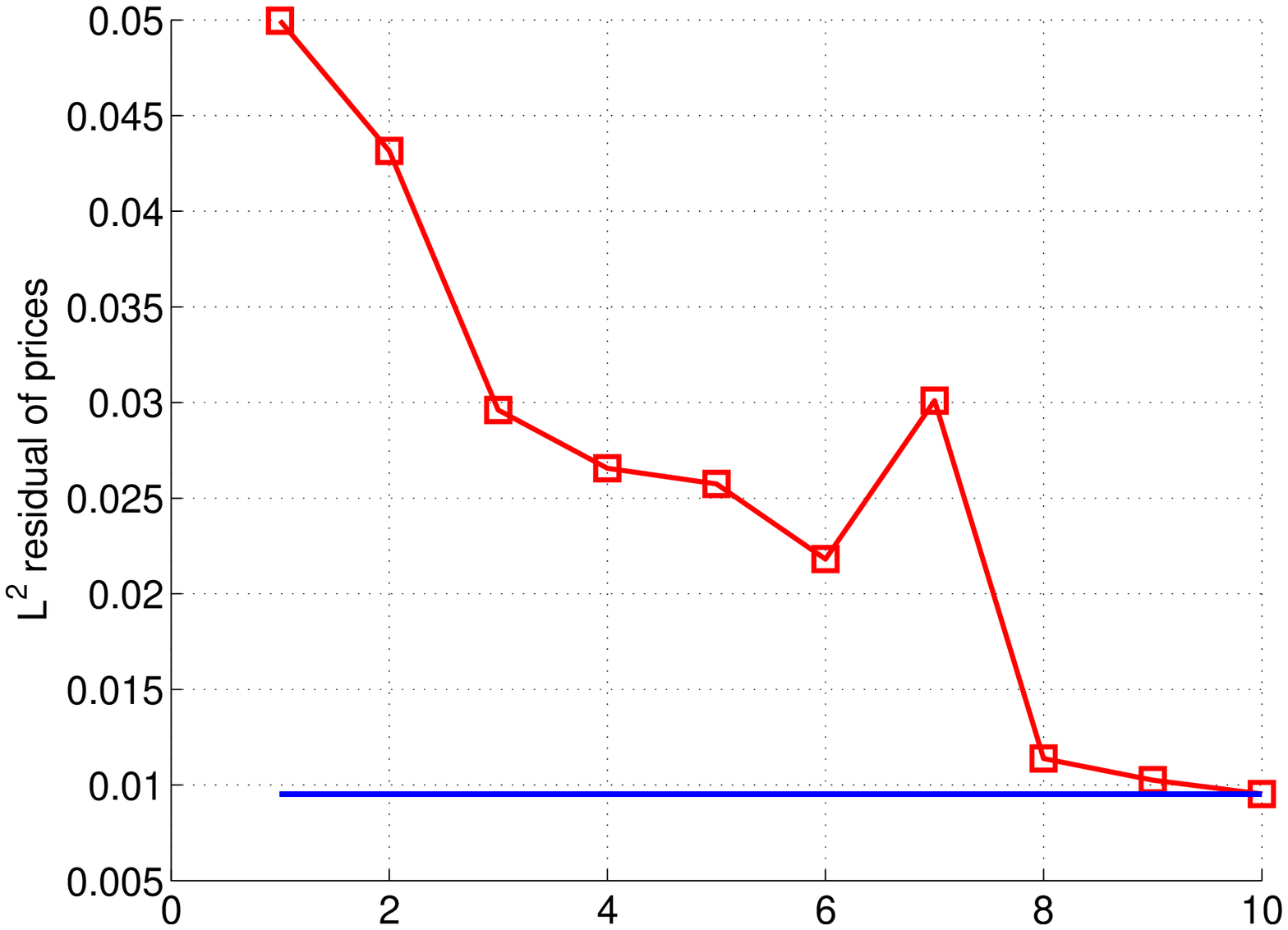}\hfill
       \includegraphics[width=0.9\textwidth]{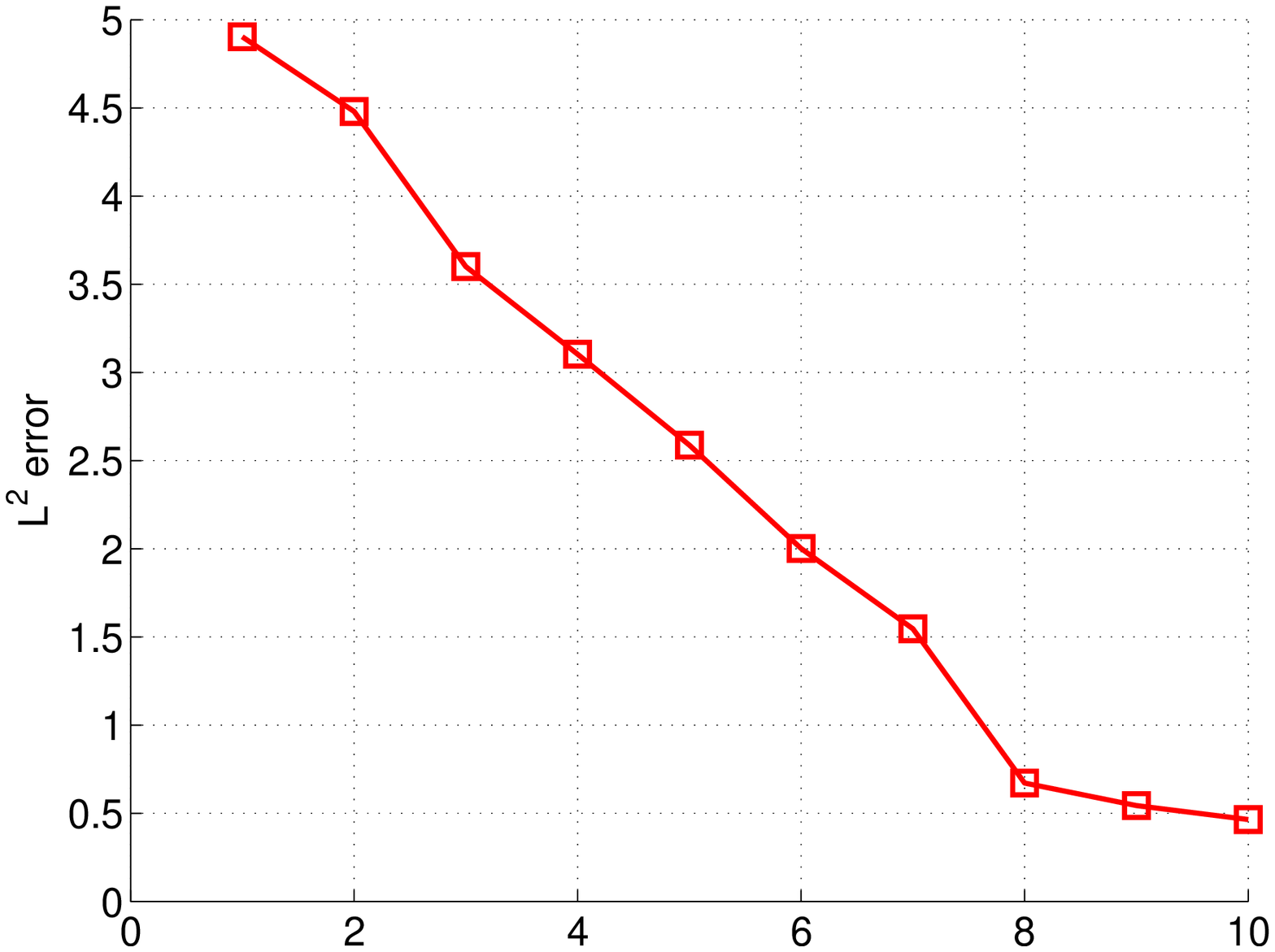}\hfill
\caption{Above: normalized $\ell_2$ data misfit (line with squares) and the 
normalized tolerance (continuous straight line). Below: normalized $\ell_2$-distance between the reconstructions and the true local volatility.}
  \label{fig:scarf01error}
\end{minipage}\hfill
\begin{minipage}{.54\textwidth}
   \centering
        \includegraphics[width=0.4\textwidth]{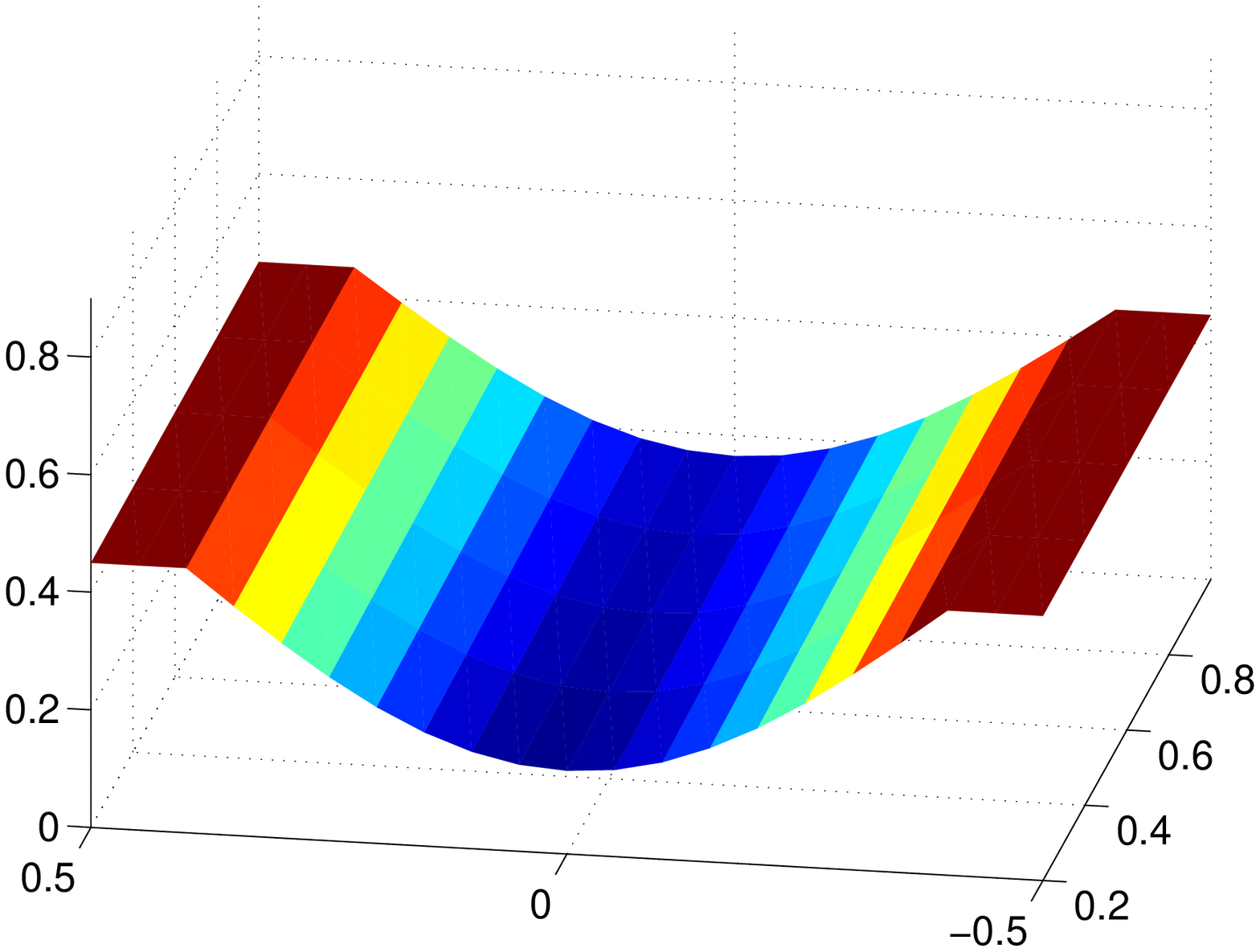}\hfill
        \includegraphics[width=0.4\textwidth]{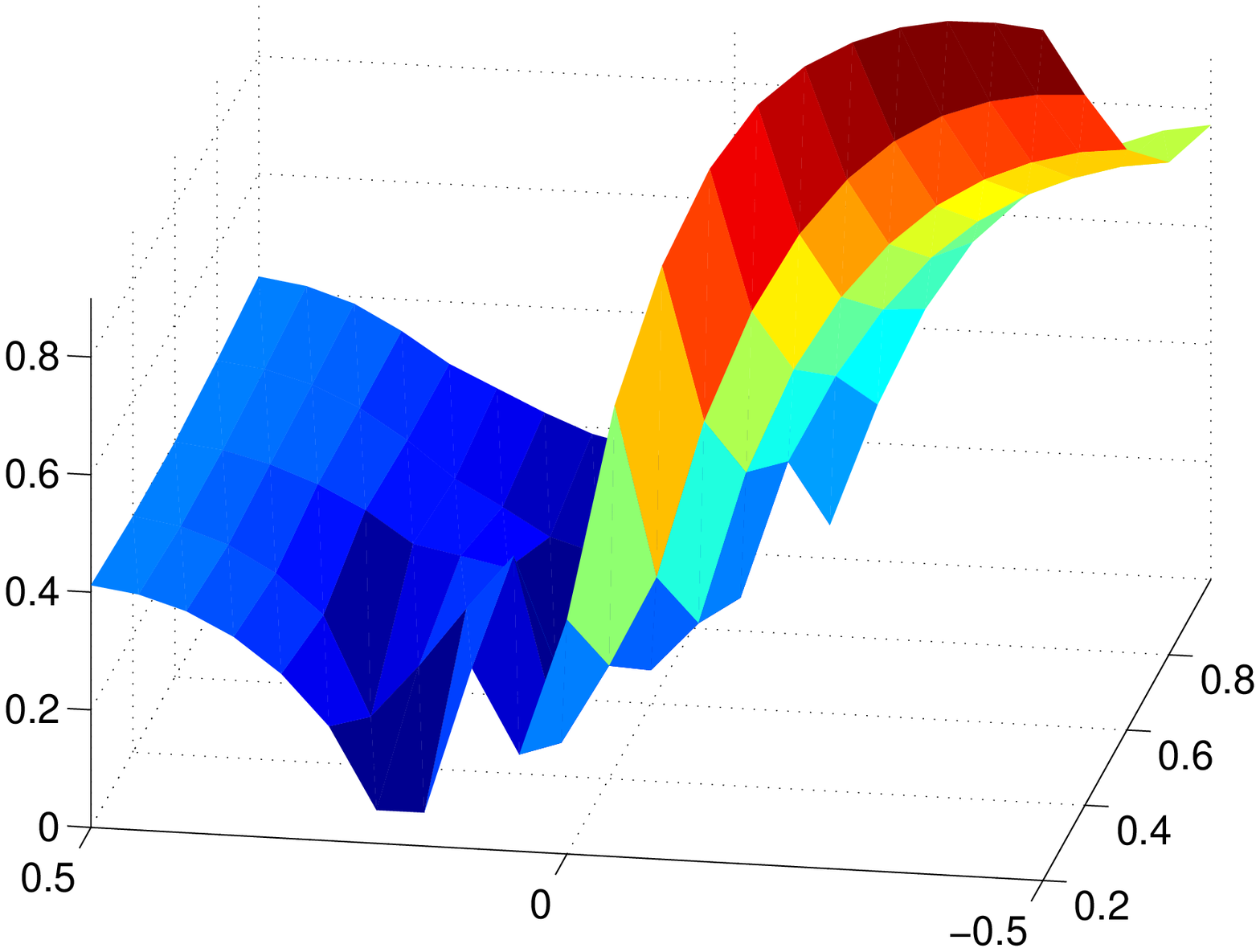}\hfill
	\includegraphics[width=0.4\textwidth]{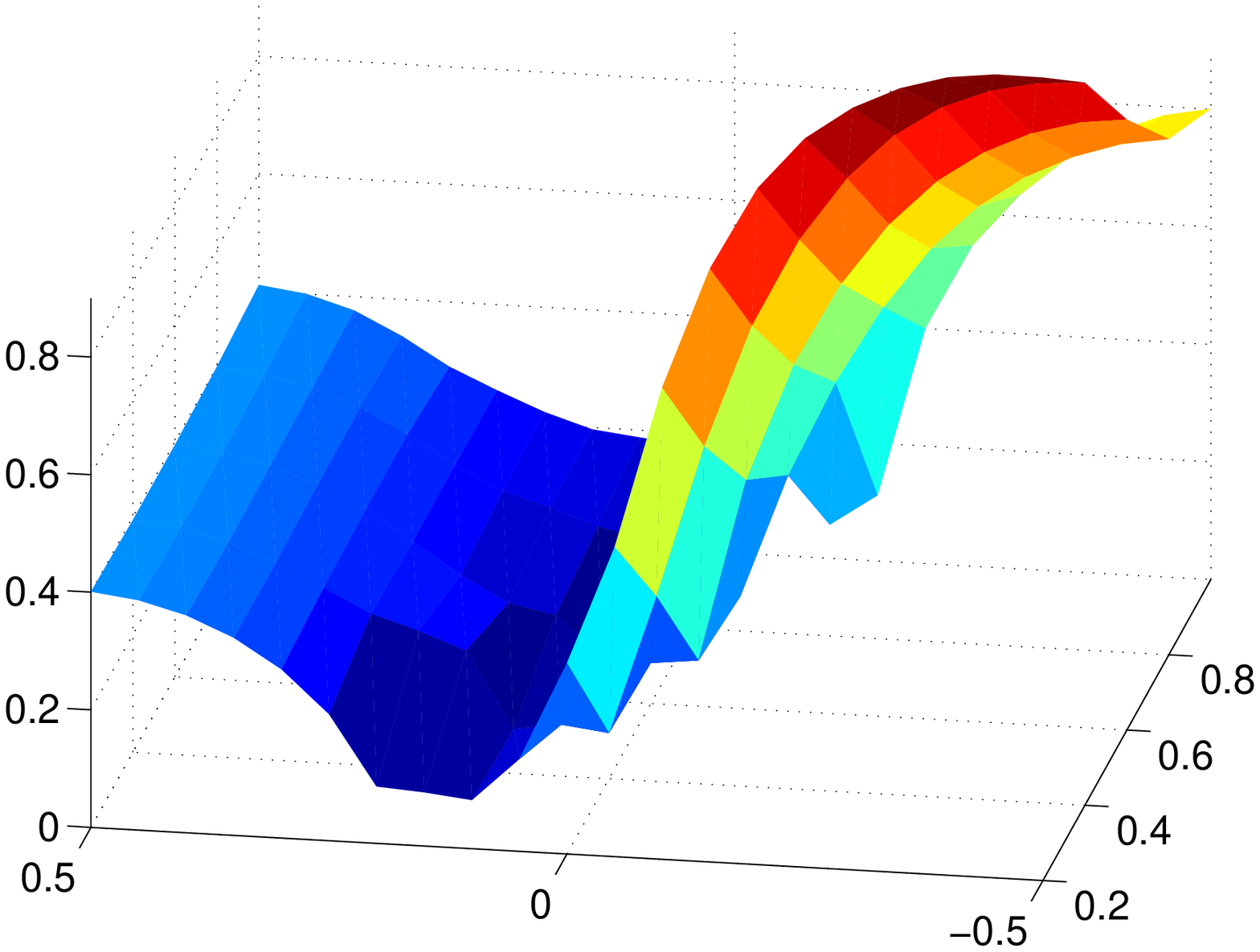}\hfill
	\includegraphics[width=0.4\textwidth]{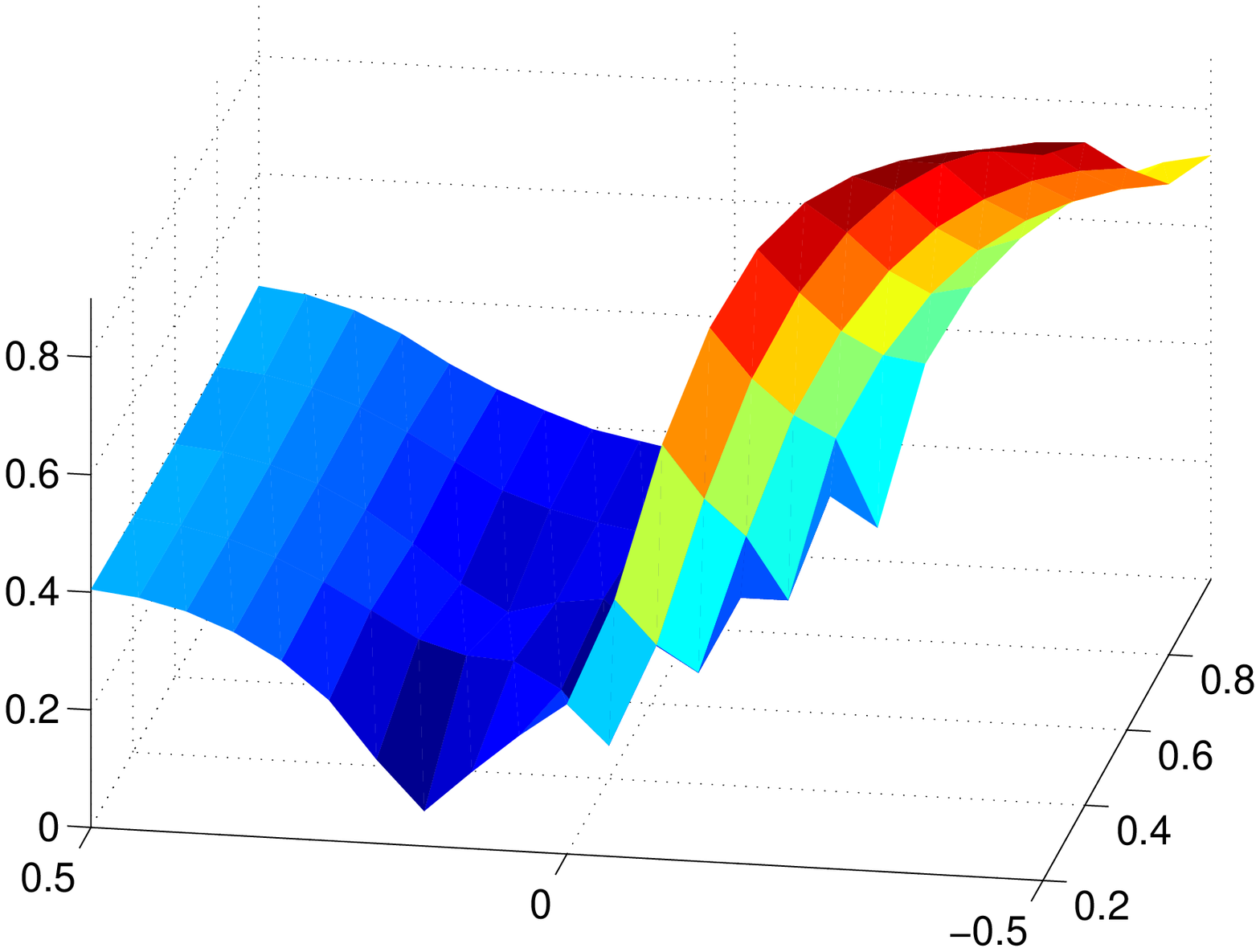}\hfill
	\includegraphics[width=0.4\textwidth]{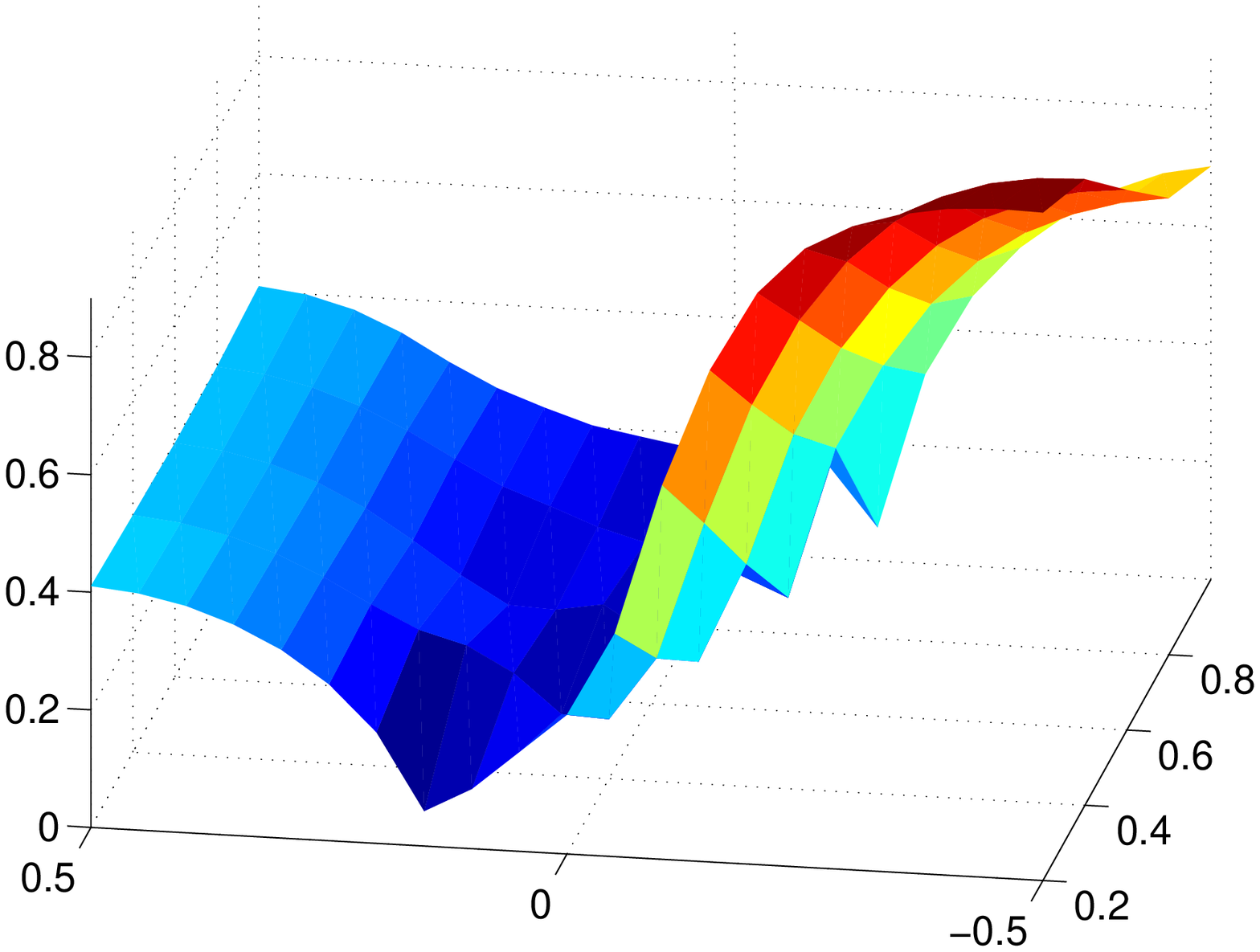}\hfill
	\includegraphics[width=0.4\textwidth]{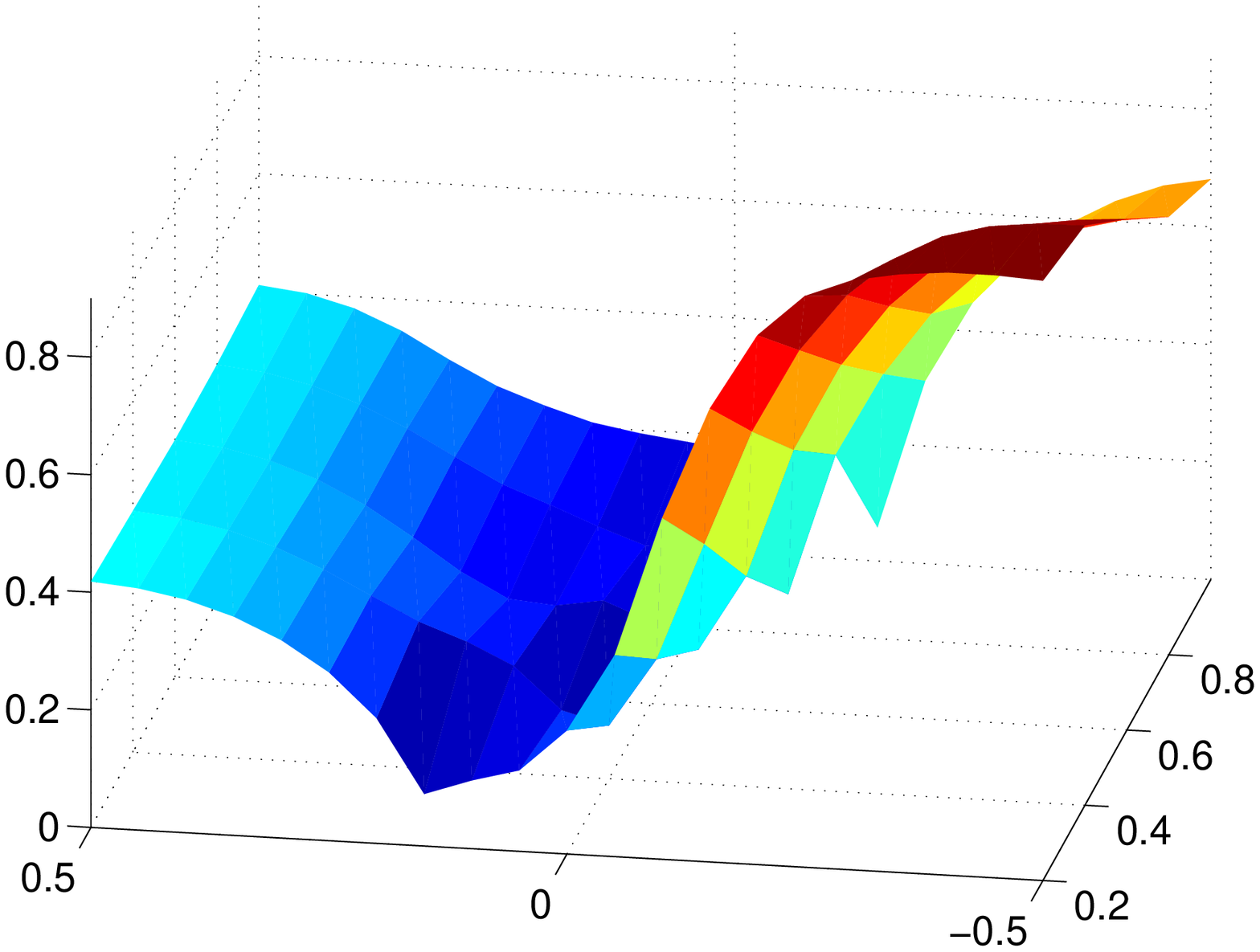}\hfill
	\includegraphics[width=0.4\textwidth]{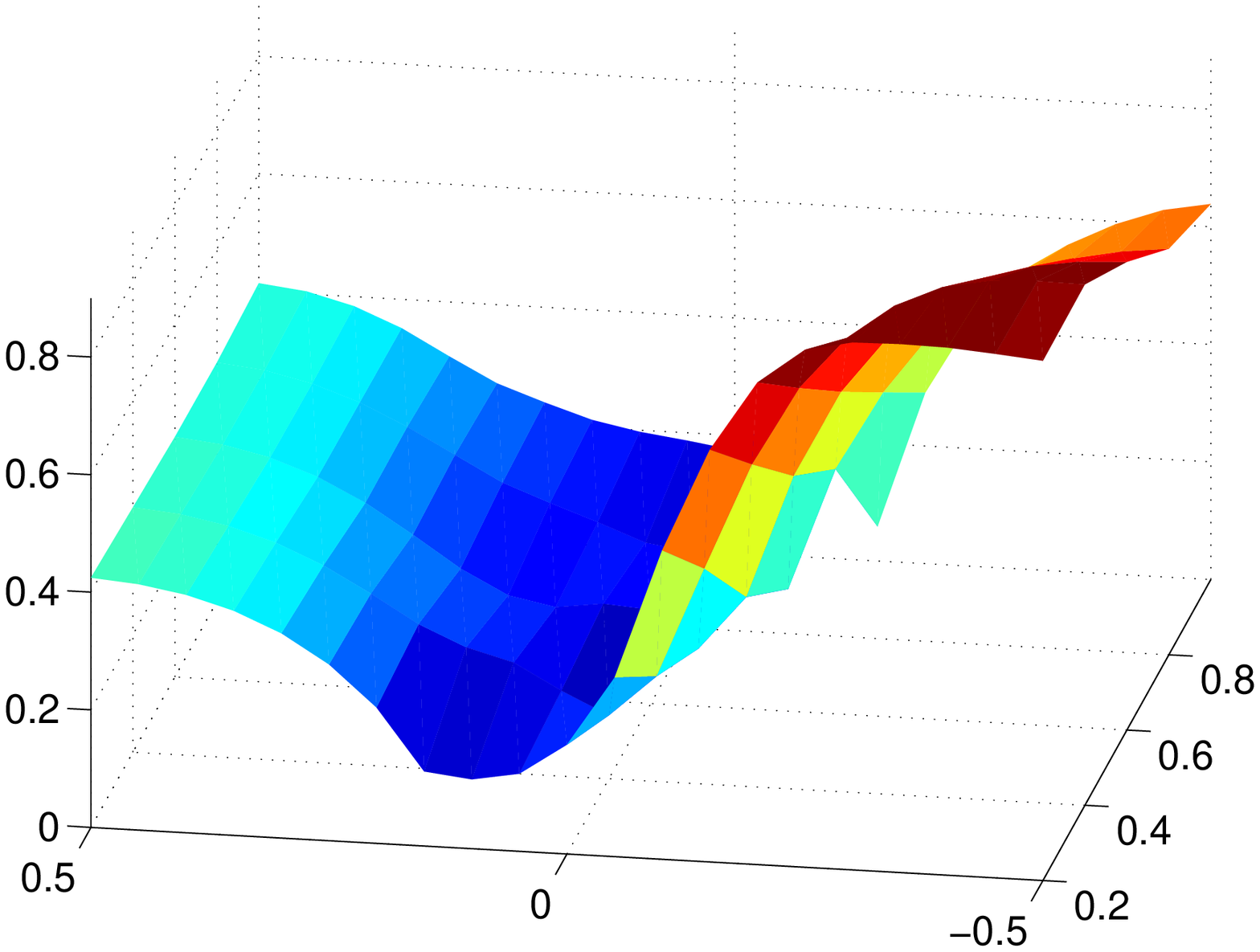}\hfill
	\includegraphics[width=0.4\textwidth]{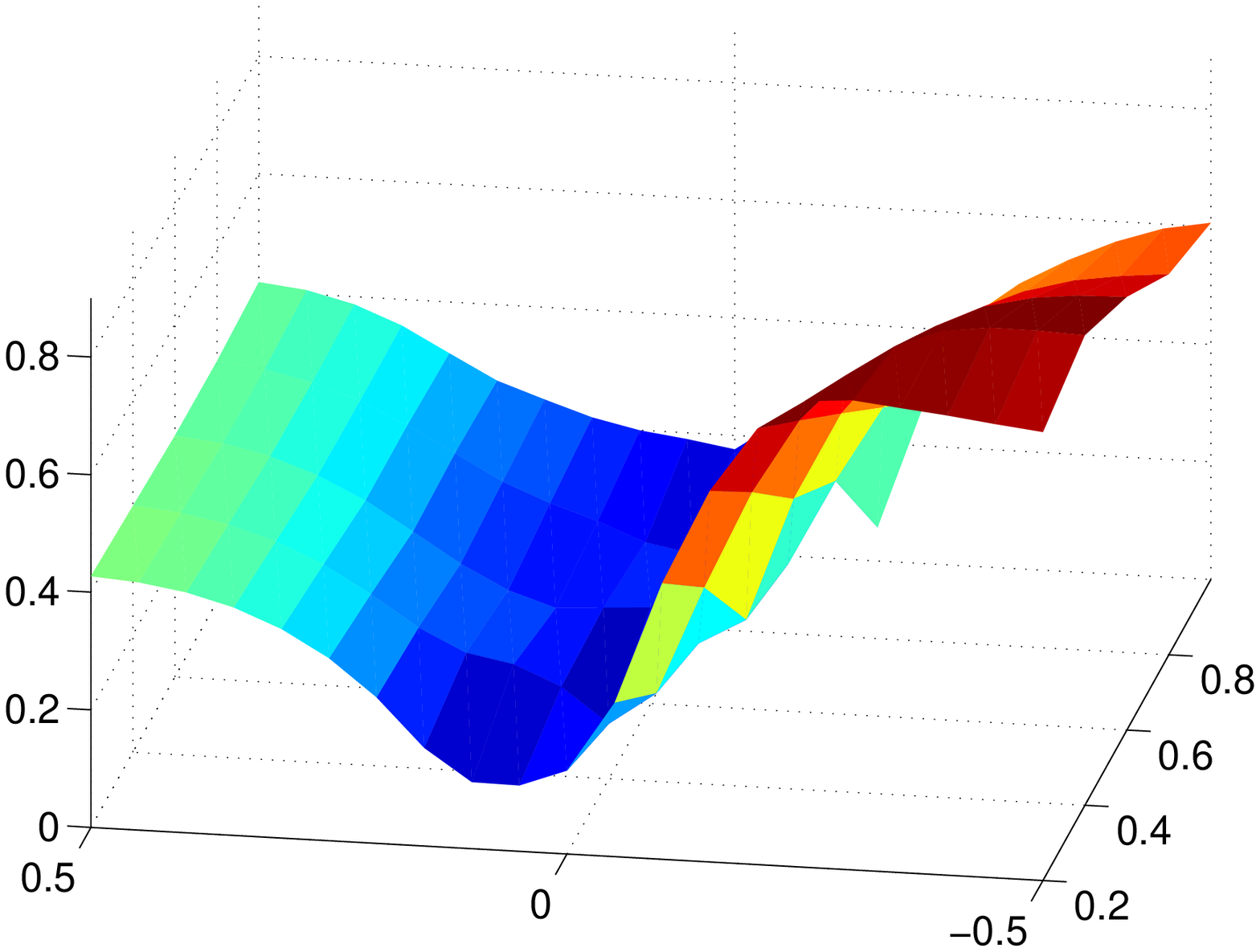}\hfill
	\includegraphics[width=0.4\textwidth]{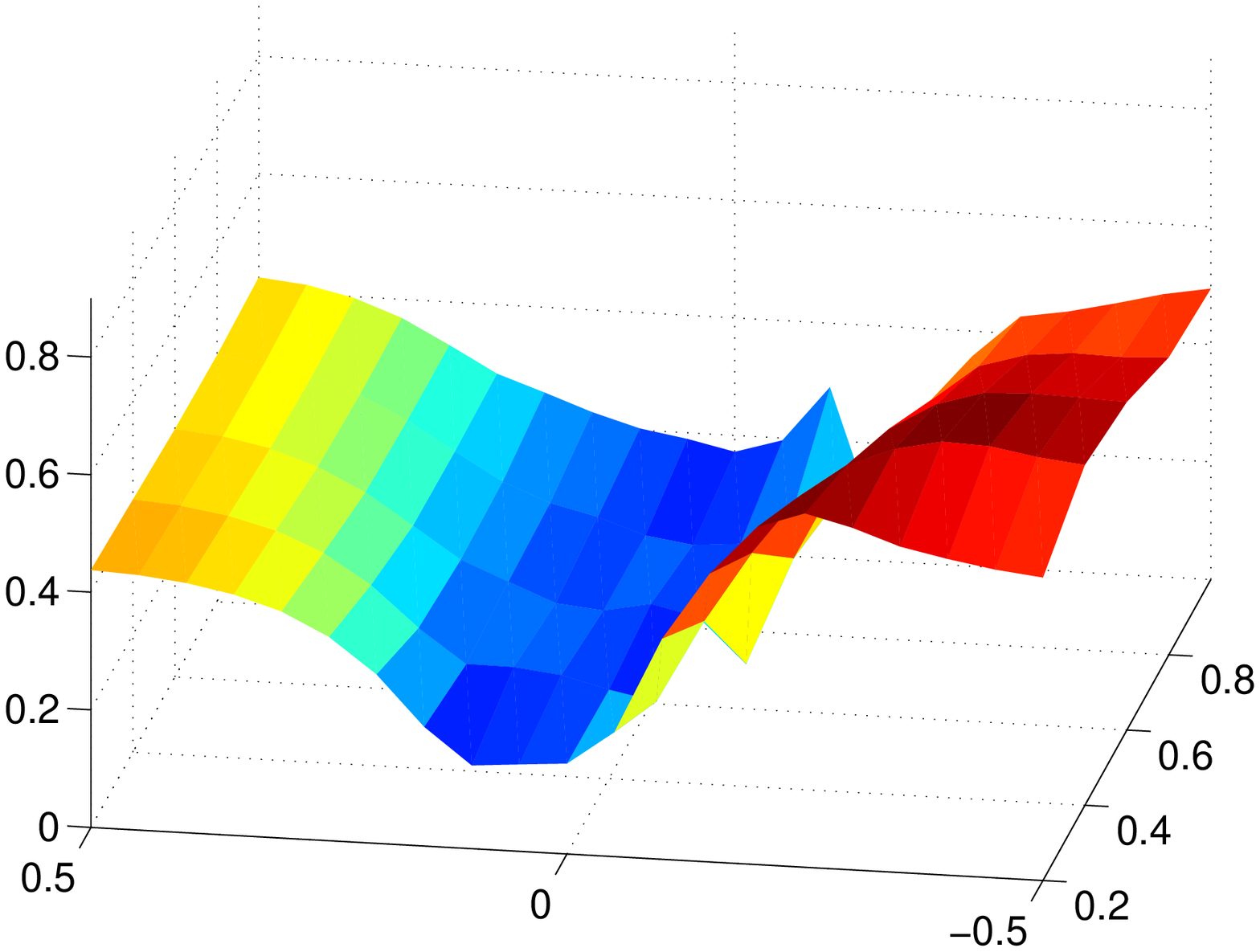}\hfill
	\includegraphics[width=0.4\textwidth]{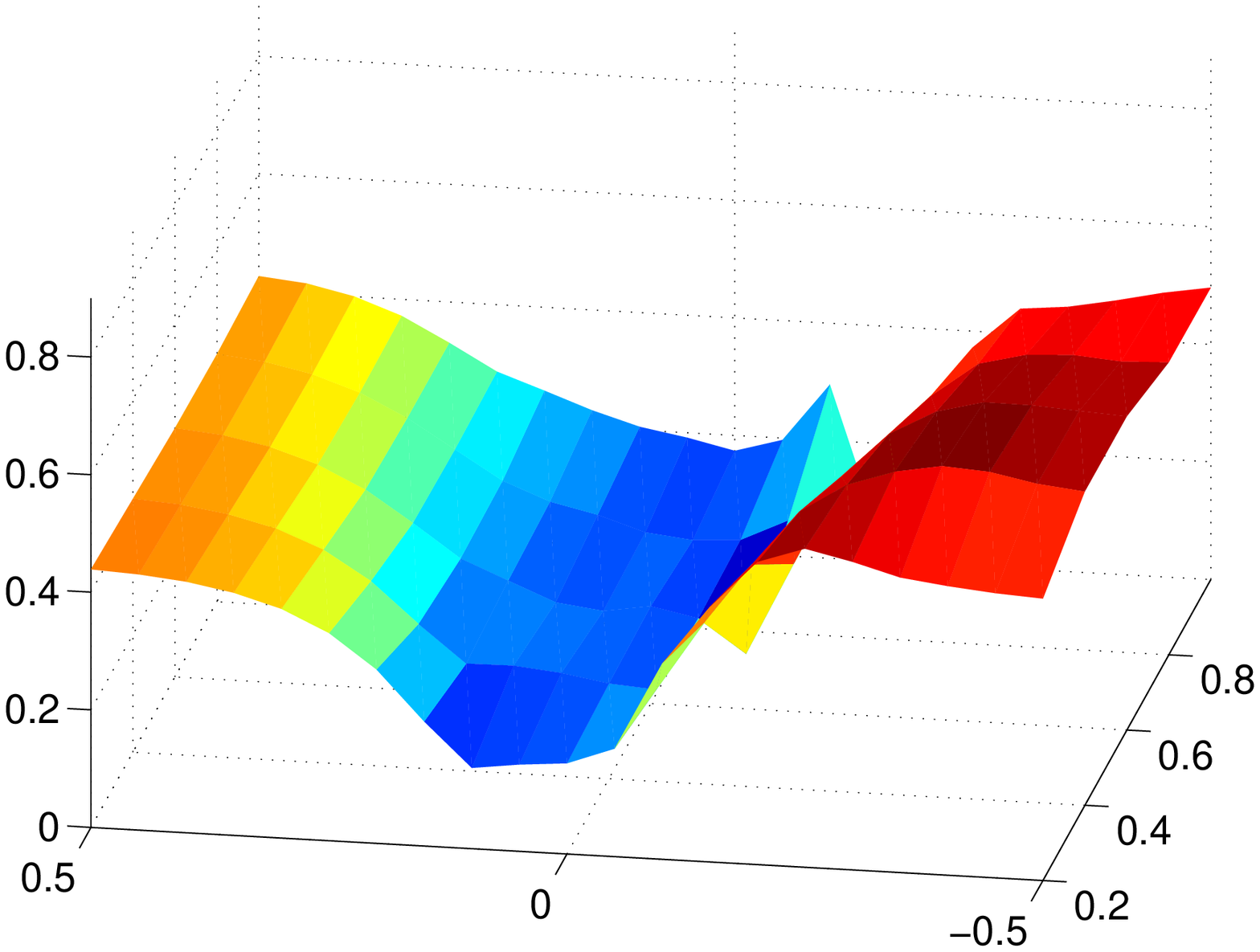}\hfill
	\includegraphics[width=0.4\textwidth]{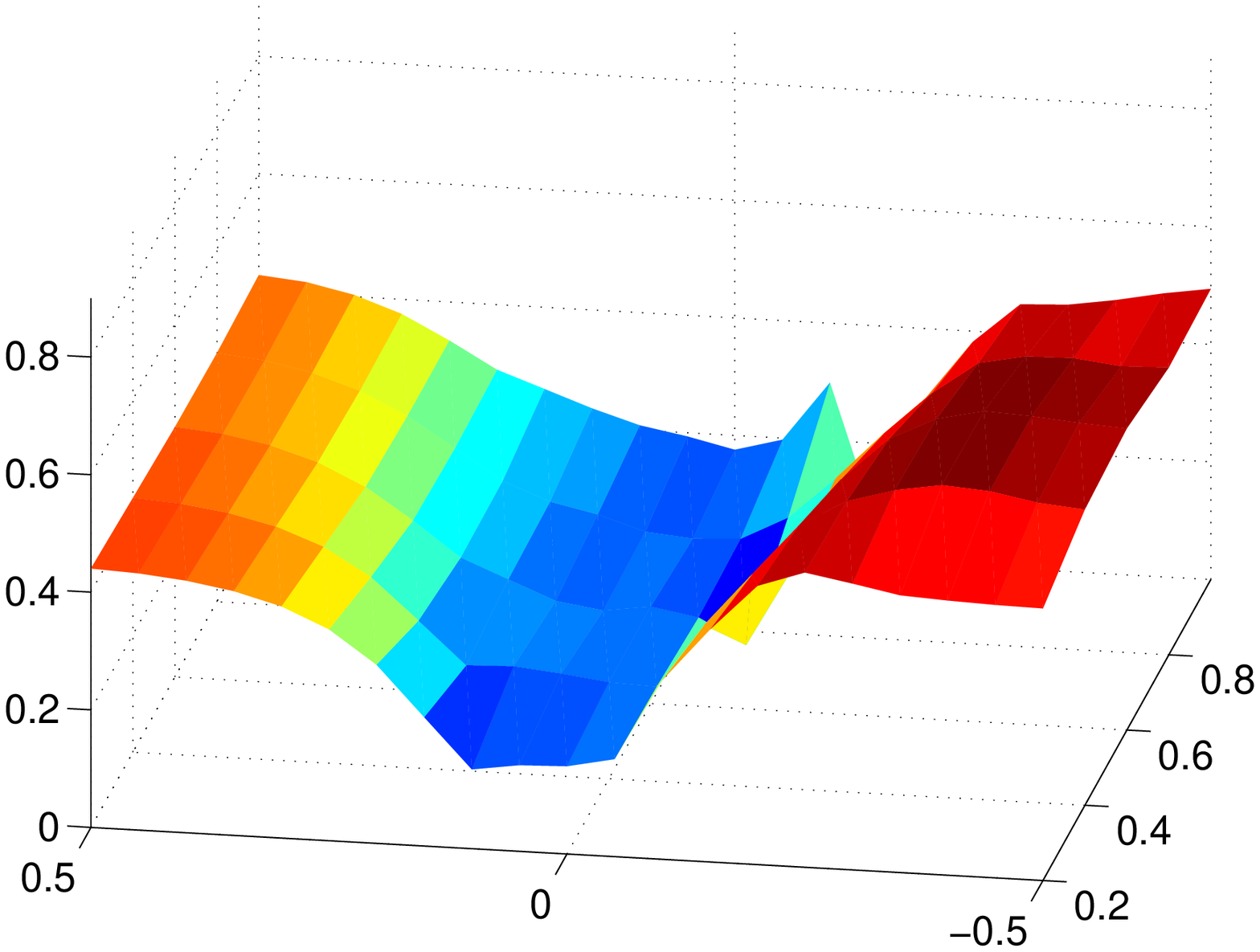}\hfill
 \caption{Original local volatility surface (top left) and its reconstructions as we adjust the values of the underlying prices $\mathscr F$.}
  \label{fig:scarf01}
\end{minipage}%
\end{figure}

The normalized $\ell_2$ 
error and data misfit are depicted in Figure~\ref{fig:scarf01error}. Figure~\ref{fig:scarf01} presents the original and reconstructions of the local volatility, as we update the values of $\mathscr F$.

After 10 steps, the distance between data and the model prices was below the tolerance, 
and we found a local volatility surface similar to the original one, even with scarce data. 
The mean of the relative error at each $(\tau_j,y_i)$ is $0.1226$, 
with standard deviation $0.1732$. We also found a set of underlying prices close to the original one, 
see Table~\ref{tab:scarf01}.

\begin{table}[!htb]
\centering
\begin{tabular}{|l|ccccc|}
\hline
 & $F_{0,\tau_1}$ & $F_{0,\tau_2}$ & $F_{0,\tau_3}$ & $F_{0,\tau_4}$ & $F_{0,\tau_5}$\\
\hline
\hline
$\mathscr{F}_{\text{true}}$ & $1.0809$ & $1.0951$ & $1.0309$ & $0.9412$ & $0.9000$\\
$\mathscr{F}^0$ & $1.0269$ & $1.0404$ & $0.9794$ & $0.8942$ & $0.8550$\\
$\mathscr{F}^{10}$ & $1.0801$ & $1.0922$ & $1.0262$ & $0.9369$ & $0.8936$\\
\hline
\end{tabular}
\caption{Values of the underlying future prices: true, initial, and after 10 steps.
The latter are clearly closer to ground truth values than the initial ones.}\label{tab:scarf01} 
\end{table}

Next we use end of the day Henry hub call option prices traded at 06-Sep-2013, 
and maturing at 29-Oct-2013, 27-Nov-2013, 27-Dec-2013, 29-Jan-2014, 26-Feb-2014, and 27-Mar-2014. 
The options were converted into European prices. To solve the inverse problem, 
we use the mesh step sizes $\Delta \tau = \tau_{\max}/254$, where $\tau_{\max}$ is the maximum time to maturity in years (with 360 days),  
and $\Delta y = 0.05$. 

We started the optimization with respect to the local volatility with the constant surface, $a^0 = a_0 \equiv 0.1013$, 
and considered the parameters $\alpha_1 = \alpha_3 = 10^{-2}\alpha_2$, $\alpha_2 = 10^{-3}$ and $\alpha_4 = \alpha_5 = \alpha_6 = 0$ 
in the functional \eqref{eq:psi}. The adjustment of the underlying future prices started with the observed ones, 
and we used the parameters $\alpha_1=\alpha_2=\alpha_3 = \alpha_6 =0$, $\alpha_4 = \alpha_5 = 1$. 
The {\em a priori} $\hat{\mathscr F}$ was taken as the set observed 
futures.

\begin{table}[!htb]
\centering
\begin{tabular}{|c|cccccc|}
\hline
Maturity & 10/29/13 & 11/27/13 & 12/27/13 & 01/29/14 & 02/26/14 & 03/27/14\\
\hline
Original & $3.62$ & $3.78$ & $3.87$ & $3.87$ & $3.83$ & $3.77$\\
Adjusted & $3.62$ & $3.82$ & $3.87$ & $3.87$ & $3.84$ & $3.77$\\
\hline
\end{tabular}
\caption{Original and adjusted Hery hub future prices for the different expiration dates.}\label{tabfut2}
\end{table}

\begin{figure}[!ht]
   \centering
           \includegraphics[width=0.45\textwidth]{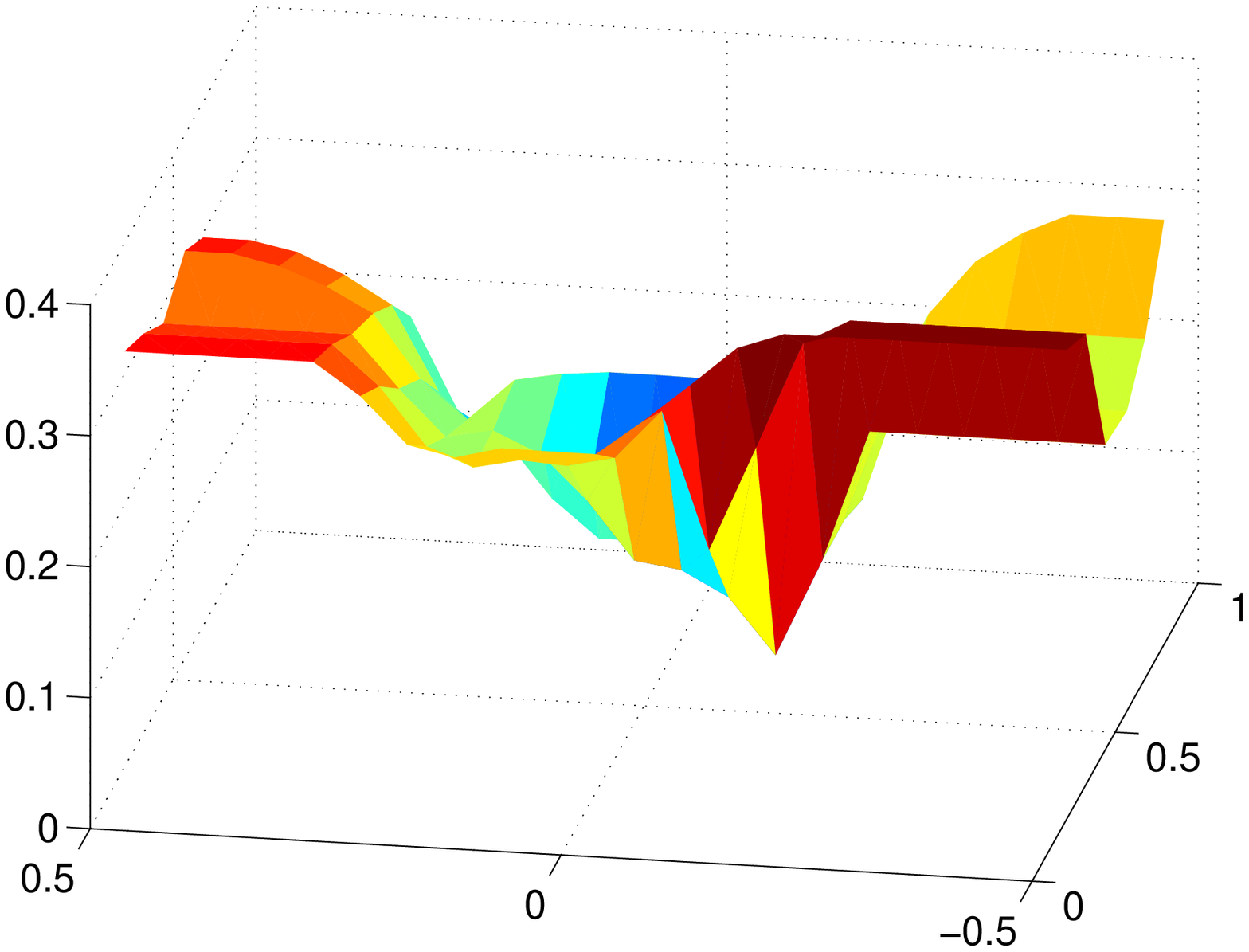}\hfill
           \includegraphics[width=0.45\textwidth]{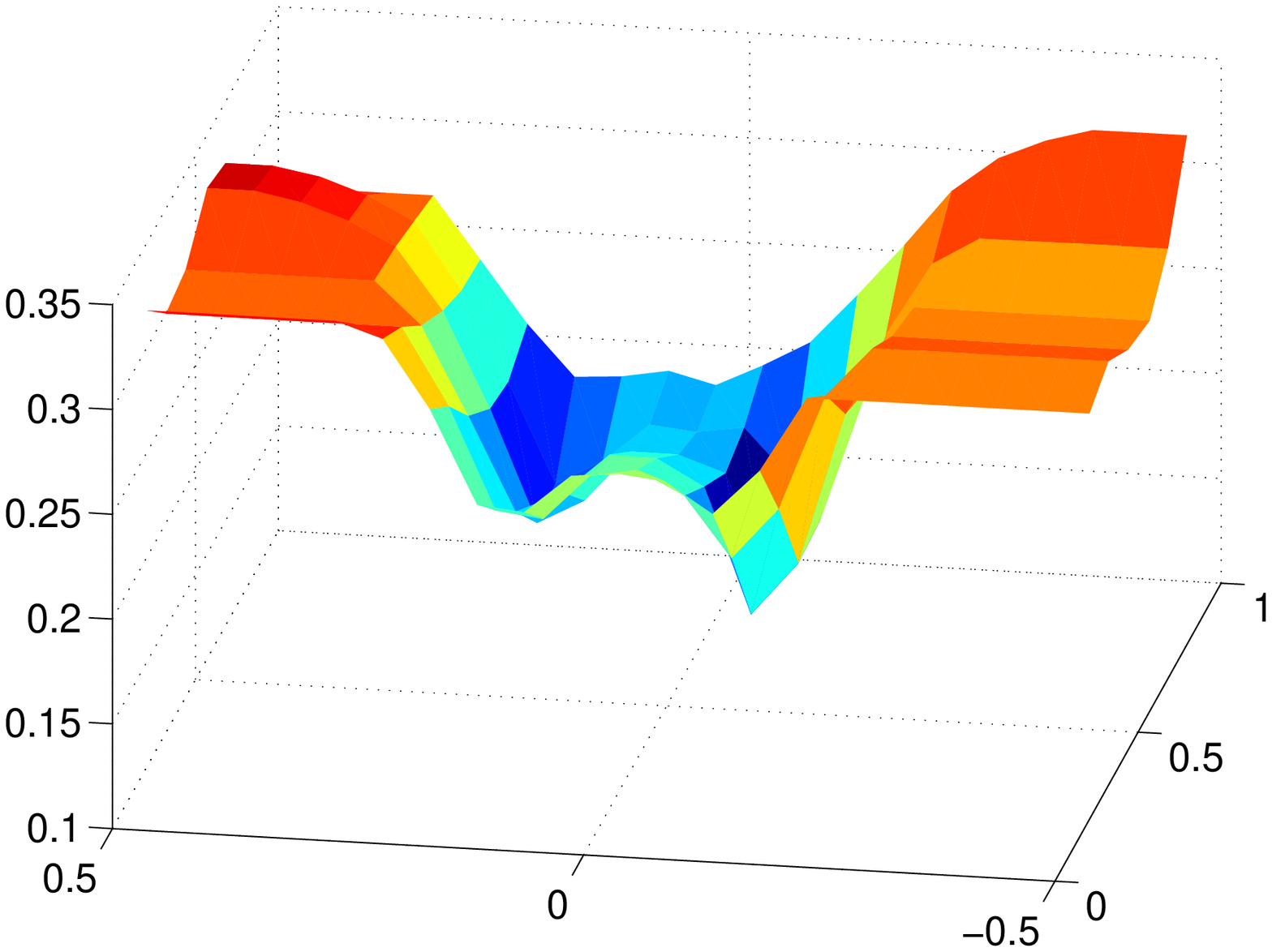}\hfill
 \caption{Reconstructed local volatility surfaces with unadjusted (left) and adjusted (right) Henry hub prices.}
  \label{fig:localvol1}
\end{figure}

\begin{figure}[!ht]
\centering
\begin{minipage}{0.49\textwidth}
  \centering
           \includegraphics[width=0.5\textwidth]{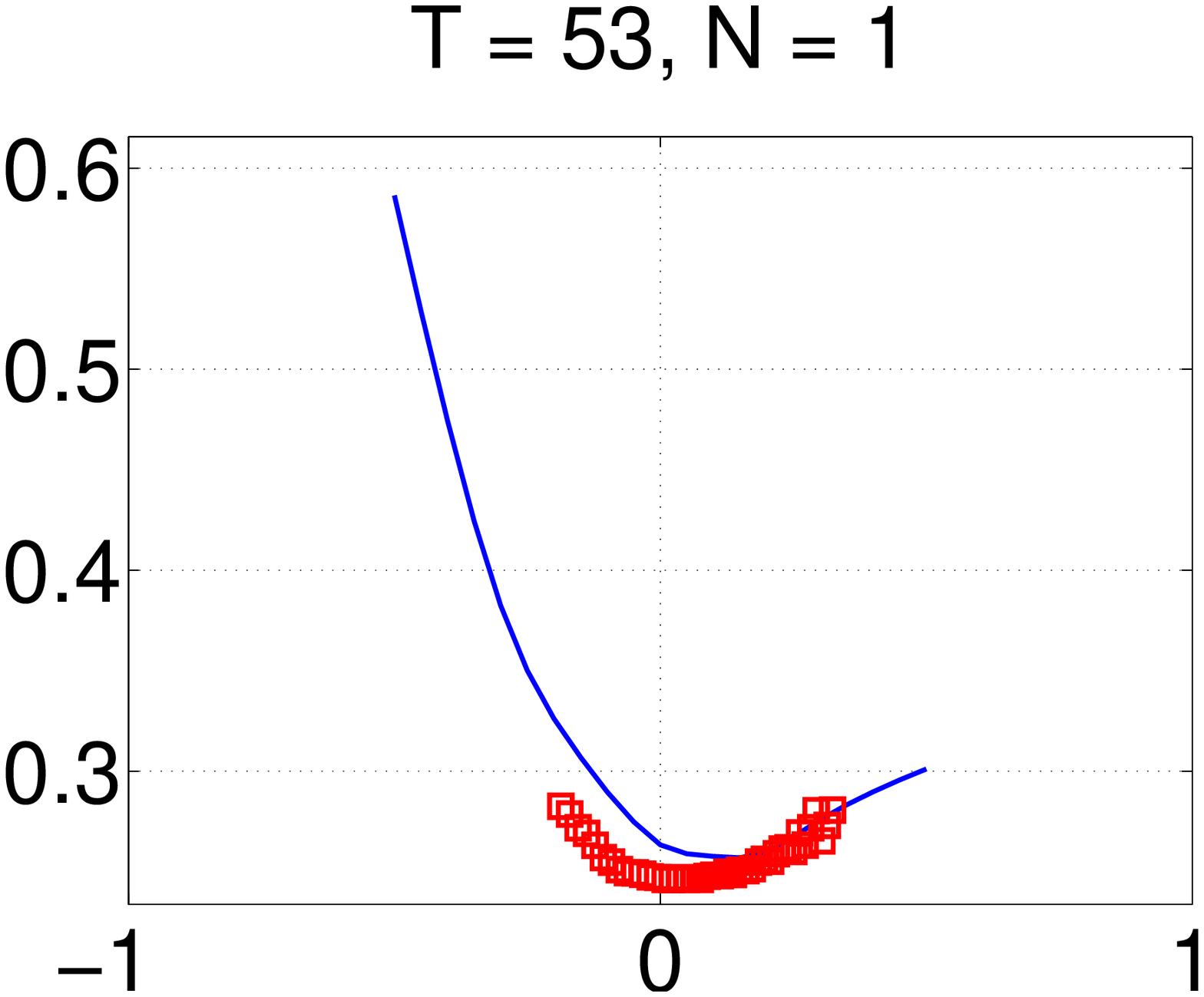}\hfill
           \includegraphics[width=0.5\textwidth]{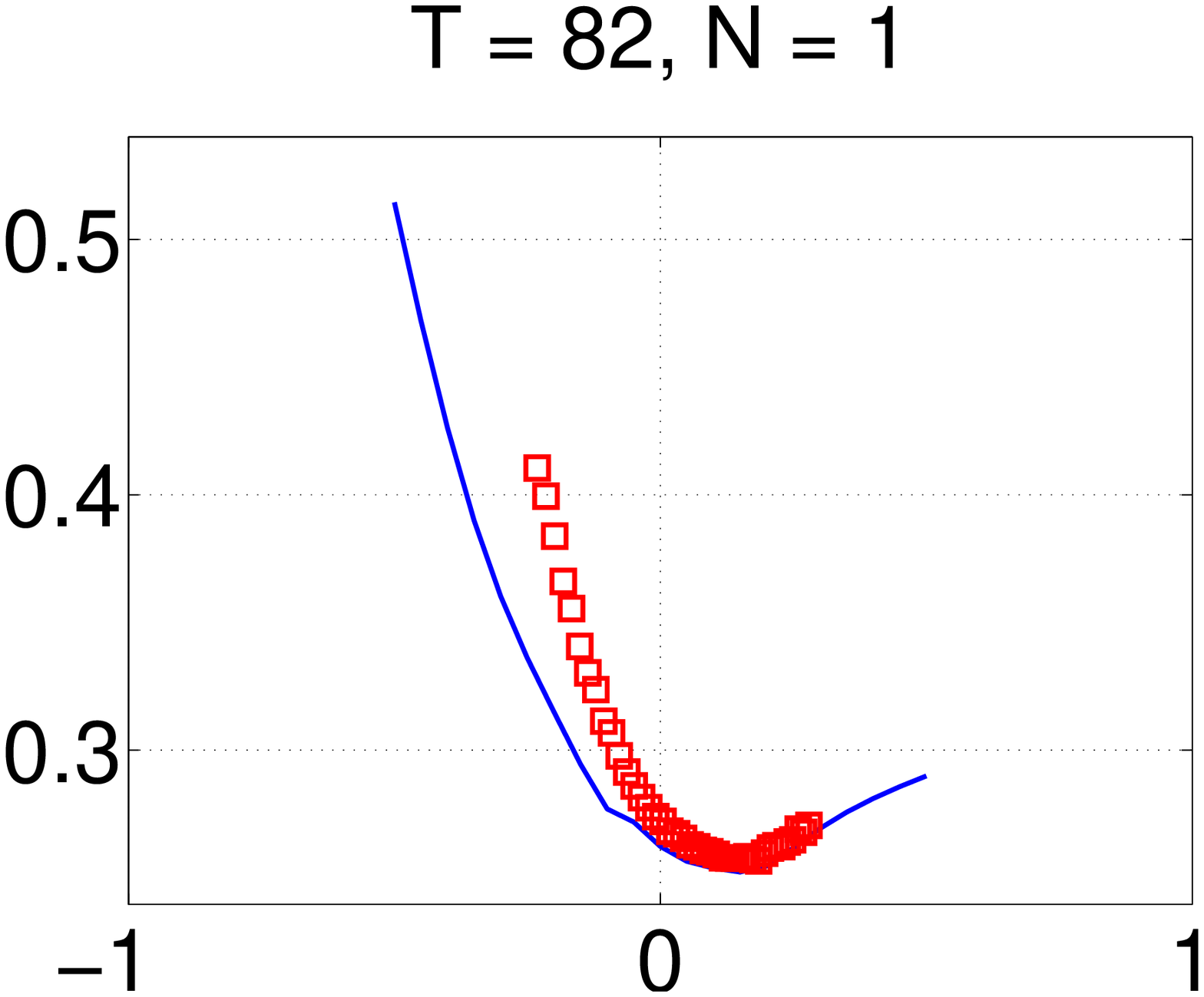}\hfill
           \includegraphics[width=0.5\textwidth]{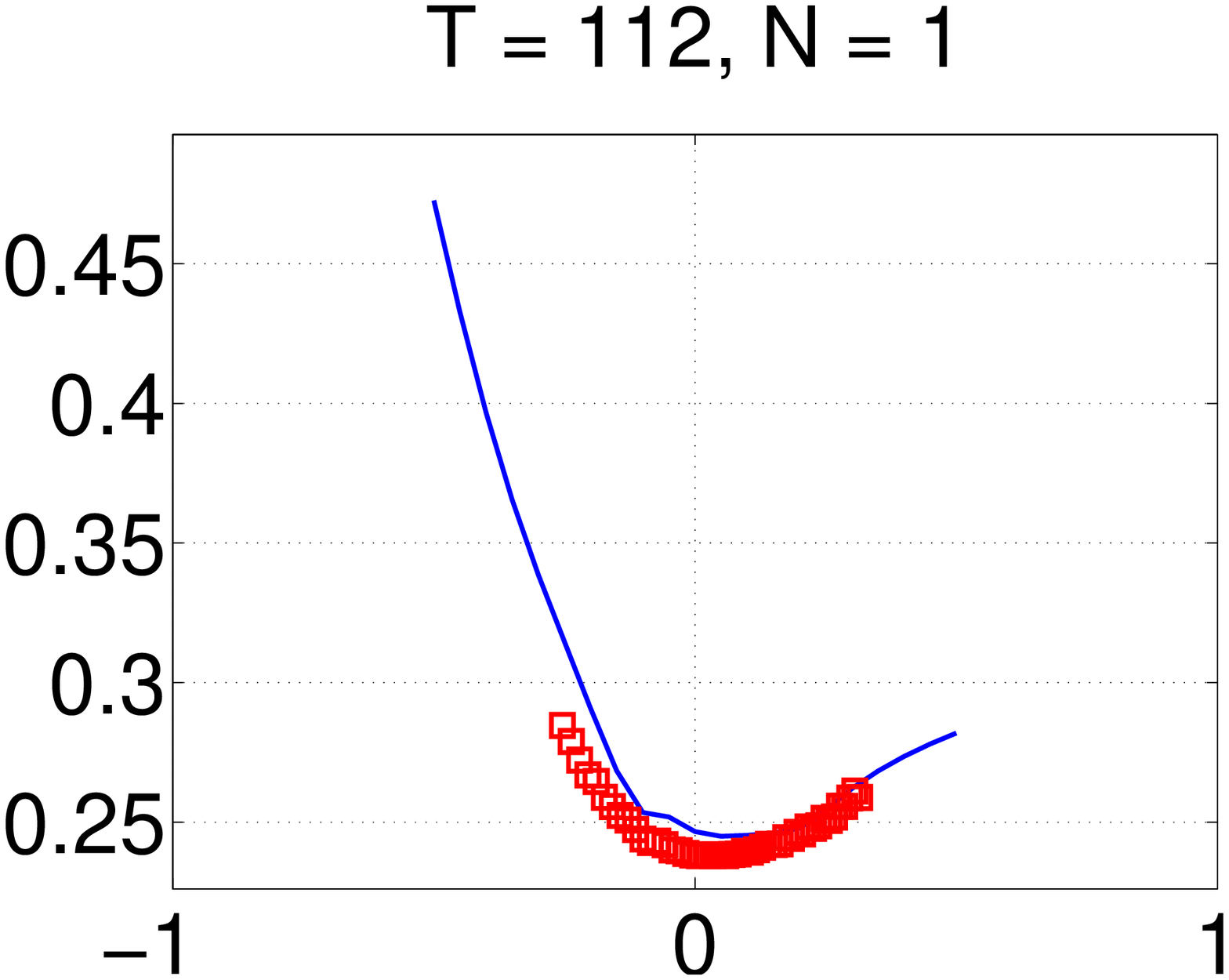}\hfill
           \includegraphics[width=0.5\textwidth]{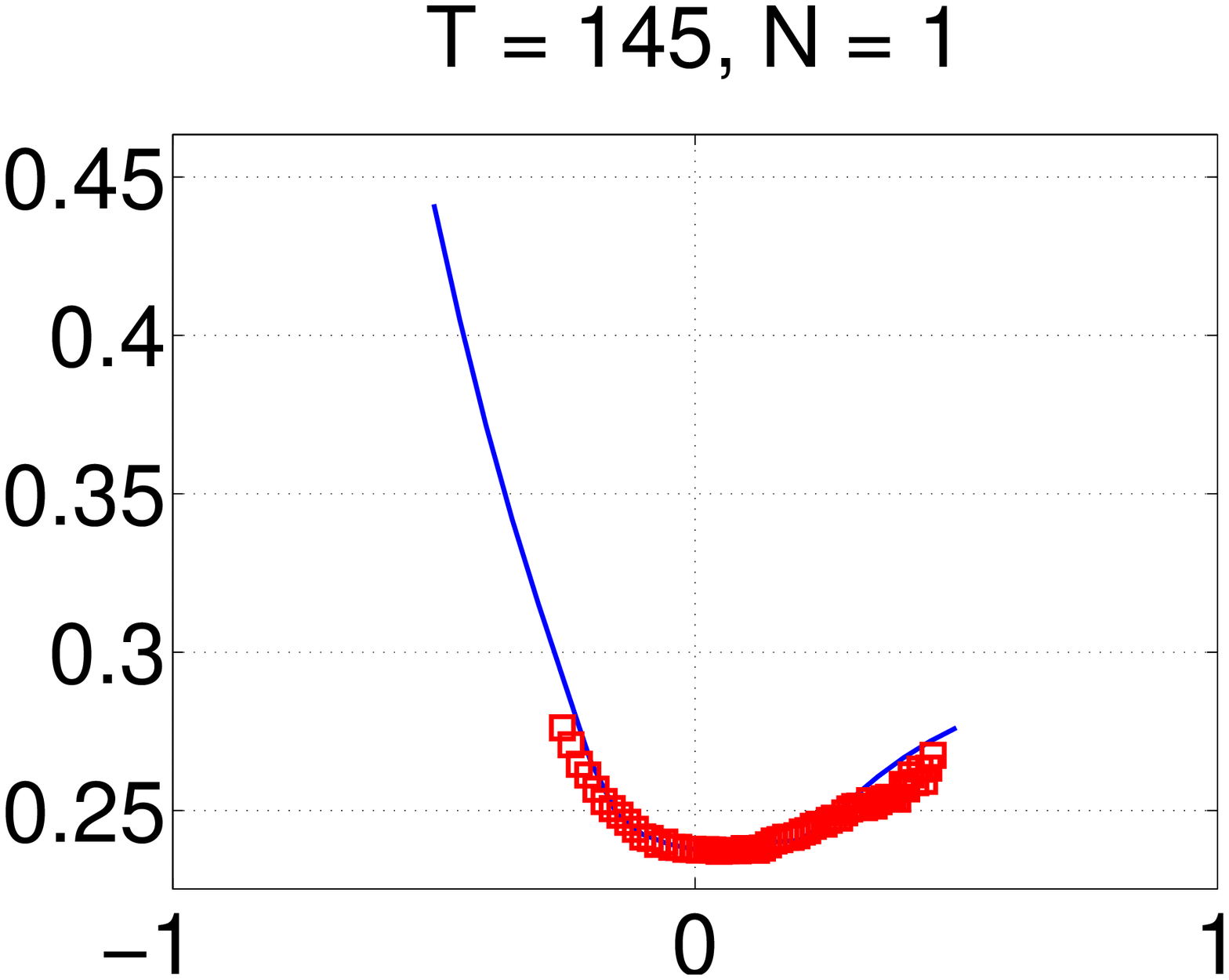}\hfill
           \includegraphics[width=0.5\textwidth]{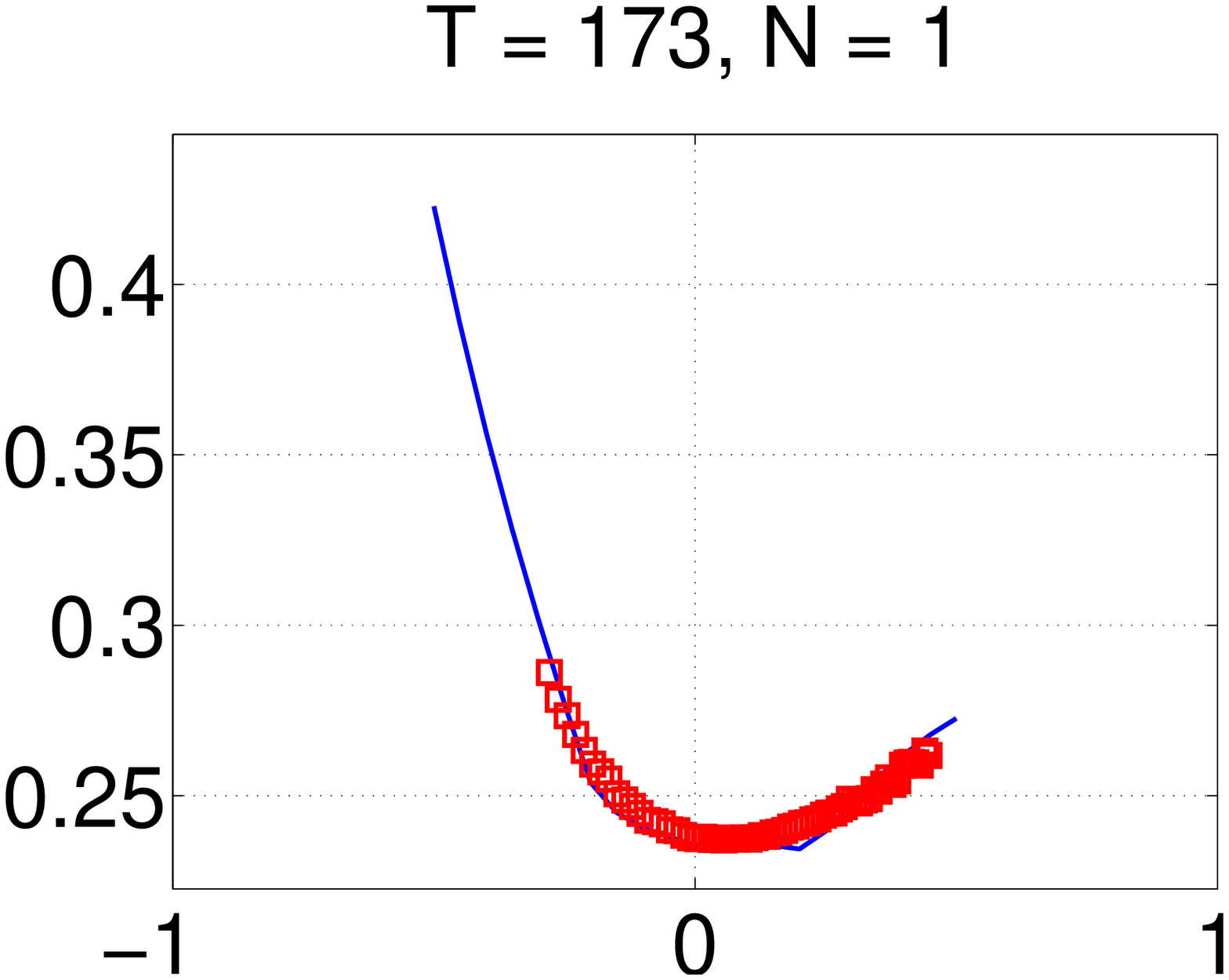}\hfill
           \includegraphics[width=0.5\textwidth]{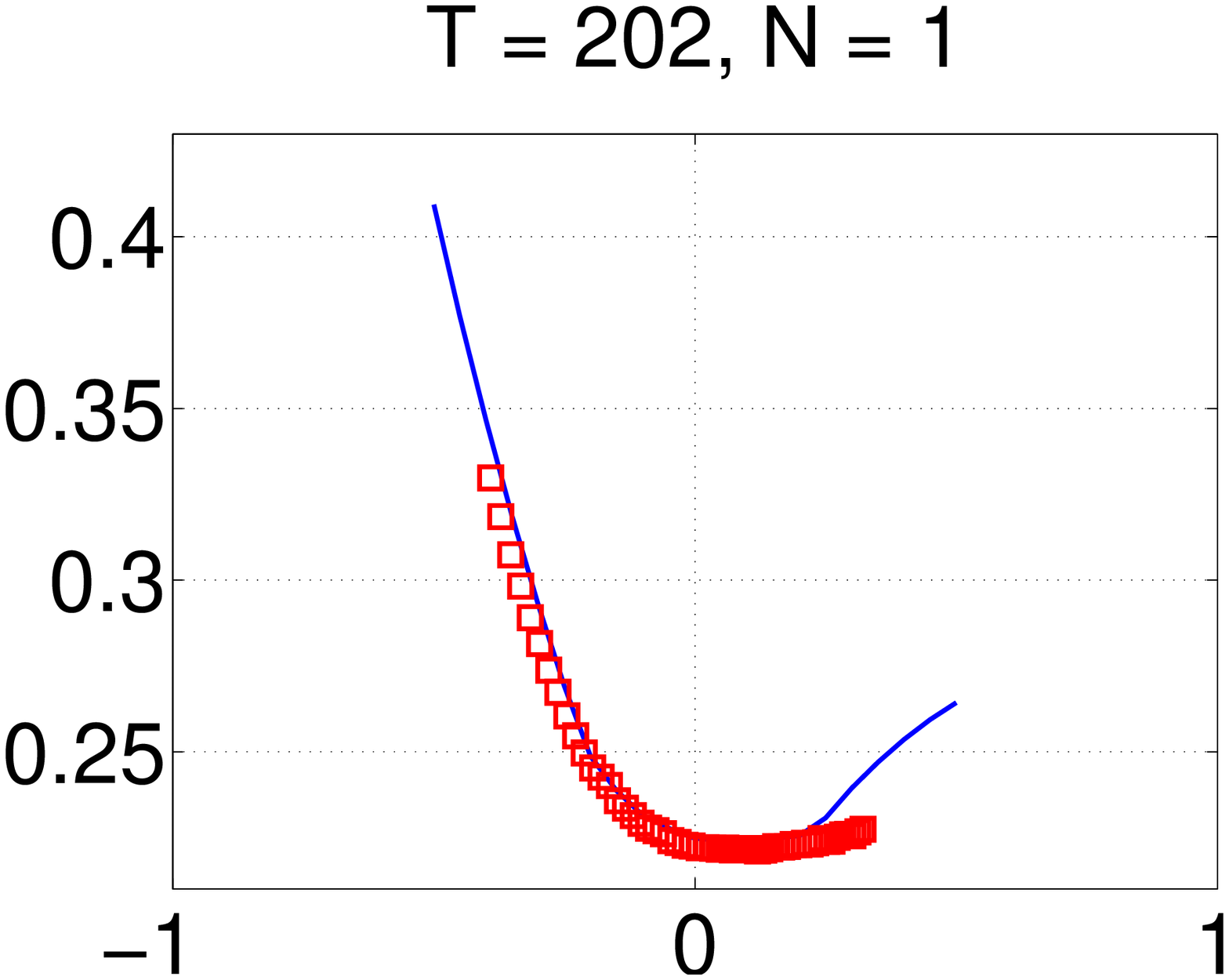}\hfill
\caption{Implied volatilities at each maturity of the market prices (squares) and of reconstructions (continuous line) with unadjusted underlying prices.}
  \label{fig:impvol11}
\end{minipage}\hfill
\begin{minipage}{0.49\textwidth}
  \centering
           \includegraphics[width=0.5\textwidth]{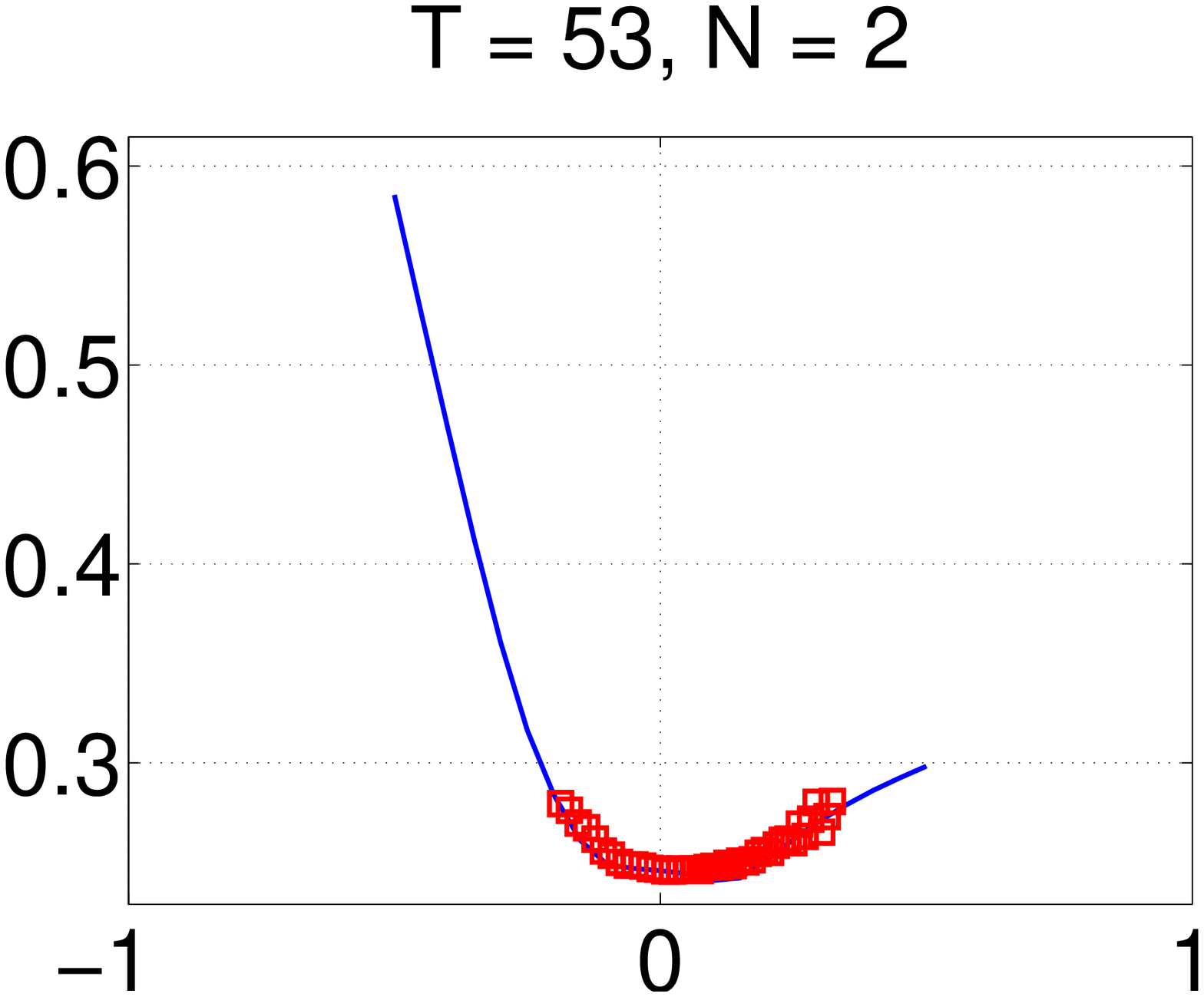}\hfill
           \includegraphics[width=0.5\textwidth]{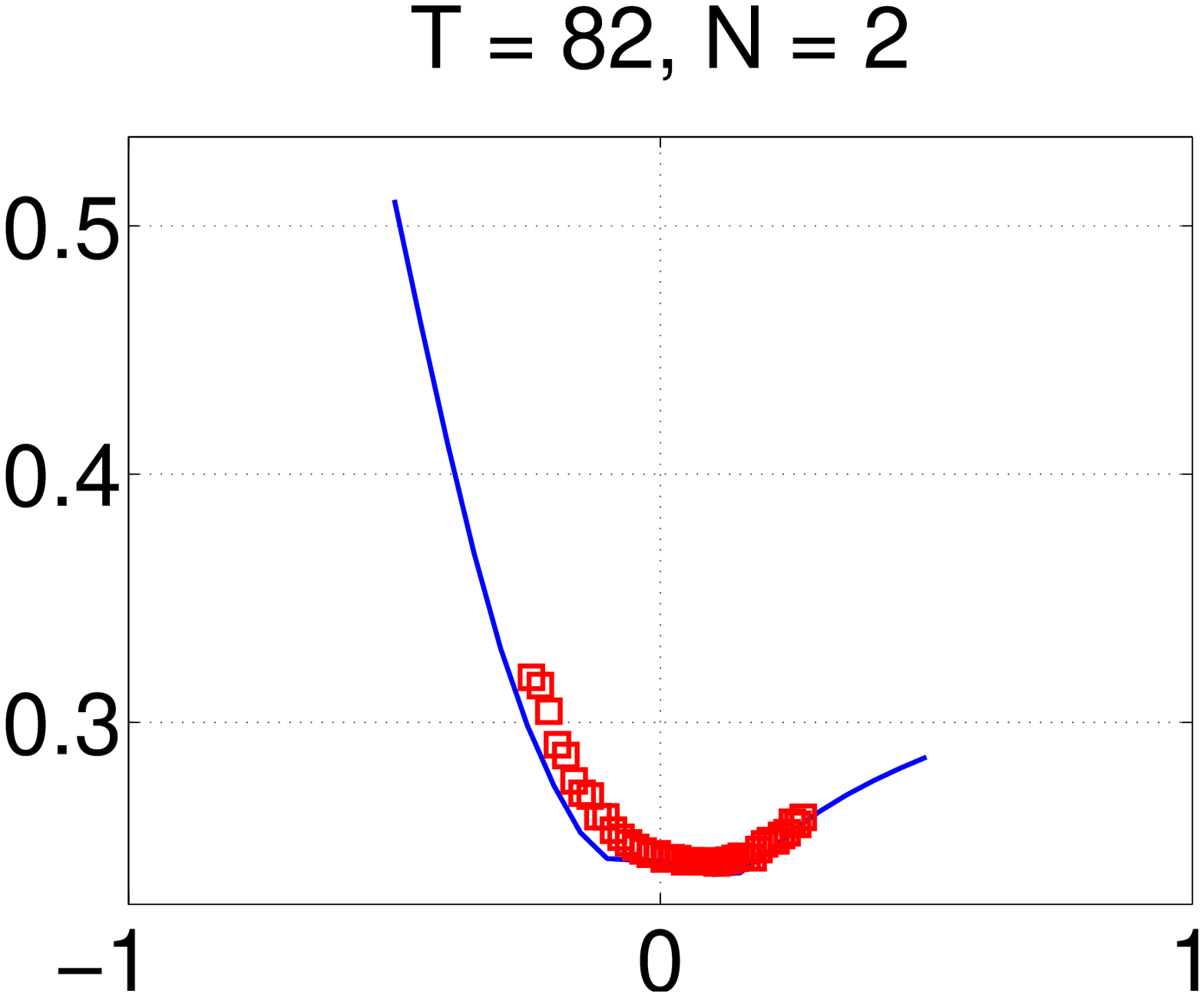}\hfill
           \includegraphics[width=0.5\textwidth]{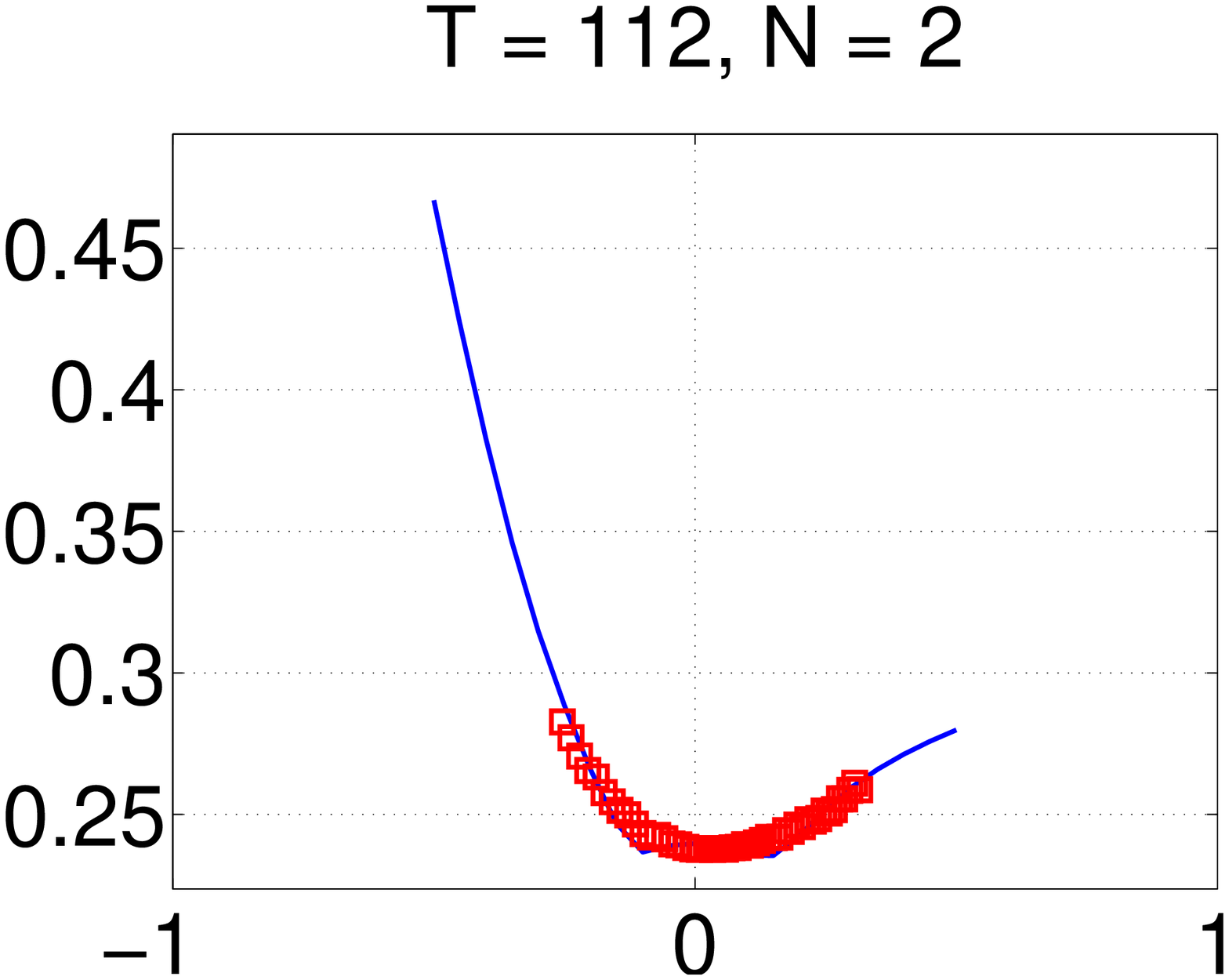}\hfill
           \includegraphics[width=0.5\textwidth]{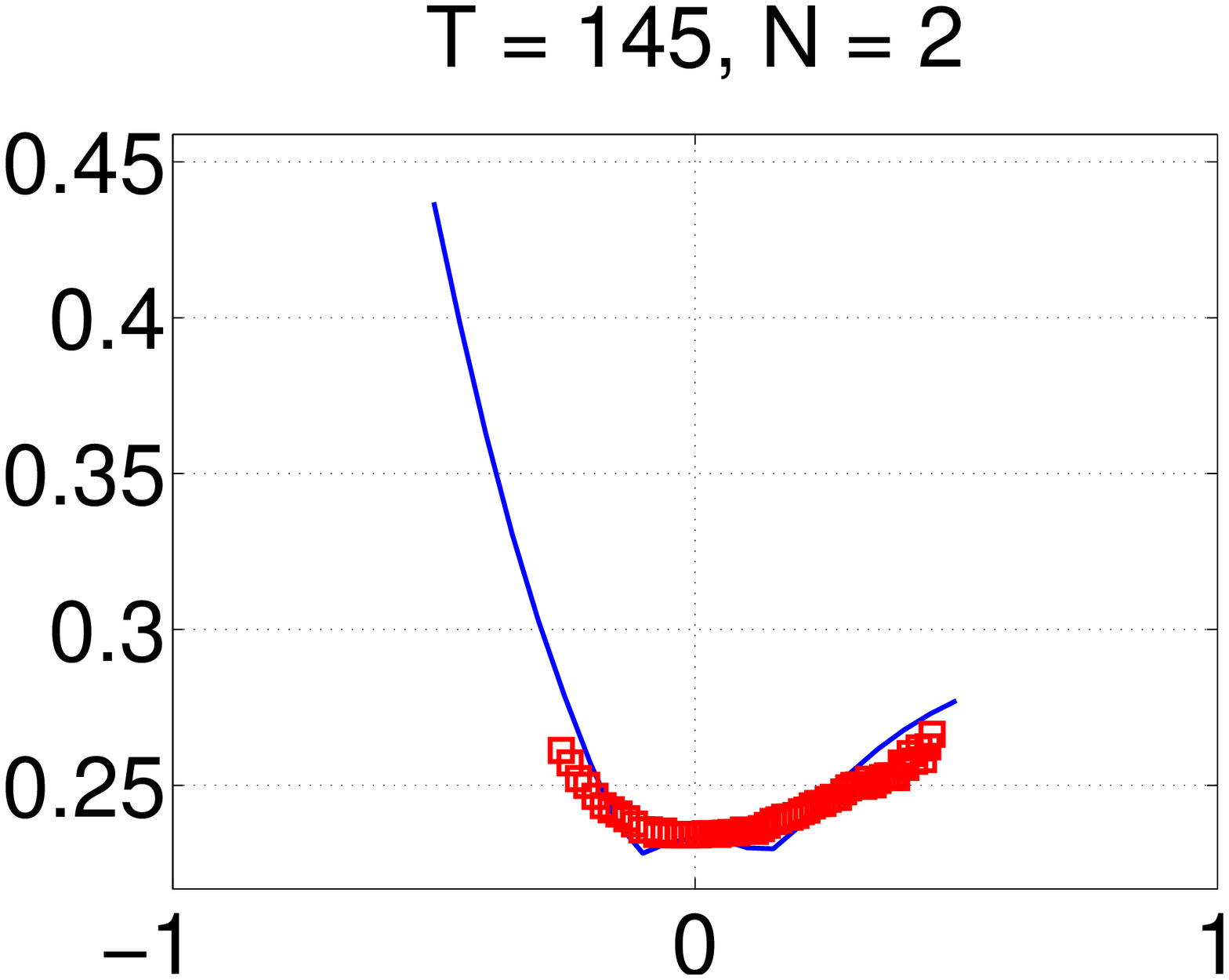}\hfill
           \includegraphics[width=0.5\textwidth]{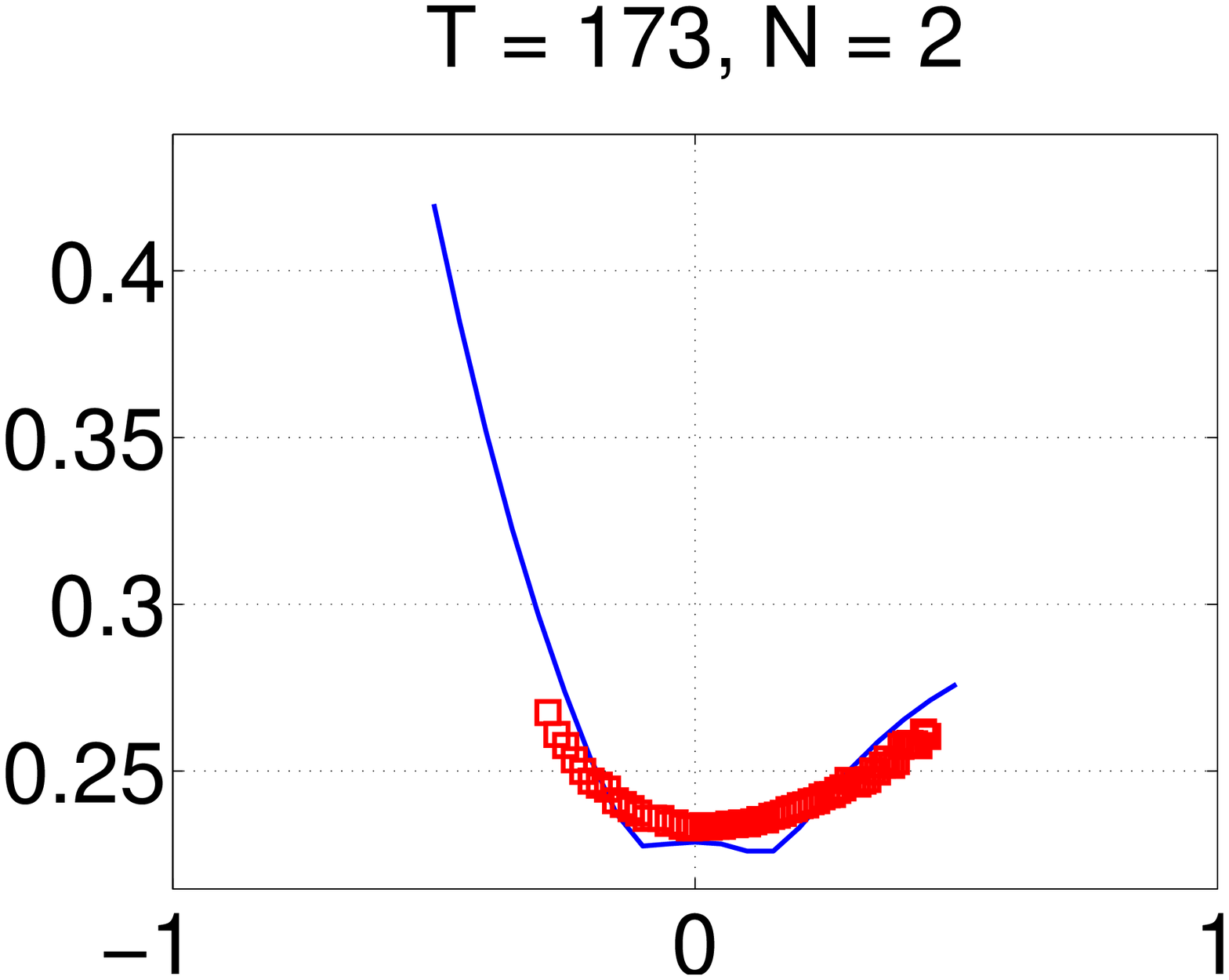}\hfill
           \includegraphics[width=0.5\textwidth]{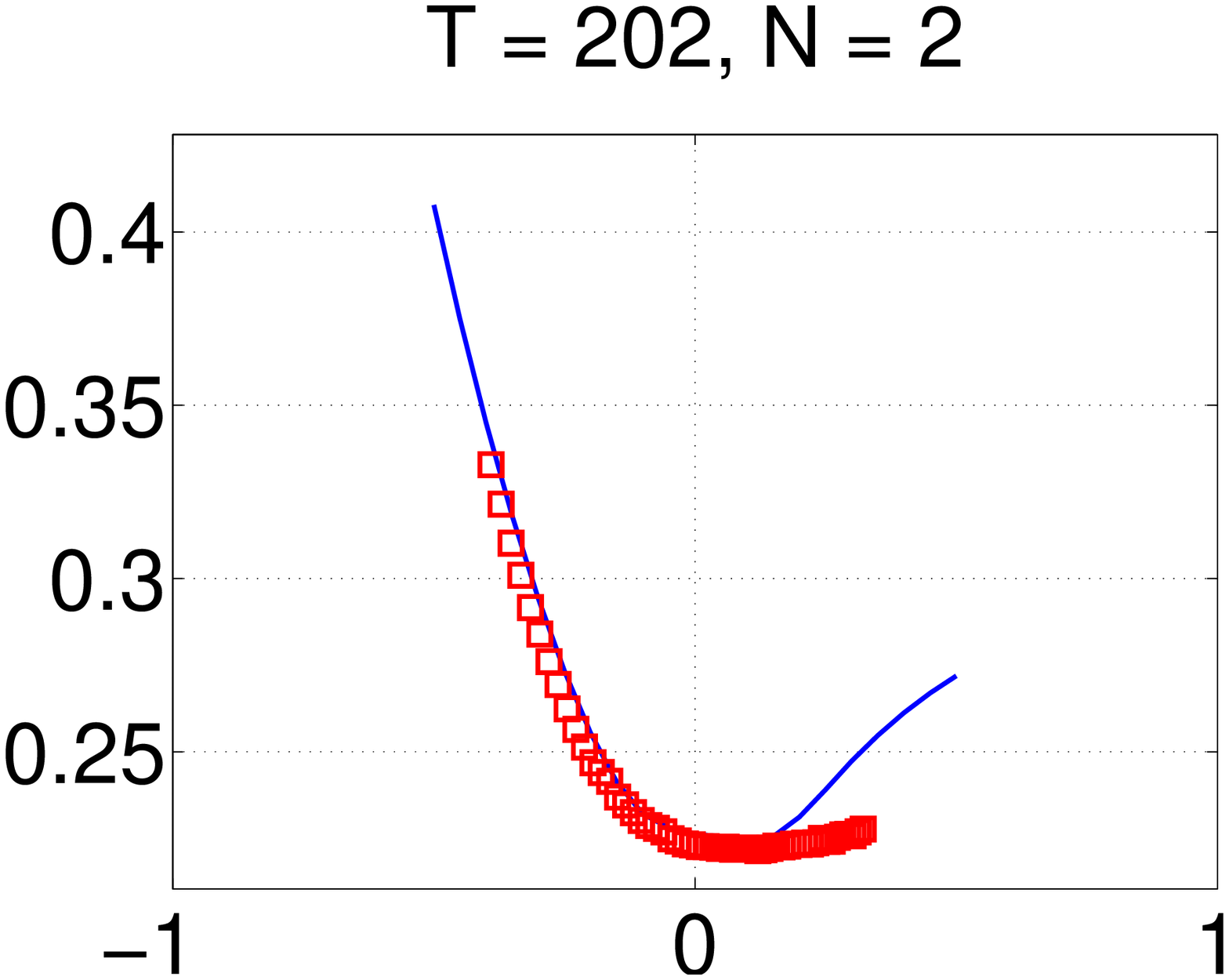}\hfill
 \caption{Implied volatilities at each maturity of the market prices (squares) and of reconstructions (continuous line) with adjusted underlying prices.}
  \label{fig:impvol12}
\end{minipage}%
\end{figure}

We set the tolerance for the normalized data misfit $R(a)$ as $0.012$. So, after two iterations, we found $R(a) = 0.01199$, and the future prices at the maturities $11/27/2013$ and $02/26/2014$, in Table~\ref{tabfut2}, changed a little bit. This was sufficient to improve the adherence to the market smile, as we can see in Figures~\ref{fig:impvol11} and \ref{fig:impvol12}, comparing the implied volatilities before and after this adjustment. This change can also be seen in the reconstructed local volatilities, where a more parsimonious surface is obtained after this correcting step. See Figure~\ref{fig:localvol1}. These results lead us to conclude that, if we include the underlying asset prices as additional unknowns, then upon adjusting them we can improve reconstruction quality. 

\subsection{The {\em Online} Approach with Synthetic Data}\label{sec:synthetic}
With this synthetic example we next illustrate that reconstructions are improved 
as we increase the call prices dataset. 
So, we first generate European call option prices with the volatility surface \eqref{vol}, using $\tau_{\max} = 0.5$, and the mesh steps $\Delta\tau = 0.005$ and $\Delta y = 0.025$, for different values of the underlying asset $S_0$. Then, we apply a relative noise as in \eqref{eq:noise}, and collect the resulting prices at the maturities $\tau_i = i\cdot 0.1$, $i=1,...,5$, and at the strikes $y_j = j\cdot 0.05$, with $j=-10,-9,...,0,1,...,10$. The values taken for the underlying asset are $S_0(l) = 0.8 + l\cdot\Delta s$, with the step sizes $\Delta s = 0.1, 0.05, 0.025$. 
Thus, we reconstruct the local volatility surface under the online setting using $1$, $5$, $9$ and $17$ sets of option prices. 
 
We started the minimization with the constant surface $\af^0 \equiv 0.08$, and stop the iterations, whenever the normalized $\ell_2$-data misfit was below $0.011$.

When using a unique set of option prices, we set $\alpha_2 = 10^{-3}$; 
when using more than one, we set $\alpha_2 = 10^{-4}$. 
The weights in \eqref{eq:psi} were taken as $\alpha_1 = 10^{-3}\alpha_2$, $\alpha_3 = 0.5\alpha_2$, $\alpha_4=\alpha_5=0$ and $\alpha_6 = 10^{-2}\Delta s^2/\Delta y^2\alpha_2$. Note that, in this example, we do not correct the underlying asset prices as in Section~\ref{sec:adjust}.

\begin{figure}[!ht]
\centering
\begin{minipage}{.44\textwidth}
  \centering
      \includegraphics[width=0.8\textwidth]{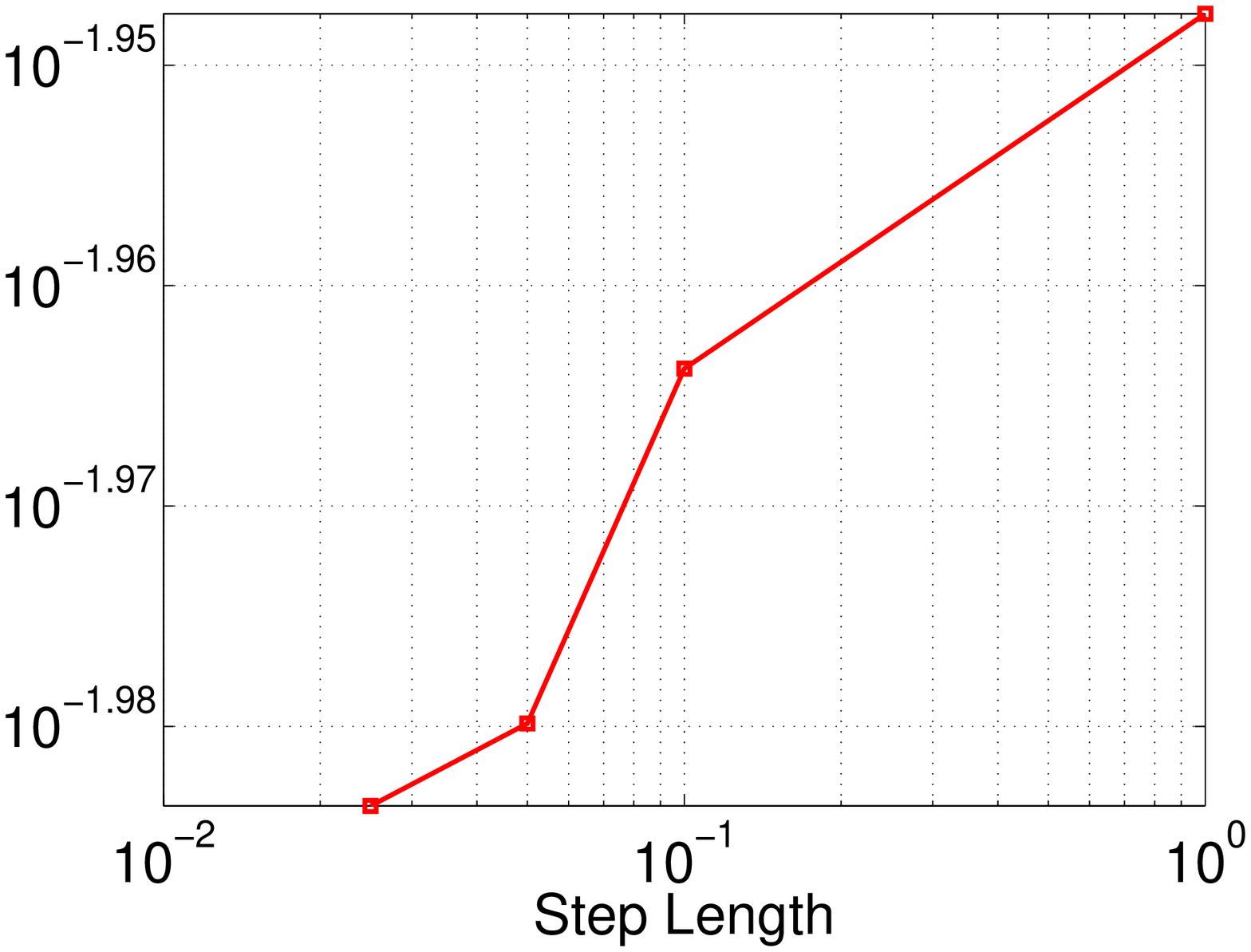}
      \includegraphics[width=0.8\textwidth]{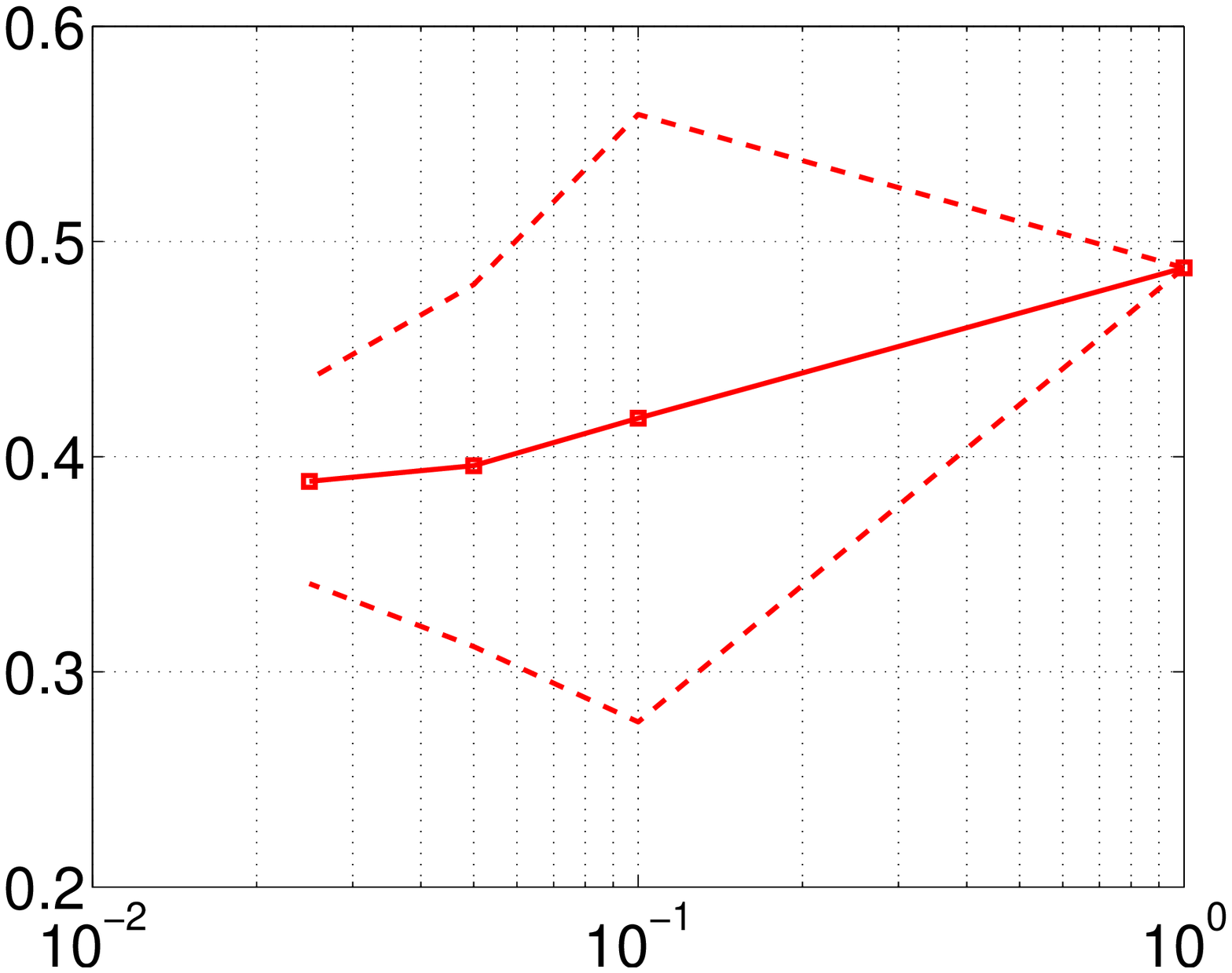}\hfill
  \caption{Above: Normalized data misfit vs $\Delta s$ (squares). Below: Mean (squares) 
	and standard deviation (dashed line) of normalized $\ell_2$-error in reconstructions.}
  \label{onlineberr}
\end{minipage}\hfill
\begin{minipage}{.54\textwidth}
  \centering
      \includegraphics[width=0.5\textwidth]{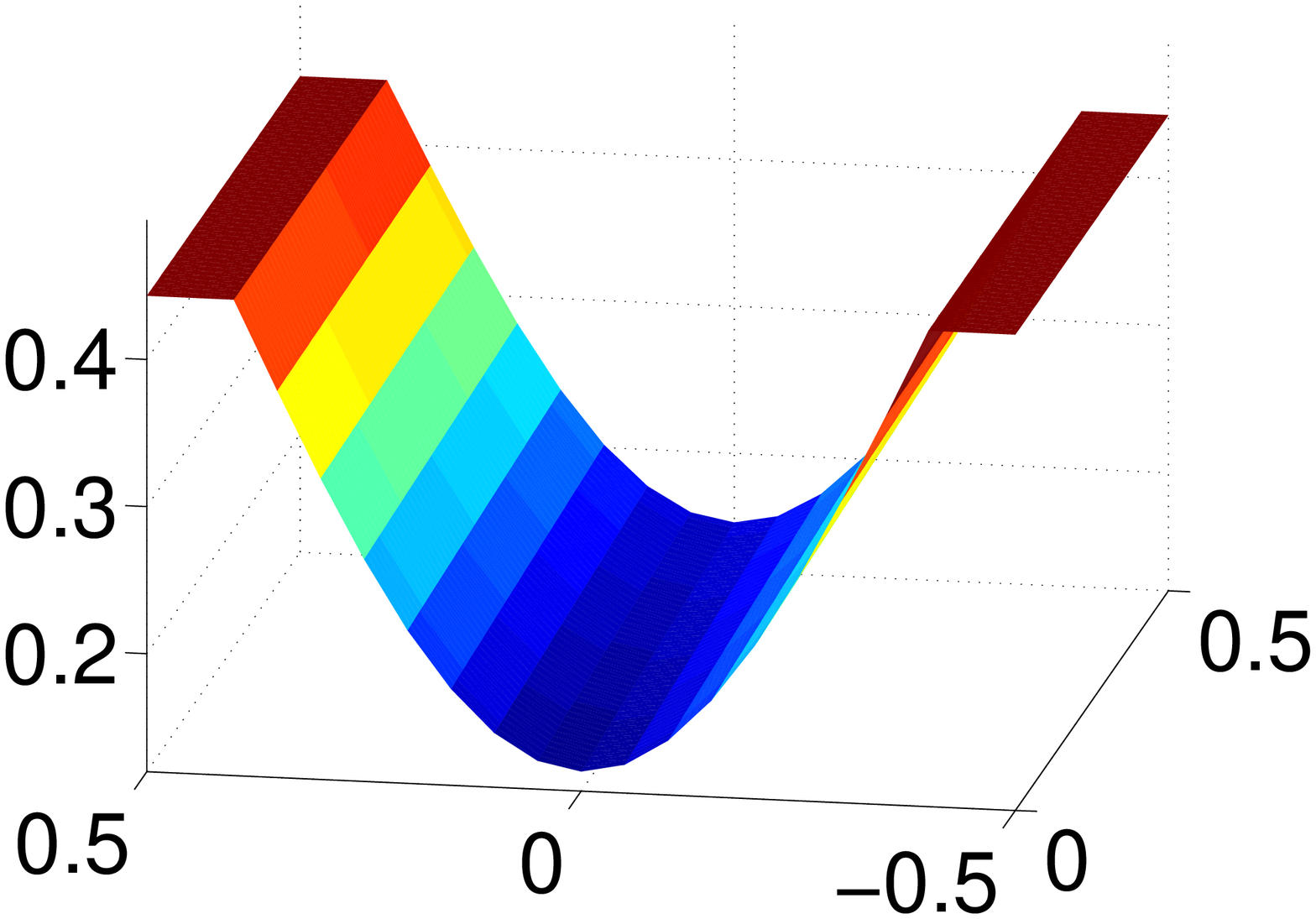}\hfill
      \includegraphics[width=0.5\textwidth]{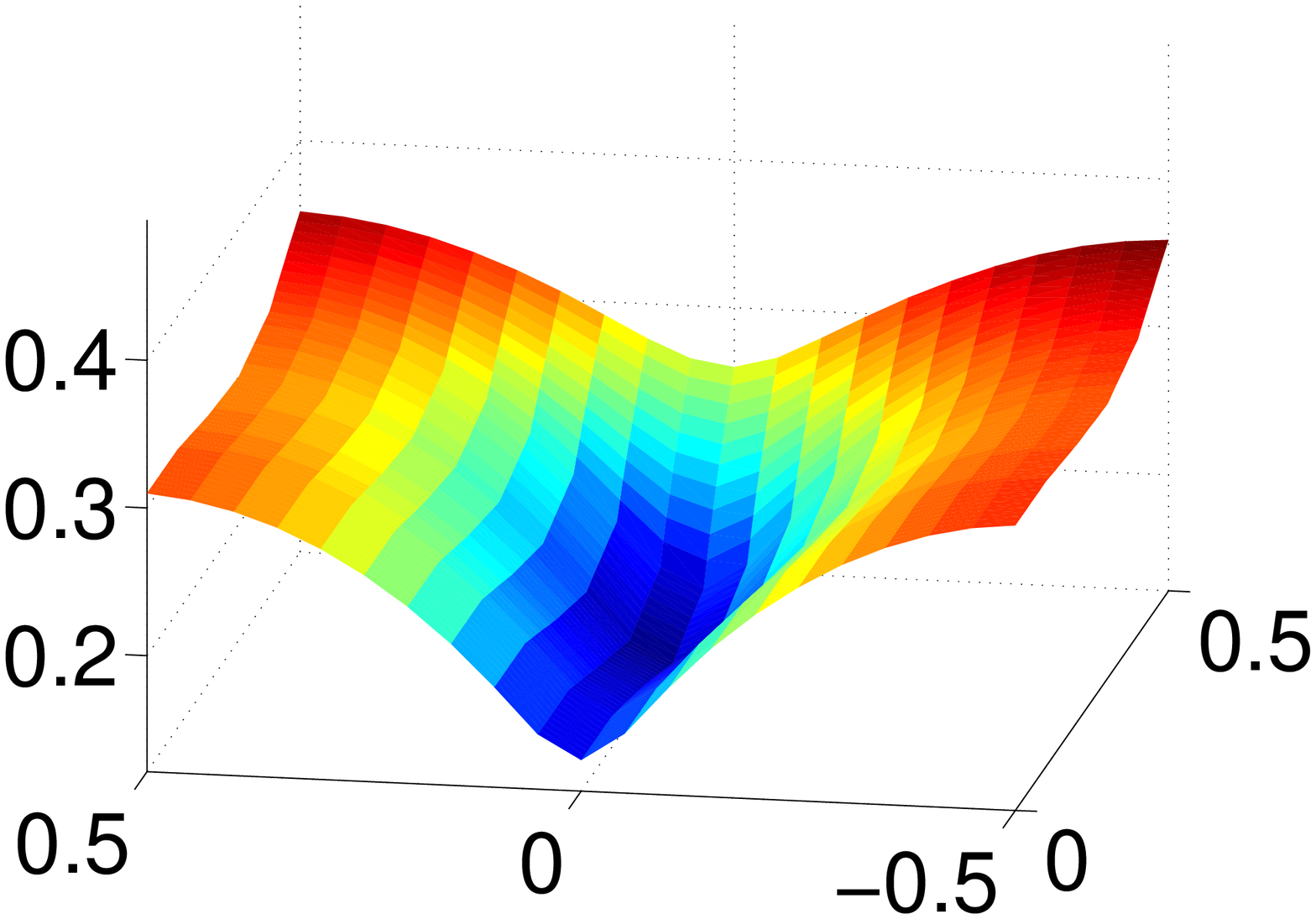}\hfill
      \includegraphics[width=0.5\textwidth]{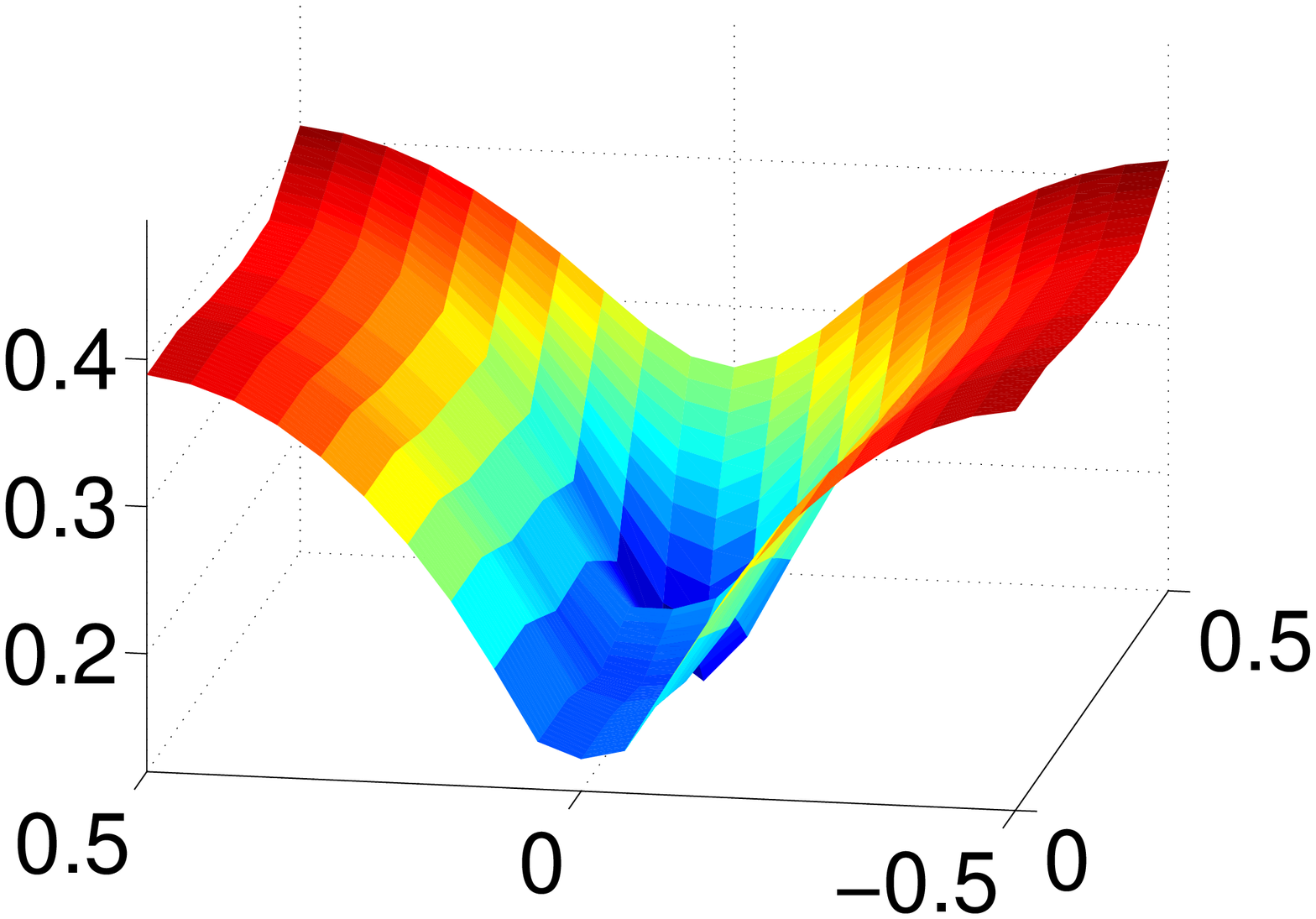}\hfill
      \includegraphics[width=0.5\textwidth]{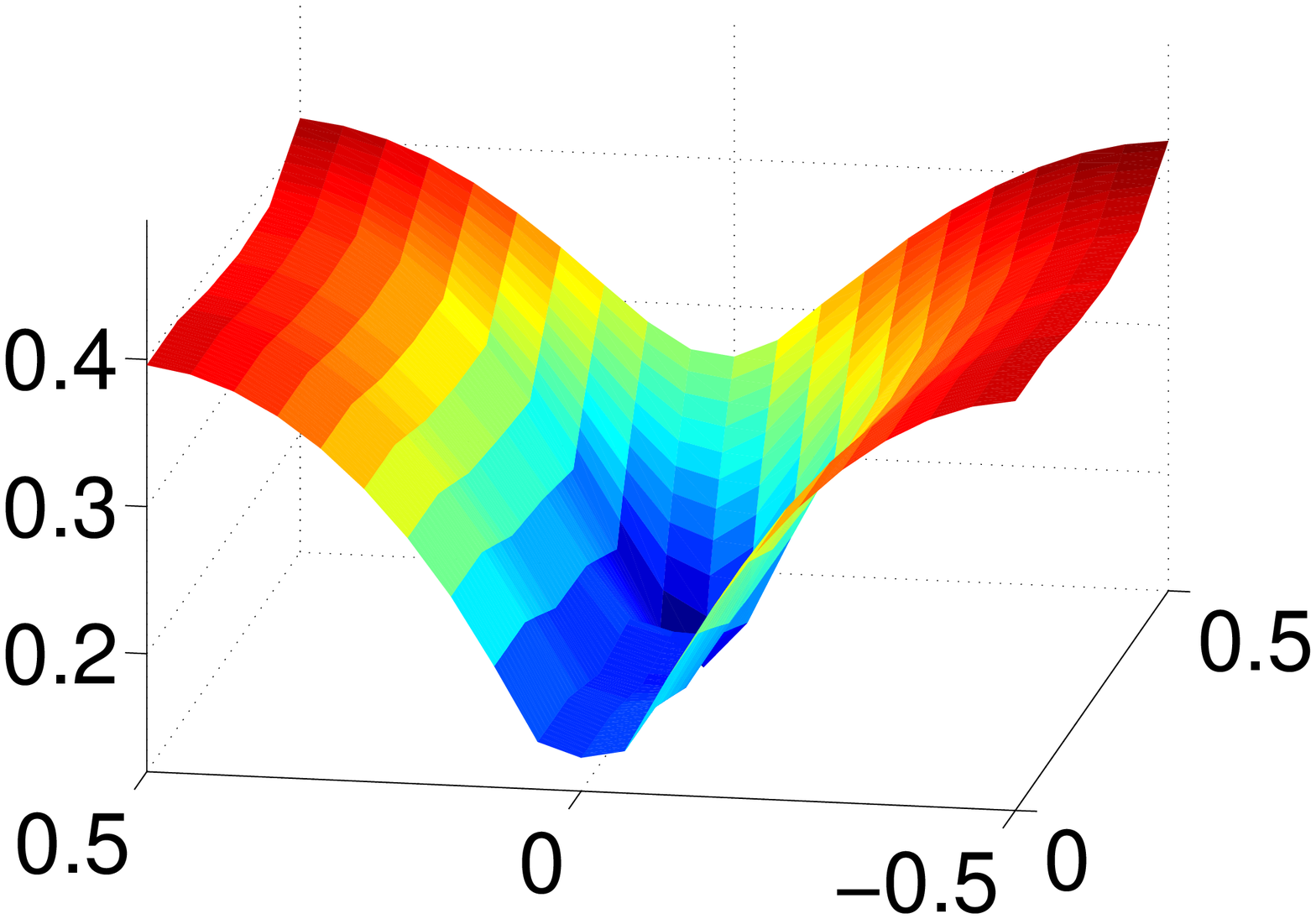}\hfill
      \includegraphics[width=0.5\textwidth]{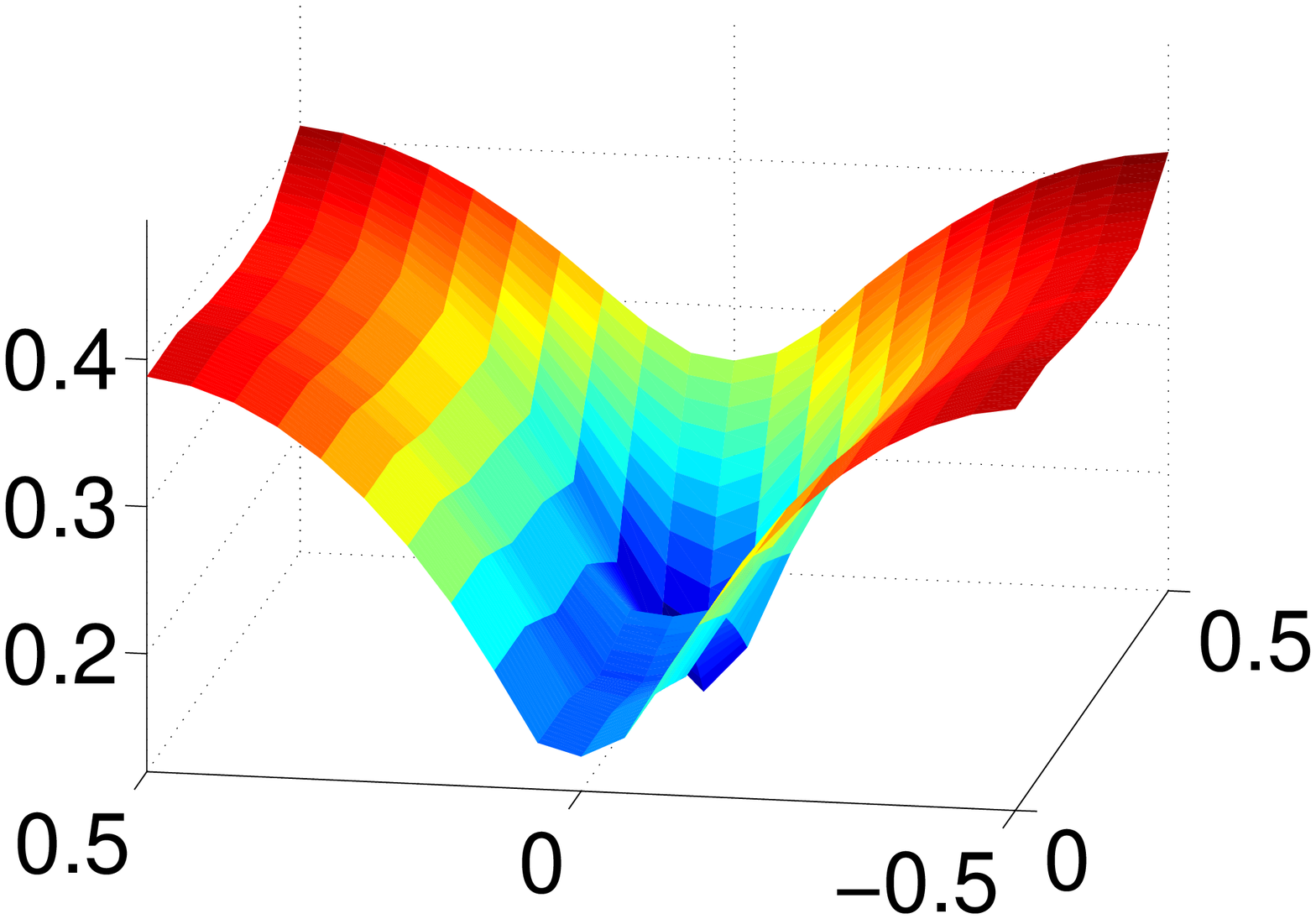}\hfill
  \caption{Left, first line: Original local volatility surface. The others: Reconstructions using different values of $\Delta s$.}
  \label{onlineb}
\end{minipage}%
\end{figure}

As can be seen in Figure~\ref{onlineberr}, both the $\ell_2$-error and the data misfit decreases as we decrease the step $\Delta s$, thus increasing the number of sets of call option prices. 
The error is $20\%$ smaller when using 17 option price sets than when using only one. 
Looking at the reconstructed local volatility in Figure~\ref{onlineb}, 
the solutions become more similar to the original one as we increase the dataset, although the calibration was never perfect, since the data is sparse and noisy.

\subsection{The {\em Online} Approach with Henry Hub Prices}\label{sec:hh}
The aim of this test is to illustrate that the online model can be 
beneficially used with market data. 
In addition, it has a smile adherence at least as good as the SVI model presented by \citep{gatheralsvi2}.

We perform some tests using the online calibration with Henry Hub call option prices, the option prices were traded at 04-Sep-2013, 05-Sep-2013, 06-Sep-2013, 09-Sep-2013 and 10-Sep-2013. We used the six first maturities, i.e., the future contracts, and the corresponding vanilla call options, maturing at 29-Oct-2013, 27-Nov-2013, 27-Dec-2013, 29-Jan-2014, 26-Feb-2014 and 27-Mar-2014. 

In this example we used the mesh step lengths $\Delta y = 0.05$, $\Delta \tau = 0.0025$ and $\Delta s = 0.05$. We also used the Tikhonov functional \eqref{tik2} with the penalization \eqref{eq:psi}, with the parameters $\alpha_1 = 2.0\times 10^{-3}\alpha_2$, $\alpha_2 = 5.0\times 10^{-5}$, $\alpha_3 = 2.0\times 10^{-2}\alpha_2$ $\alpha_4=\alpha_5 = 0$, and $\alpha_6 = 2.0\times 10^{-3} \cdot \Delta s^2/\Delta y^2\alpha_2$. We set the {\em a priori} local volatility surface as the constant surface $\mathcal A_0 \equiv 6.13\times 10^{-2}$. We also used it to initialize the minimization algorithm. 

The interest rate and the convenience yield were taken as $0.0325$. Since the call options are American, we converted them into European prices using the methodology of Section~\ref{sec:transformation}.

The data was given in the sparse mesh defined by transforming the market strikes into log-moneyness, and considering the time to maturity in years. Since the time mesh has different lengths, as the underlying asset evolves, we extended the surfaces until $\tau = 0.8$, repeating the set of prices 
corresponding to the larger time to maturity.

We stopped the iterations, whenever the data misfit \eqref{eq:residual2} was below the tolerance, taken as $tol = 0.024$. 
\begin{figure}[!ht]
\centering
\begin{minipage}{.49\textwidth}
\begin{minipage}{1\textwidth}
  \centering
      \includegraphics[width=0.6\textwidth]{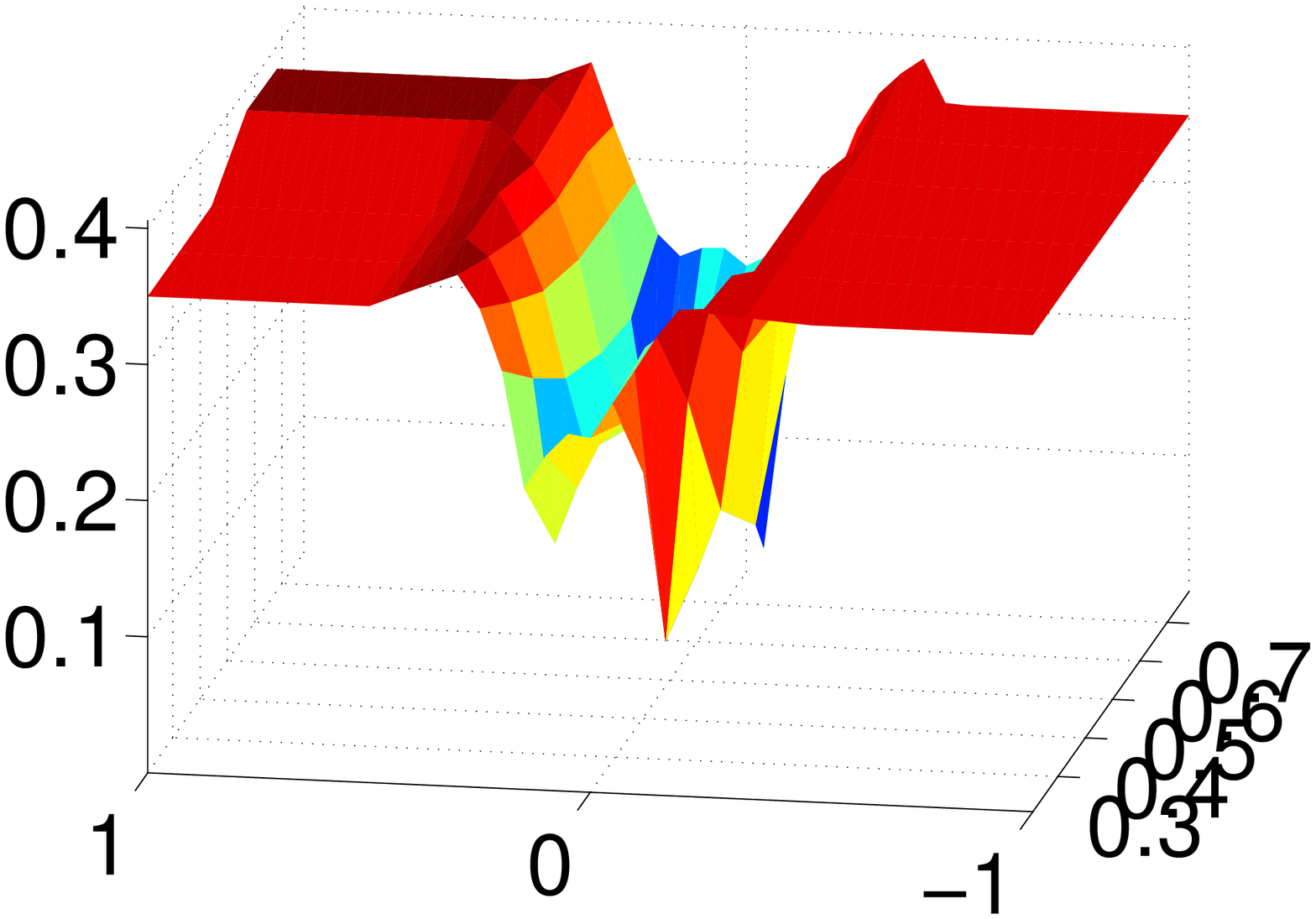}
\caption{Reconstructed Henry Hub local volatility surface. Trading date: 06-Sep-2013.}
  \label{hhonline_volsurf}
\end{minipage}\hfill
\begin{minipage}{1\textwidth}
\centering
      \includegraphics[width=0.48\textwidth]{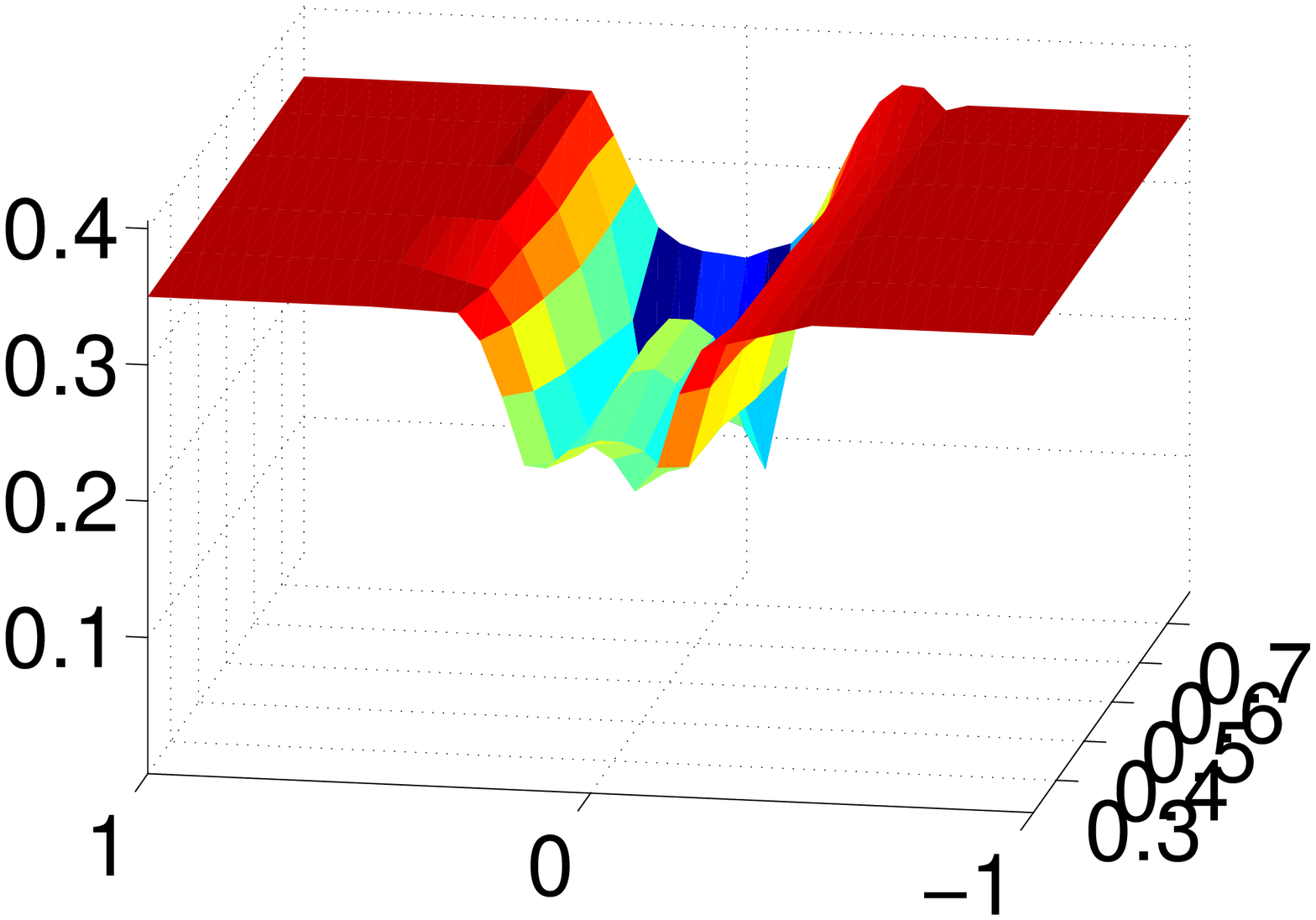}\hfill
      \includegraphics[width=0.48\textwidth]{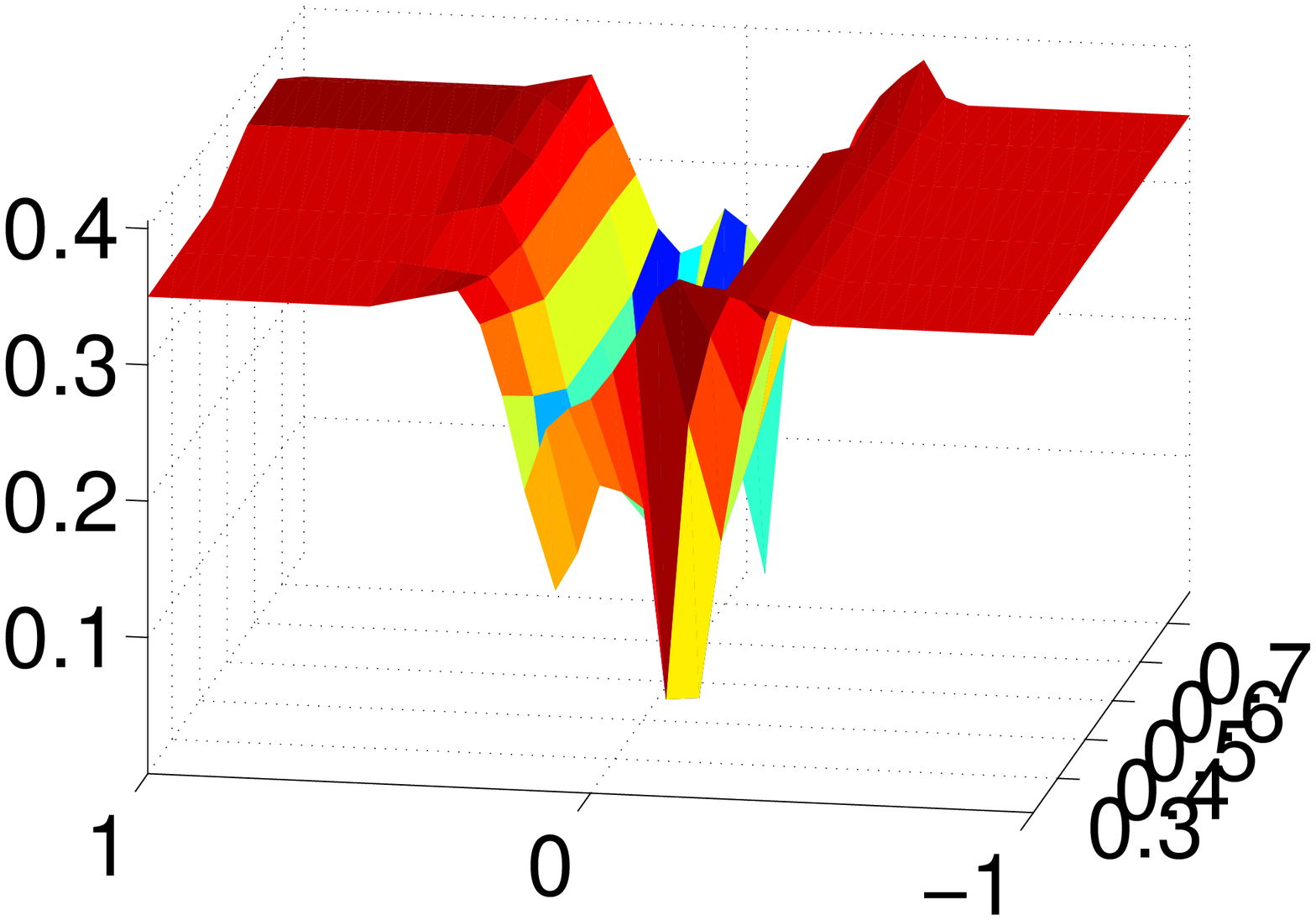}\hfill
      \includegraphics[width=0.48\textwidth]{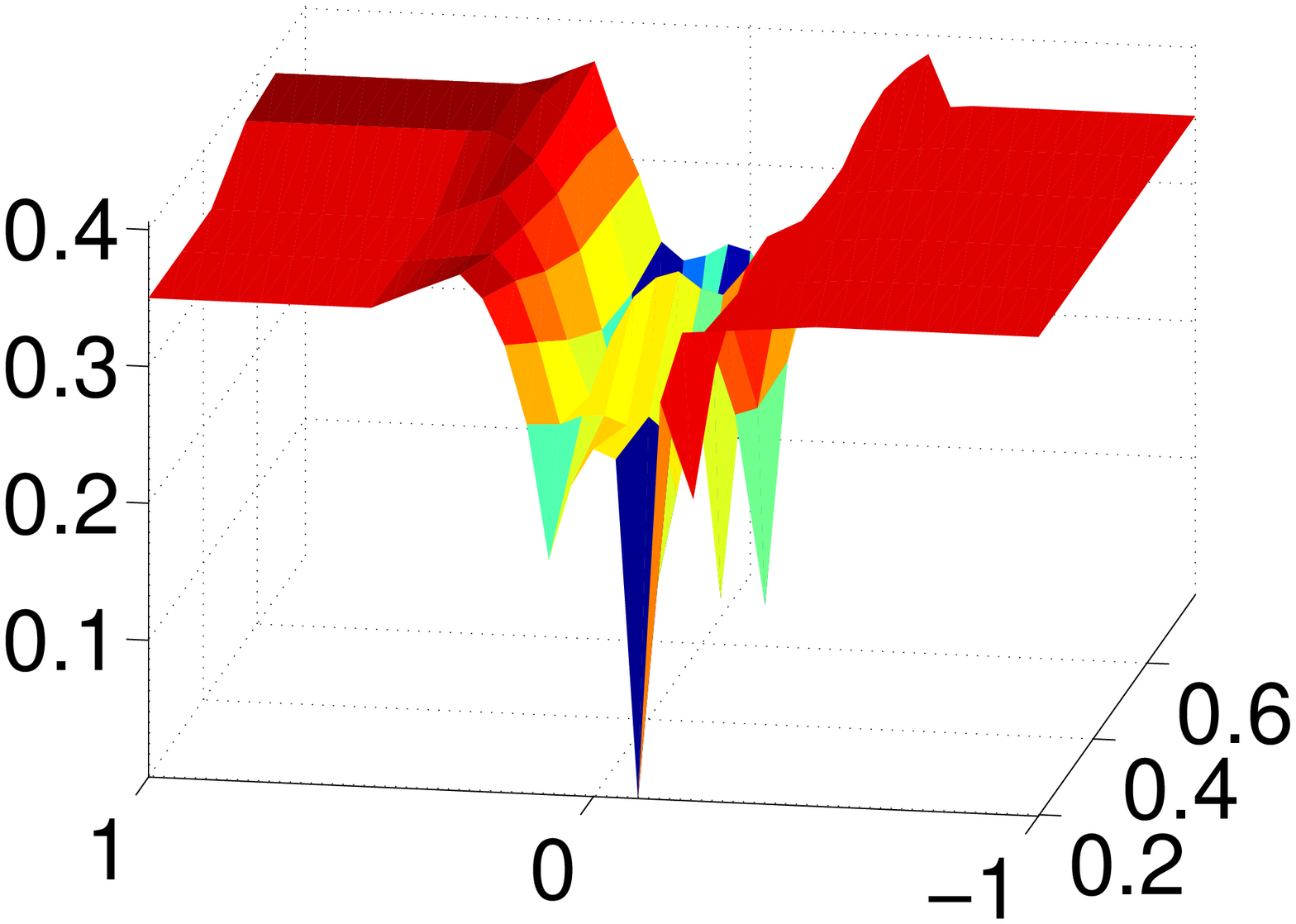}\hfill
      \includegraphics[width=0.48\textwidth]{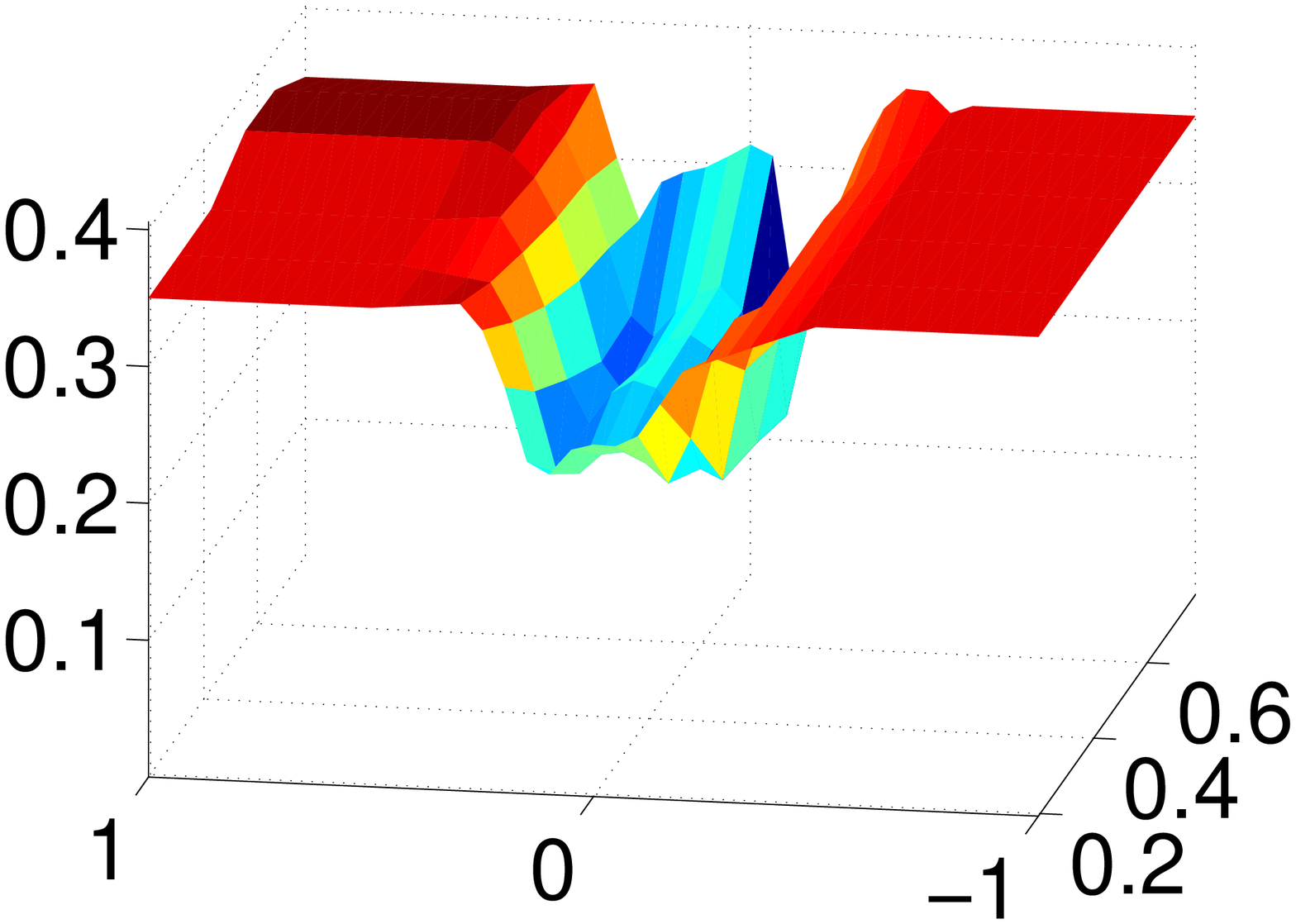}\hfill
\caption{Reconstructions of Henry Hub local volatilities. Trading dates: 04-Sep-2013, 05-Sep-2013, 09-Sep-2013 and 10-Sep-2013.}
  \label{hhonline_volsurfd}
\end{minipage}%
\end{minipage}%
\begin{minipage}{.48\textwidth}
  \centering
      \includegraphics[width=0.475\textwidth]{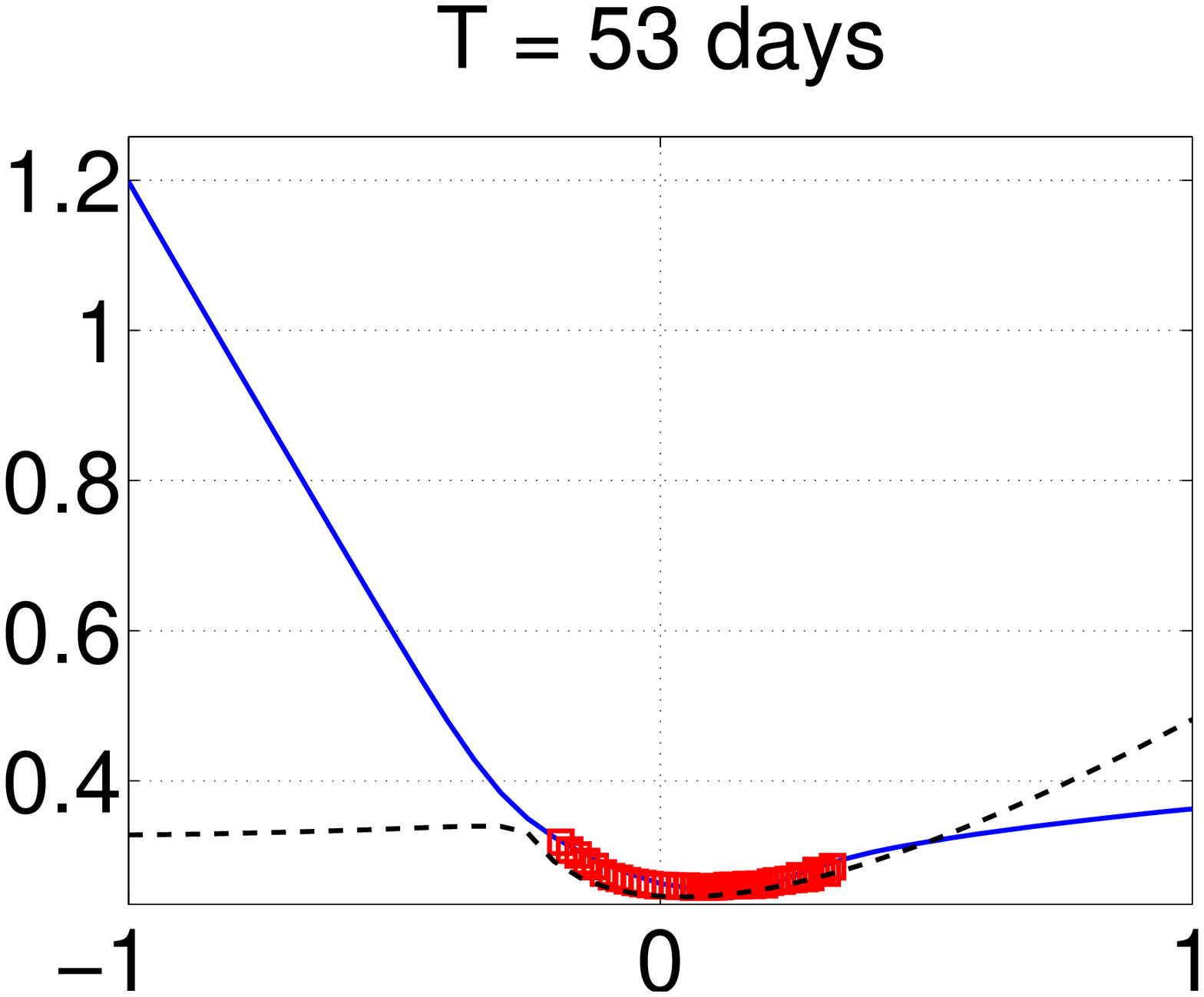}\hfill
      \includegraphics[width=0.475\textwidth]{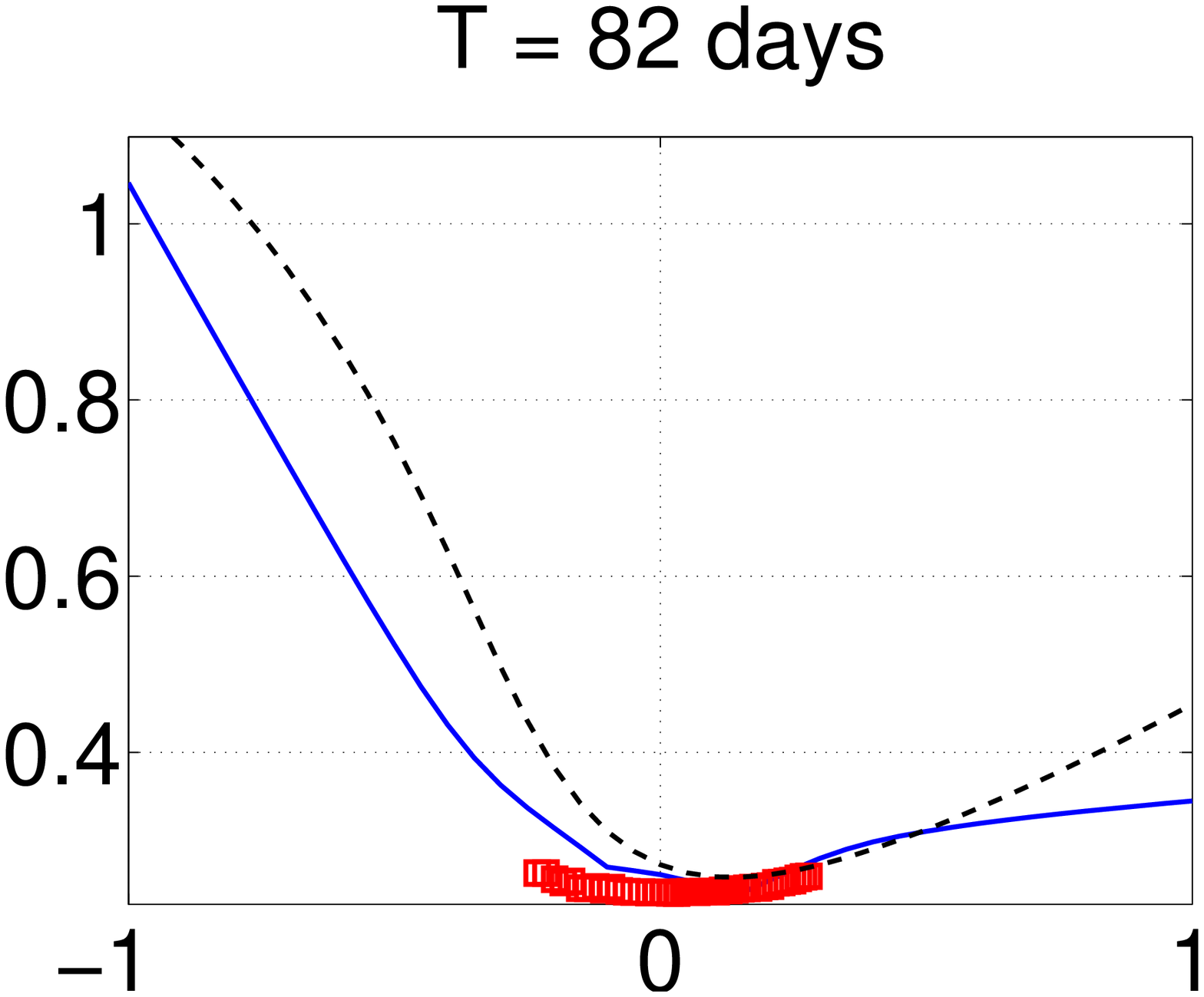}\hfill
      \includegraphics[width=0.475\textwidth]{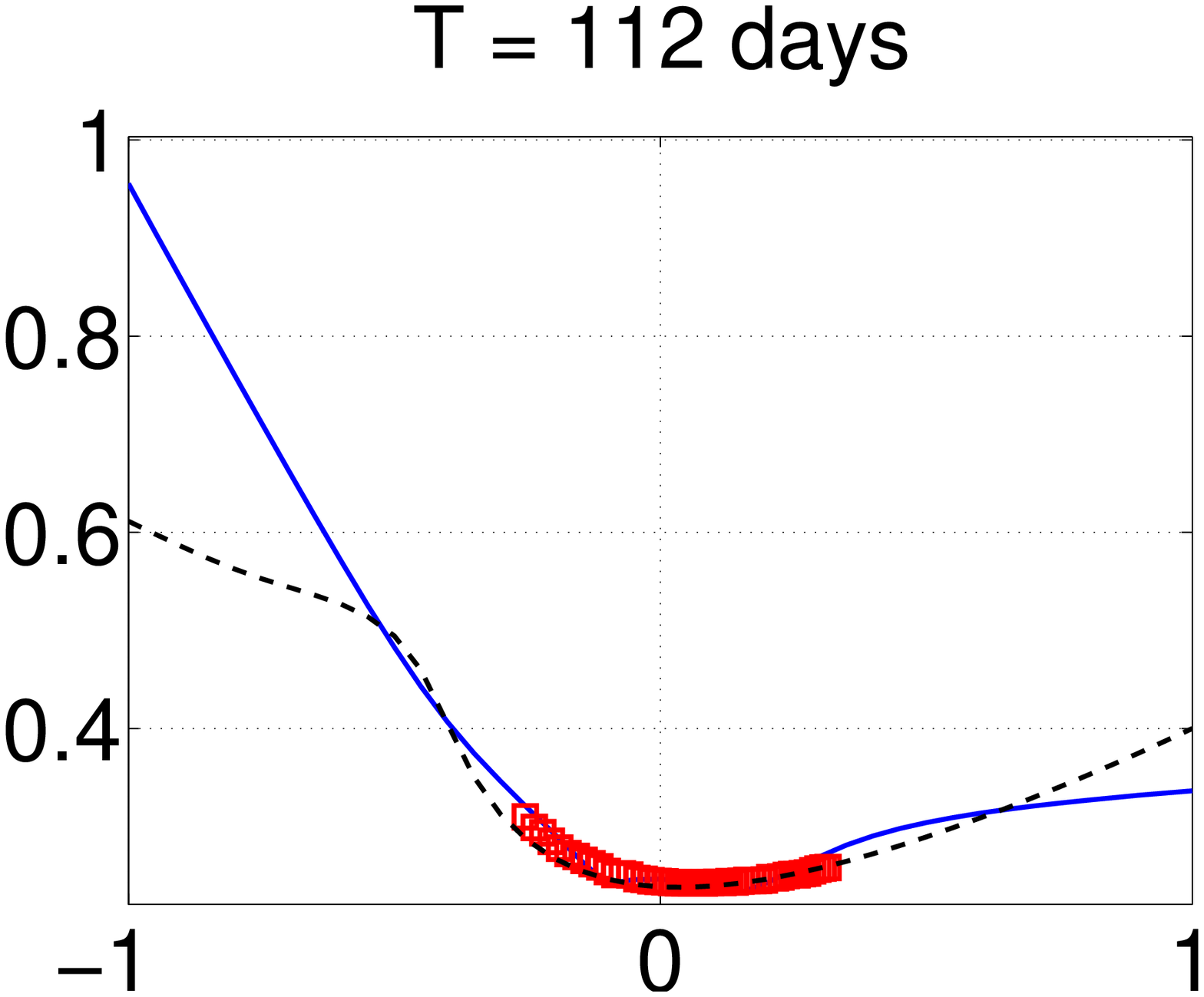}\hfill
      \includegraphics[width=0.475\textwidth]{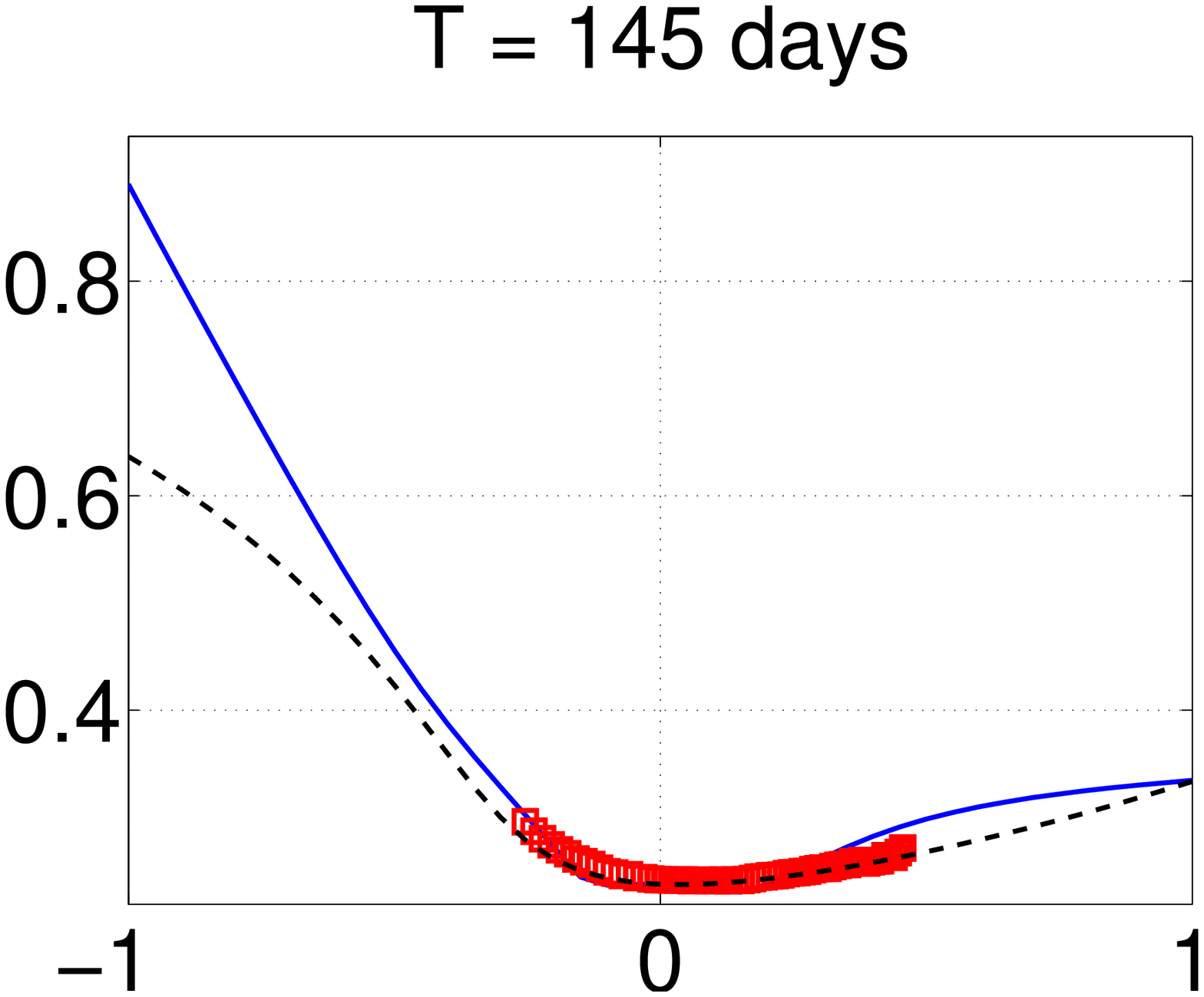}\hfill
      \includegraphics[width=0.475\textwidth]{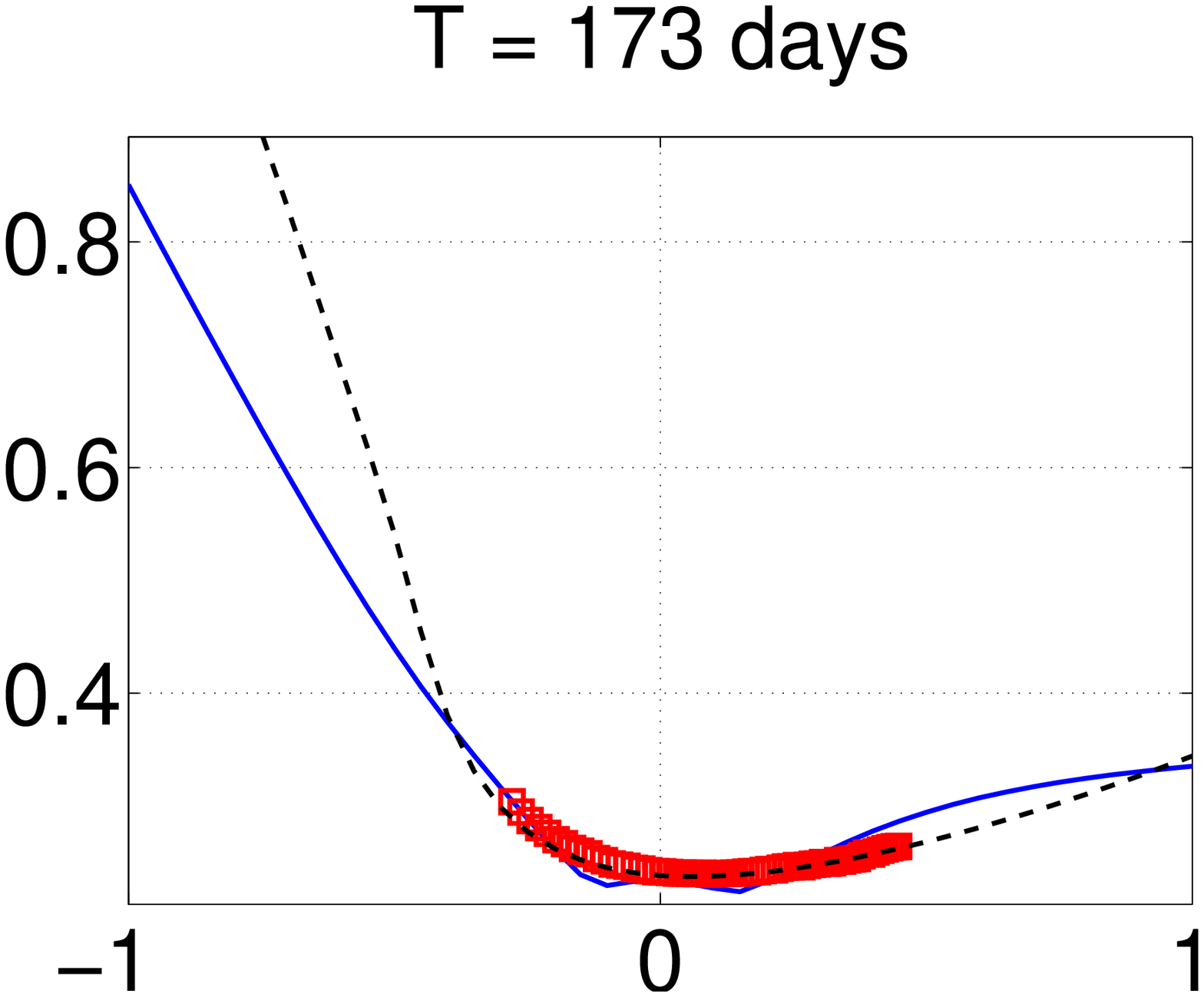}\hfill
      \includegraphics[width=0.475\textwidth]{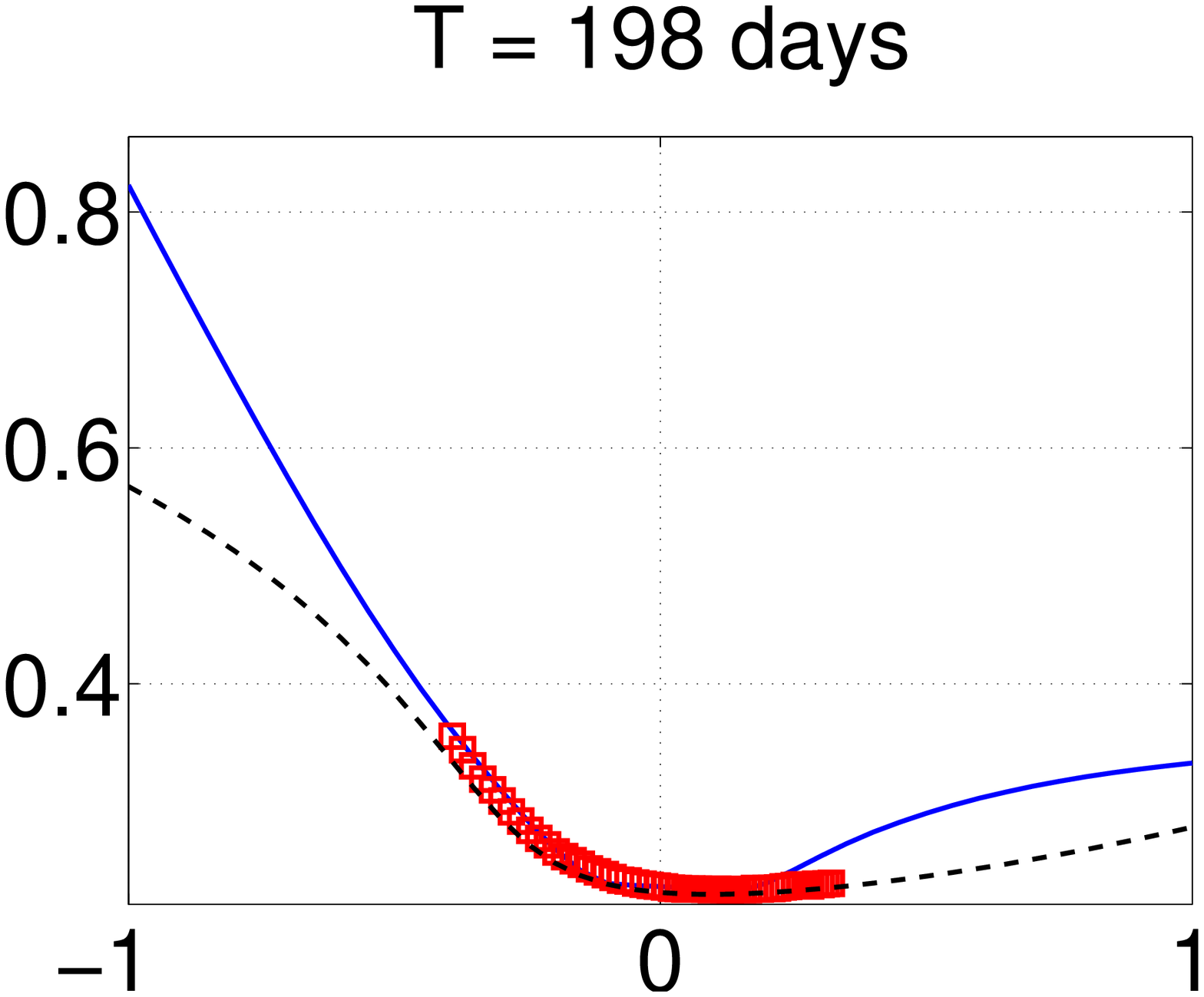}\hfill
  \caption{Implied volatilities of market data (squares), SVI (dashed line), and of reconstructions.}
  \label{hhonline_impvol}
\end{minipage}%
\end{figure}

Figure~\ref{hhonline_volsurf} presents the calibrated local volatility surface corresponding to the prices traded at 06-Sep-2013, obtained under online calibration. 
Figure~\ref{hhonline_impvol} presents a comparison between the market implied volatilities, the ones corresponding to the local volatility surfaces of Figure~\ref{hhonline_volsurf}, and the implied volatilities obtained by using the natural form of the SVI model presented by \citep{gatheralsvi2}. 

As in the previous examples, the iterations started with the constant local volatility surface $\af_0 \equiv 0.08$.

Figure~\ref{hhonline_volsurfd} presents the reconstructed local volatility surfaces for the dates 04-Sep-2013, 05-Sep-2013, 09-Sep-2013 and 10-Sep-2013, when we use all the five surfaces of price in online calibration. These local volatilities were obtained jointly with the one of Figure~\ref{hhonline_volsurf}.

As can be observed in Figures~\ref{hhonline_volsurf}-\ref{hhonline_volsurfd}, the local volatility surfaces at different levels of the underlying assets are similar, which implies that its evolution is well-behaved. In addition, Figure~\ref{hhonline_impvol} shows that the calibrated surfaces have a smile adherence similar to the SVI.

\subsection{The {\em Online} Approach with WTI Prices}\label{sec:wti}

We now repeat the test we made with Light Sweet Oil (WTI) call option prices, the option prices were traded at 09-Oct-2013, 10-Oct-2013, 11-Oct-2013, 14-Oct-2013 and 15-Oct-2013. We used the six first maturities, i.e., the future contracts, and the corresponding vanilla call options, maturing at 18-Oct-2013, 16-Nov-2013, 17-Dec-2013, 16-Jan-2014, 15-Feb-2014 and 18-Mar-2014.

We used the same methodology presented in Section~\ref{sec:hh}, but now, $\alpha_2 = 2.0\times 10^{-2}$, $\alpha_3 = 10^{-2}\alpha_2$, and the tolerance was taken as $tol = 0.01$.
\begin{figure}[!ht]
\centering
\begin{minipage}{.49\textwidth}
\begin{minipage}{1\textwidth}
  \centering
      \includegraphics[width=0.6\textwidth]{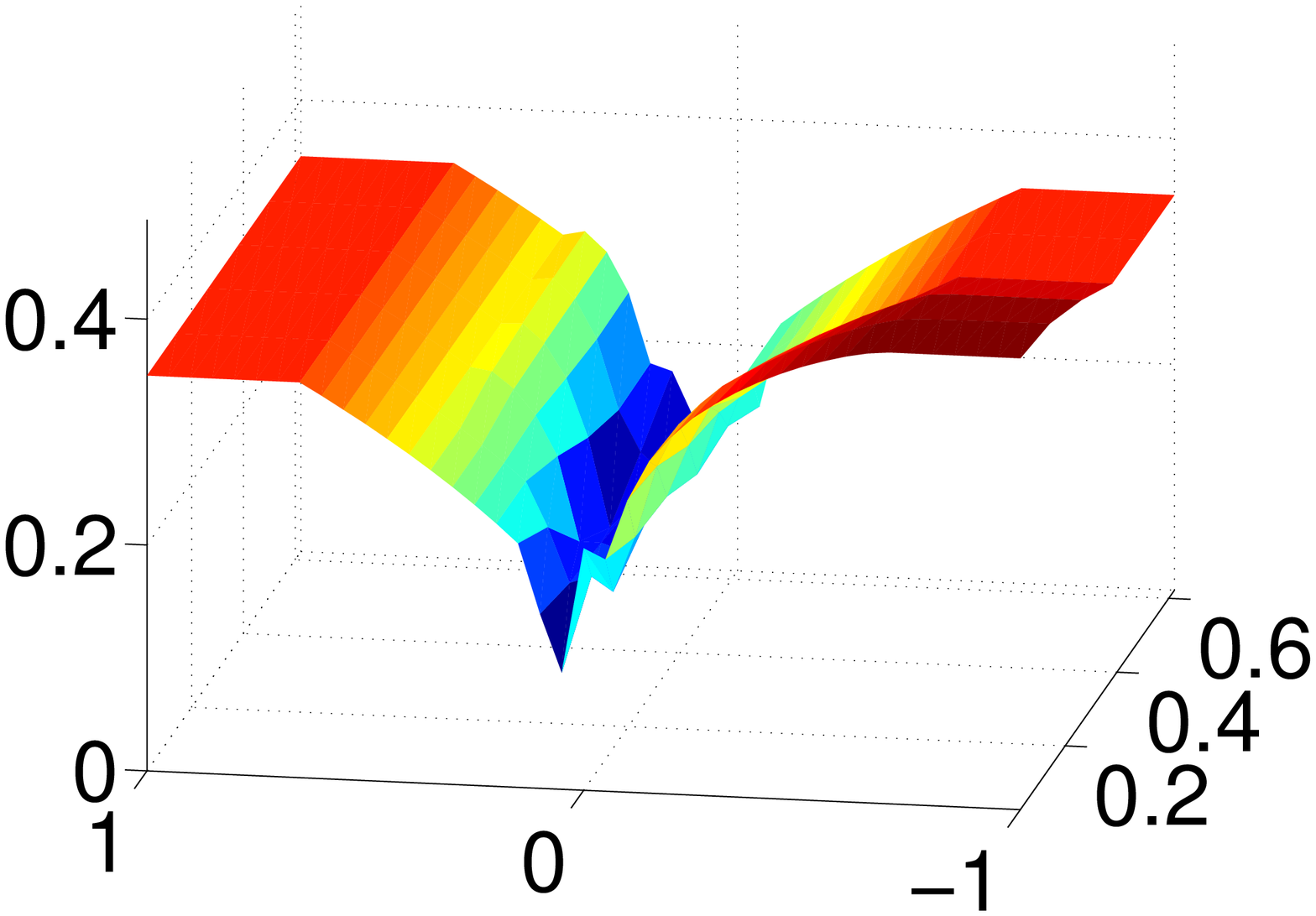}
\caption{Reconstructed WTI local volatility surface. Trading date: 06-Sep-2013.}
  \label{wtionline_volsurf}
\end{minipage}\hfill
\begin{minipage}{1\textwidth}
\centering
      \includegraphics[width=0.48\textwidth]{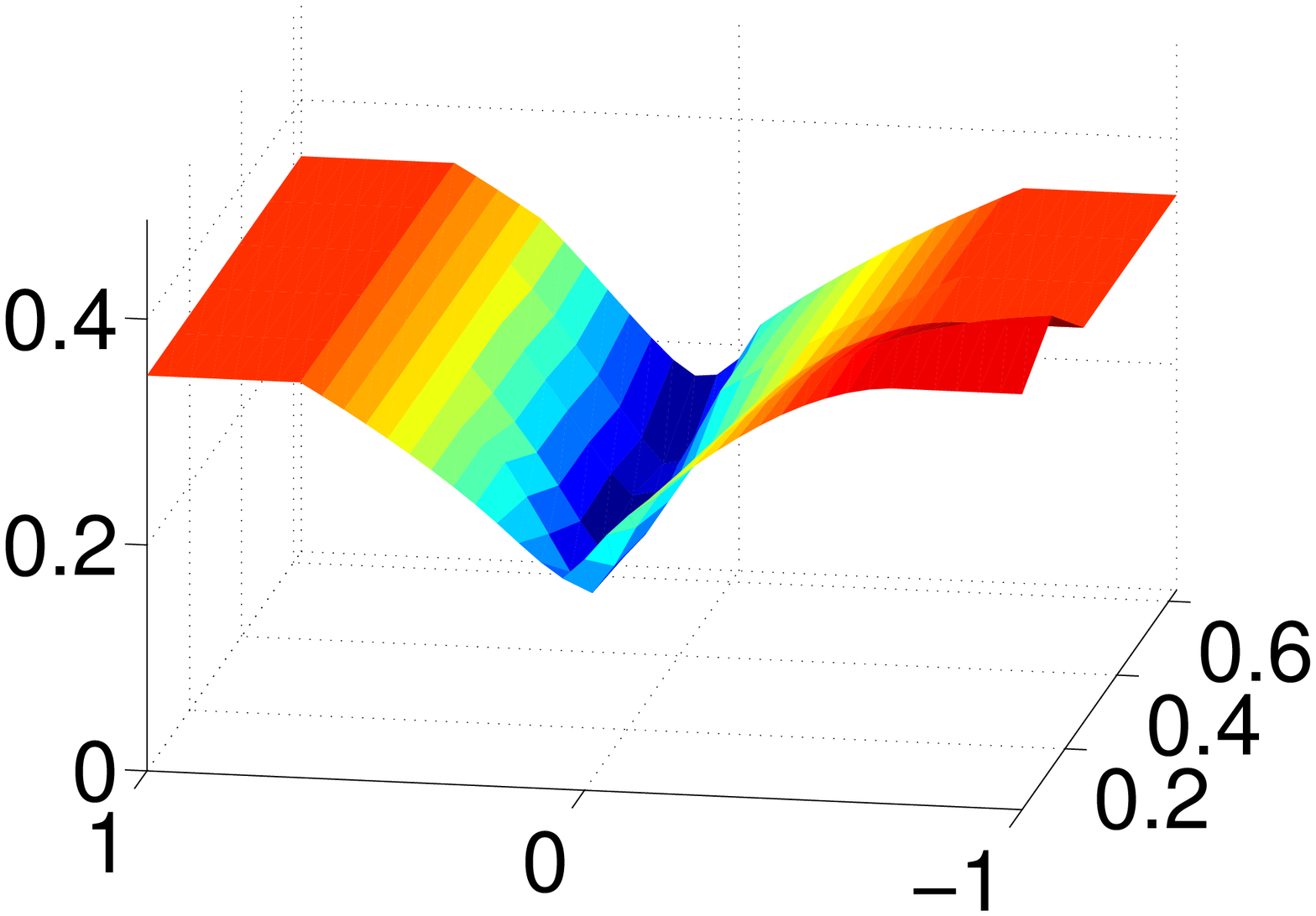}\hfill
      \includegraphics[width=0.48\textwidth]{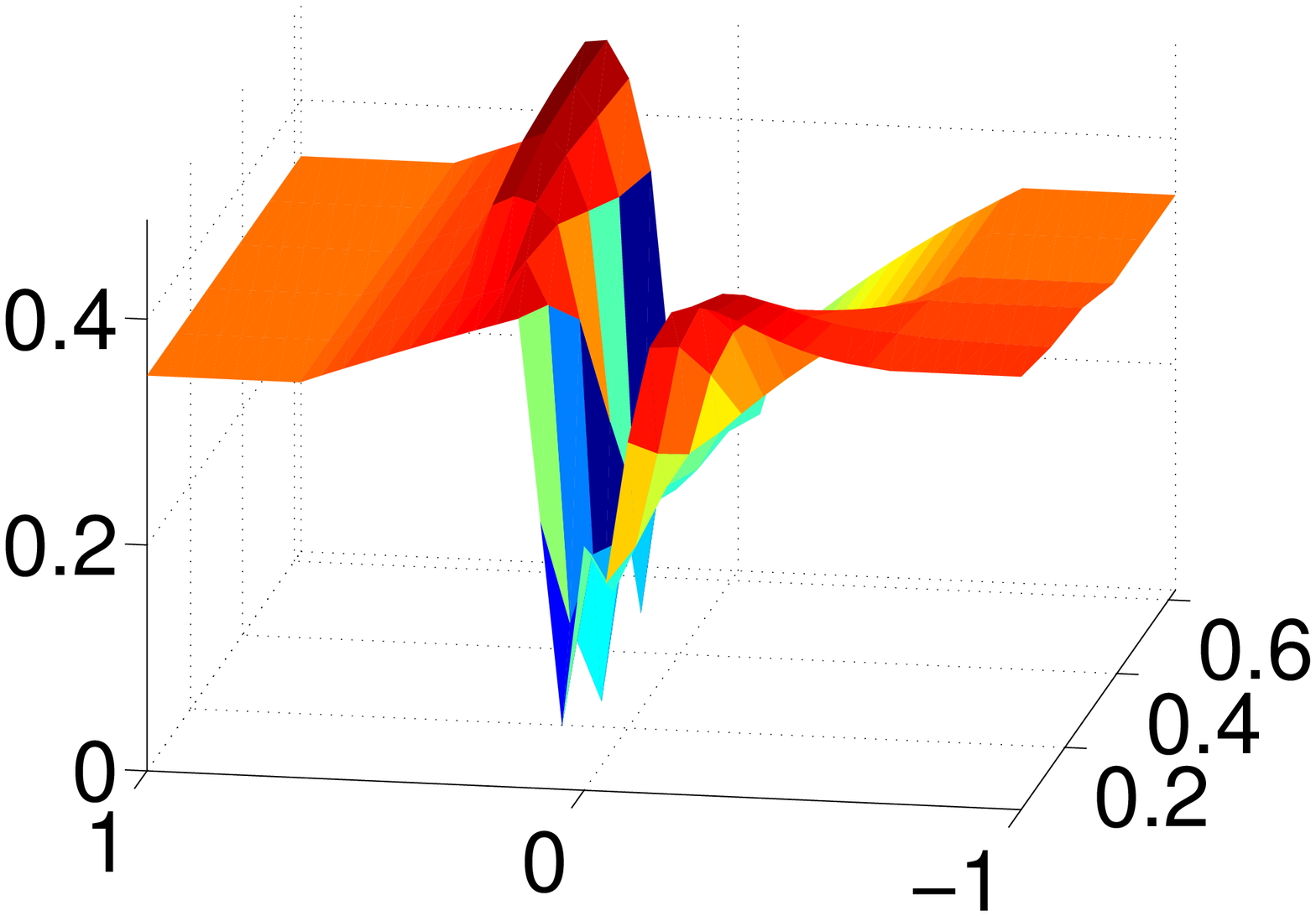}\hfill
      \includegraphics[width=0.48\textwidth]{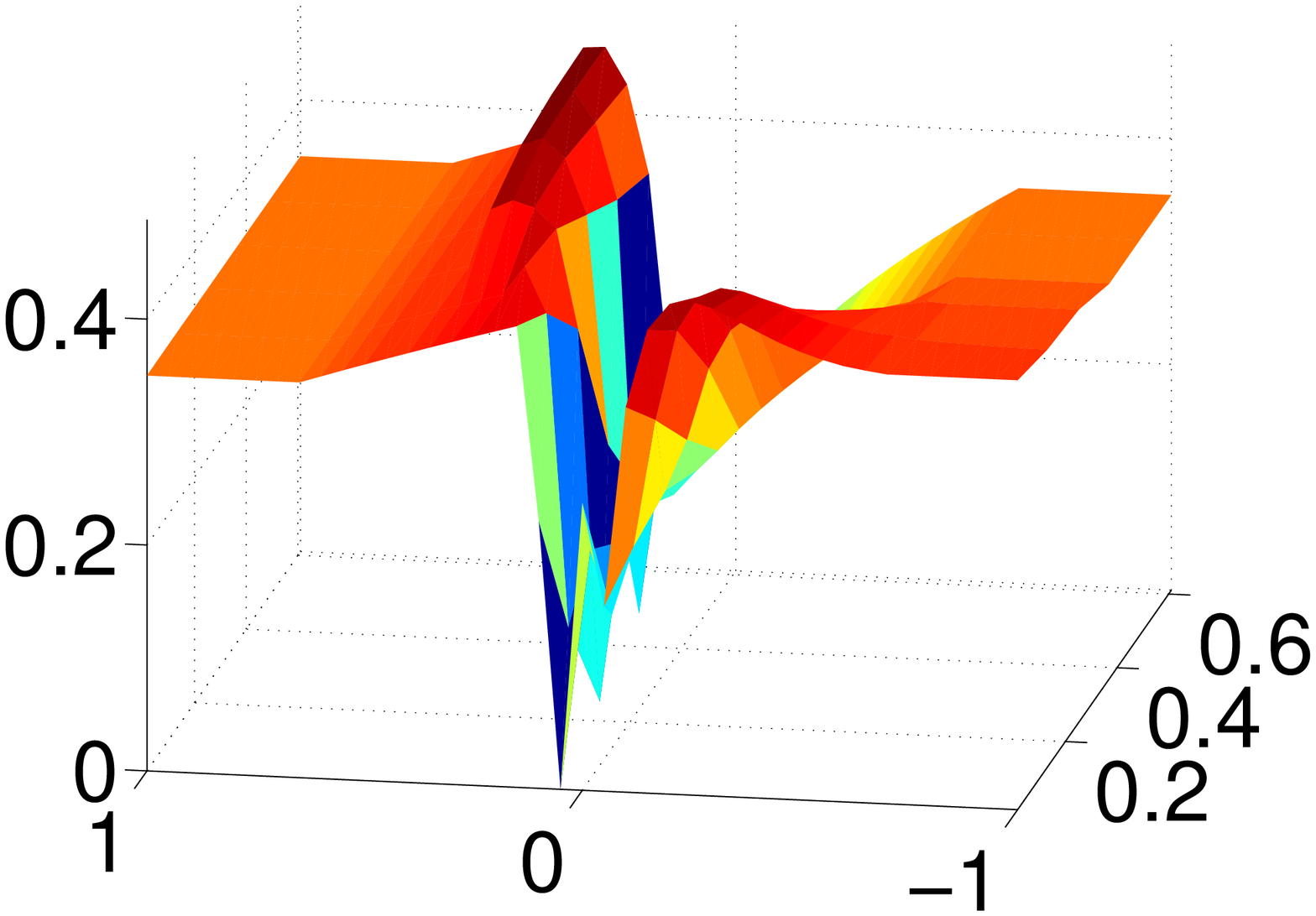}\hfill
      \includegraphics[width=0.48\textwidth]{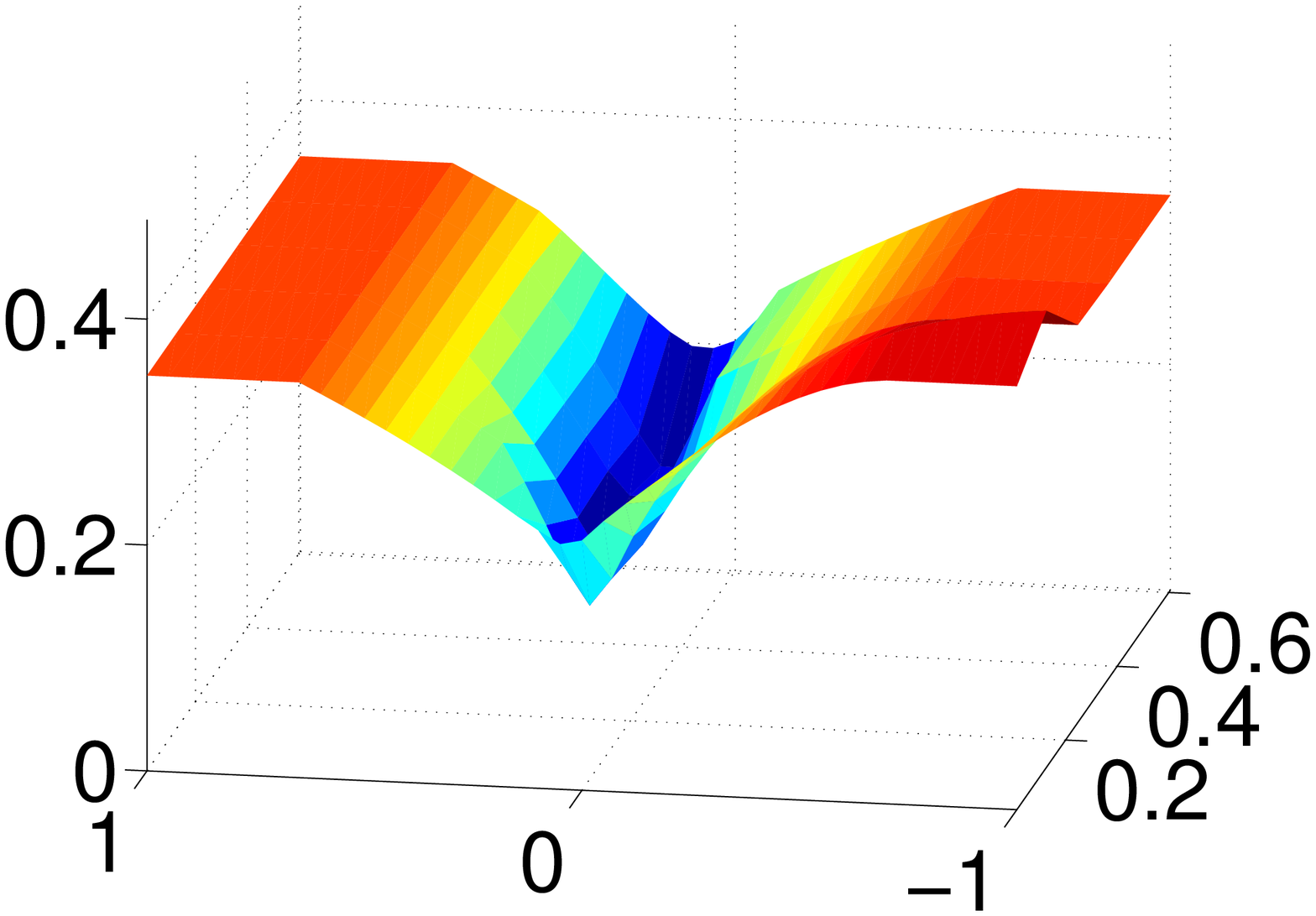}
\caption{Reconstructions of WTI local volatilities. Trading dates: 04-Sep-2013, 05-Sep-2013, 09-Sep-2013 and 10-Sep-2013.}      
  \label{wtionline_volsurfd}
\end{minipage}
\end{minipage}\hfill
\begin{minipage}{.49\textwidth}
  \centering
      \includegraphics[width=0.475\textwidth]{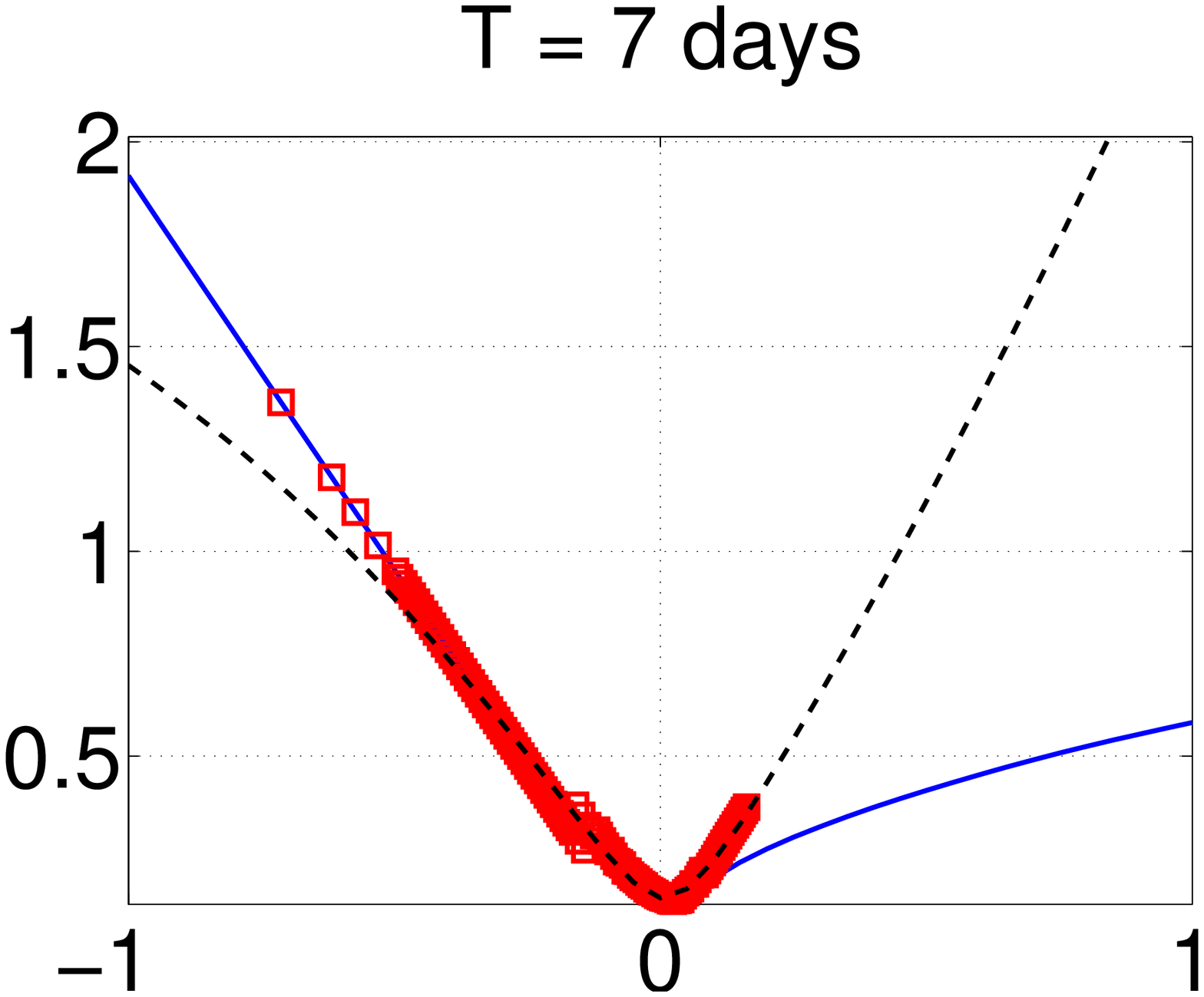}\hfill
      \includegraphics[width=0.475\textwidth]{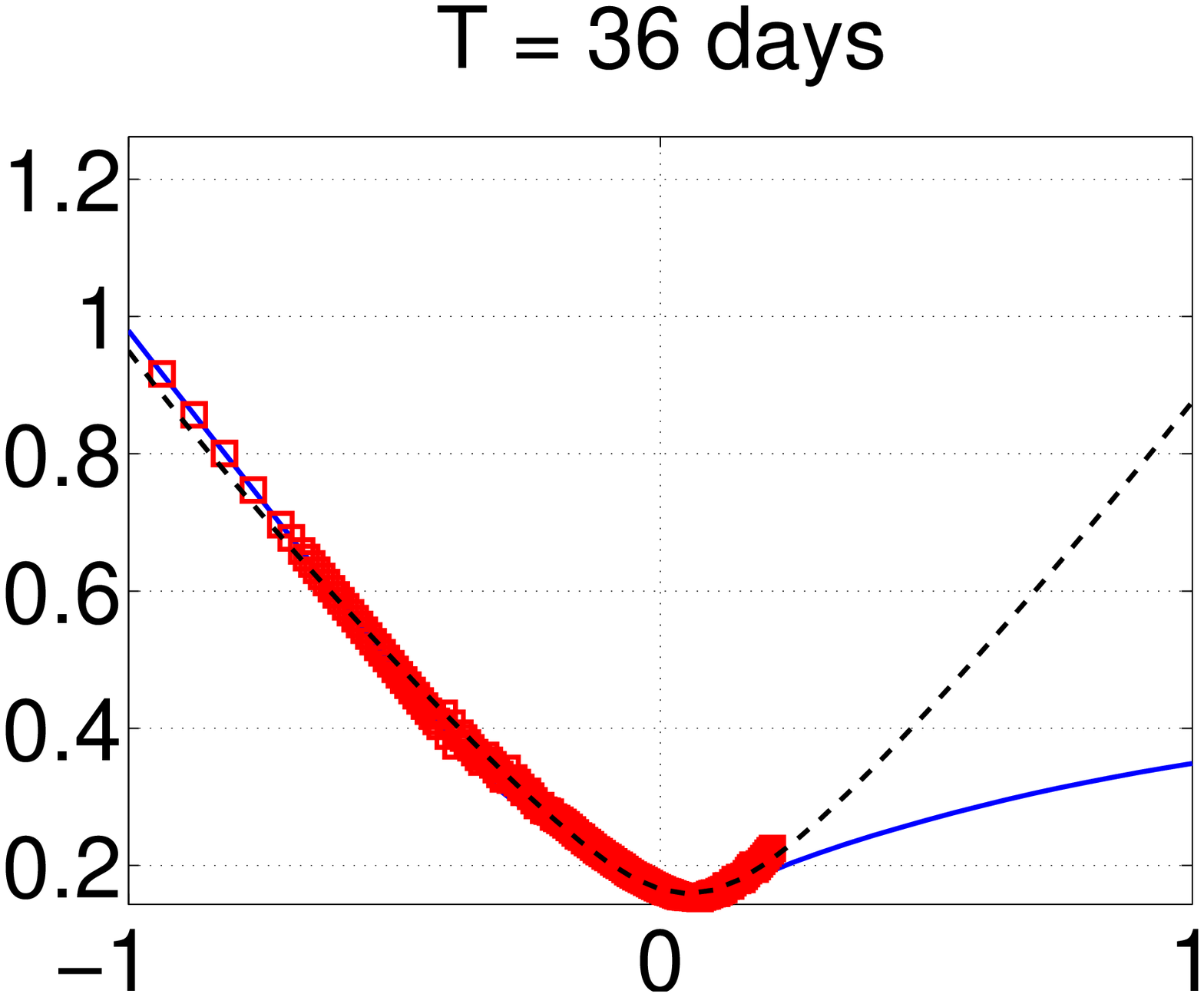}\hfill
      \includegraphics[width=0.475\textwidth]{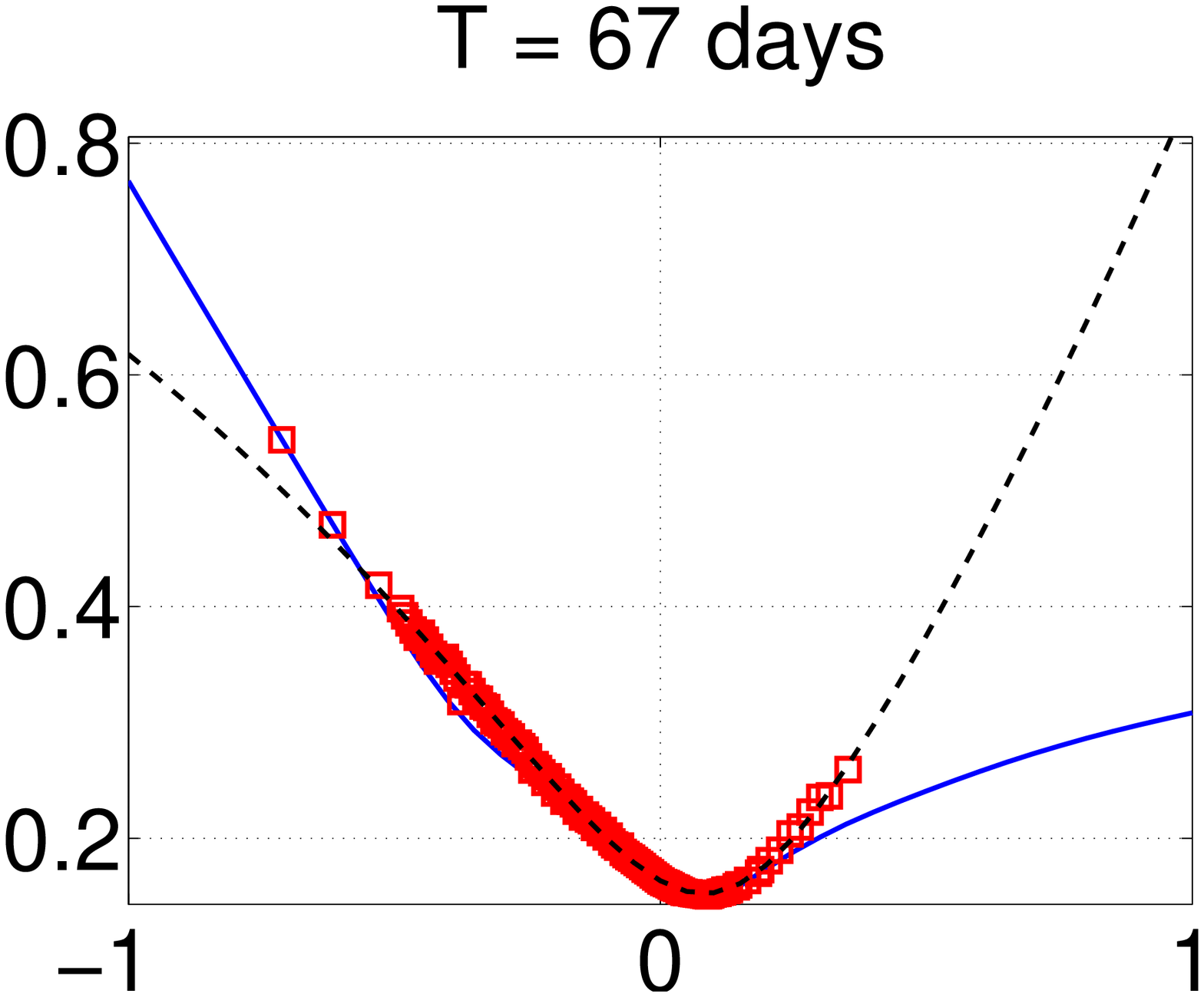}\hfill
      \includegraphics[width=0.475\textwidth]{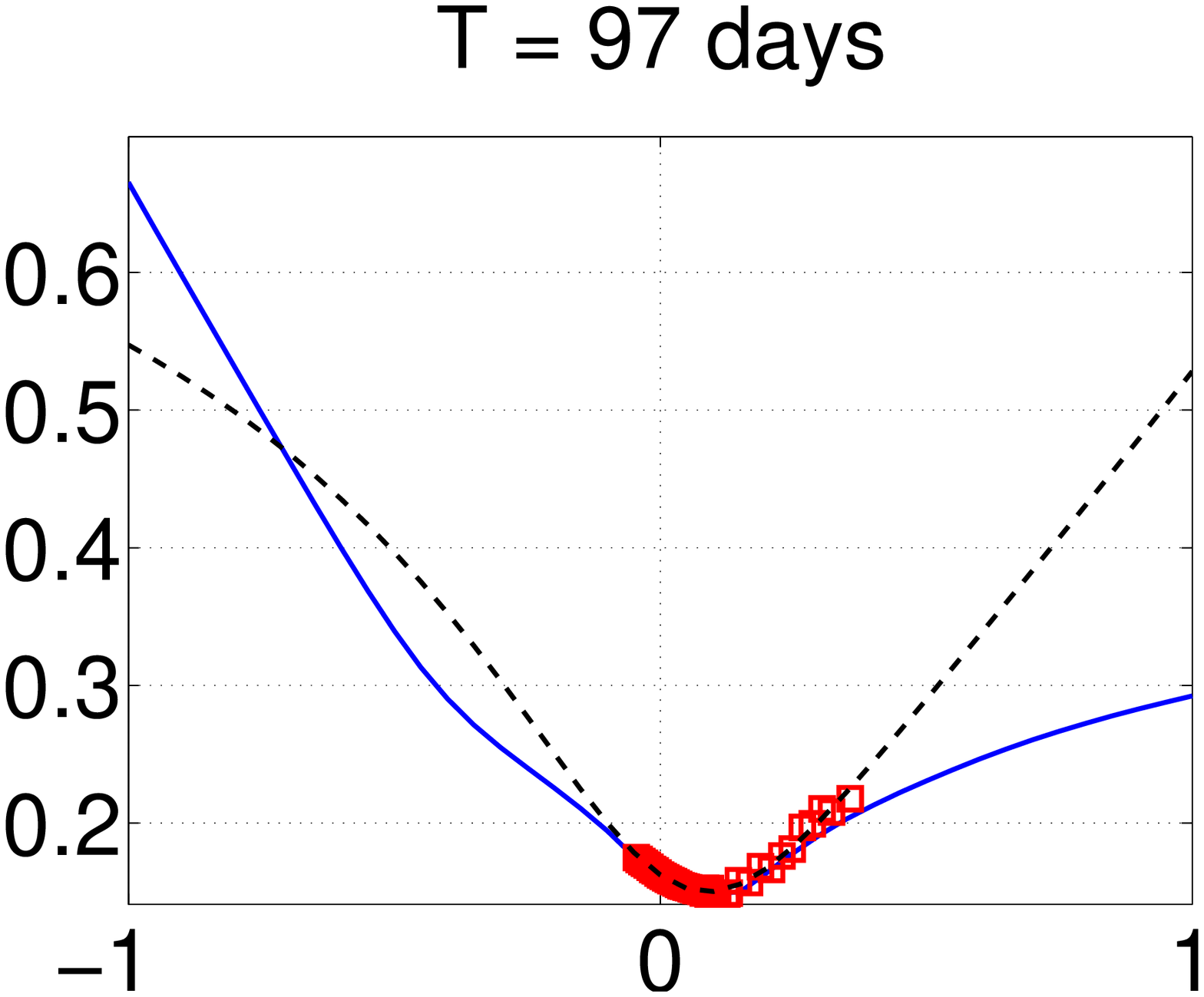}\hfill
      \includegraphics[width=0.475\textwidth]{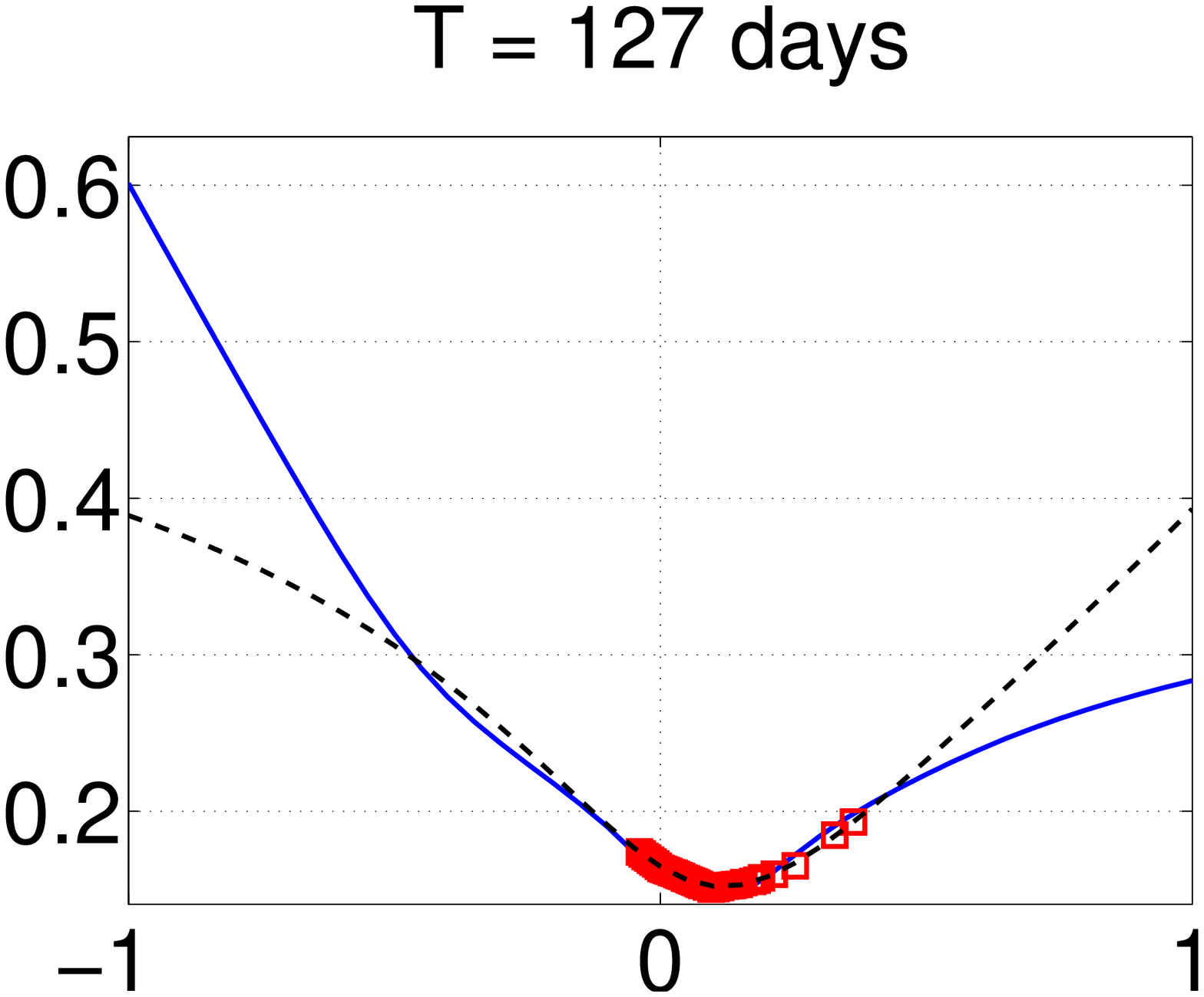}\hfill
      \includegraphics[width=0.475\textwidth]{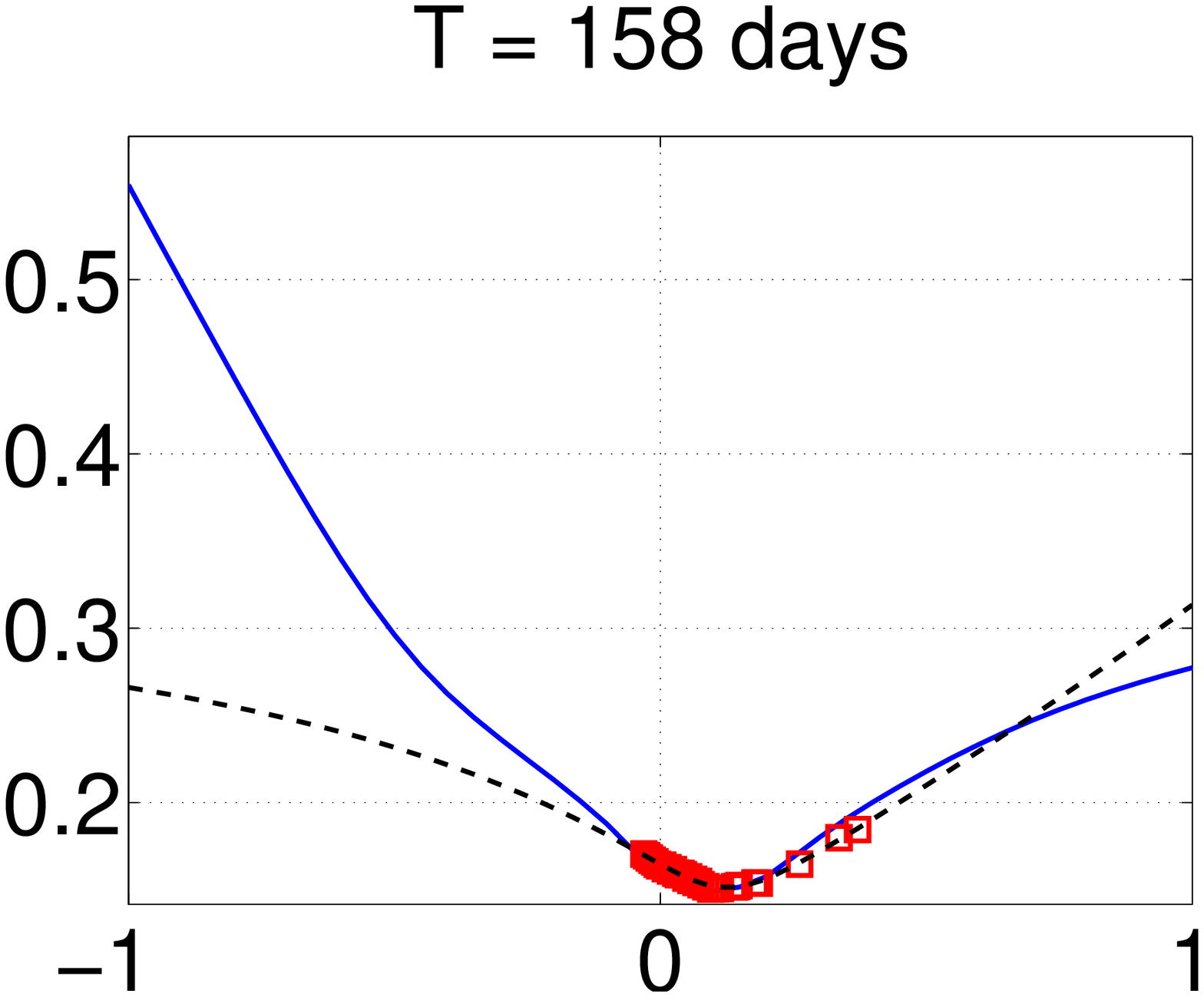}\hfill
  \caption{Implied volatilities of market data (squares), SVI (dashed line), and of reconstructions.}
  \label{wtionline_impvol}
\end{minipage}%
\end{figure}

With WTI prices we can draw similar conclusions as with Henry Hub options. Figures~\ref{hhonline_volsurf}-\ref{hhonline_volsurfd} shows that the evolution of the local volatility surface with respect to the index $s$ is well-behaved. Moreover, the calibrated surfaces have a smile adherence similar to the SVI, as we can see in Figure~\ref{hhonline_impvol}.

\subsection{Exotic Option Evaluation: an Illustration}
\label{exotic}
The aim of the present example is to illustrate the accuracy of the
local volatility model in pricing so-called exotic derivatives. 

We first generate synthetic European call prices with a model different from Dupire's. 
For this task we choose the Heston model, since it is well-known and simple to implement. Recall that, under this model, a stock price $S_t$ follows the dynamics
\begin{equation}
 \begin{array}{rcl}
   dS_t &=& \mu S_tdt + \sqrt{V_t}S_tdW_t^1, \quad 0\leq t \leq T_{\max}\\
   \\
   dV_t &=& \kappa(\theta - V_t)dt + \sigma \sqrt{V_t}dW_t^2,
 \end{array}
 \label{eq:heston}
\end{equation}
where $V_t$ is the stochastic variance. 
The Brownian motions $W^1_t$ and $W^2_t$ are assumed to have correlation $\rho$. 
For more details, see Chapter~2 of~\citep{volguide}. We evaluate the European call option data solving the associated pricing PDE with the parameters $\mu = r = 0.035$, $\kappa = 2$, $\theta=0.04$, $\sigma = 0.2$, $\rho = 0.1$, $V_0 = 0.2$ and $S_0 = 1$, at the maturities $\tau_i = i\cdot 0.1$, with $i=1,2,...,15$, and strikes  $y_j=j\cdot0.05$, with $j = -10,-9,...,0,1,...,10$. Then, we calibrate the local volatility surface, as well as the implied volatilities.

We consider for this test 
{\em European Asian call options} with srike $K$, maturity $T_{\max}$, and payoff
 $$
 A(T_{\max}) := \max\left\{0,\displaystyle\frac{1}{N}\sum_{j=0}^N S_{t_j} - K\right\},
 $$
 where $t_j = j\cdot \Delta t$ and $\Delta t = T_{\max}/N$. 
An Asian option is an example of the so-called path-dependent derivatives, whose evaluation
requires knowledge of the distribution of the process $S_t$ in a time-interval. 
Since the variance of $S_t$ also evolves with time, models with constant volatility, such as Black-Scholes, tend to lose precision when this time interval increases. 

The price at time $t=0$ of the Asian option is approximated by a simple Monte Carlo integration
$$
A(0) = \text{e}^{-rT_{\max}}\mathbb{E}\left[A(T_{\max})\right] \approx \text{e}^{-rT_{\max}}\displaystyle\frac{1}{N_{r}}\sum_{j=0}^{N_{r}}A(T_{\max})^{(j)},
$$
where $A(T_{\max})^{(j)}$ is the $j$-th realization of the random variable $A(T_{\max})$
and $N_r$ is the number of realizations. 
To generate the realizations $A(T_{\max})^{(j)}$, we use 
the Heston, Dupire and Black-Scholes models, setting the number of realizations to 
$N_{r}=10000$. The SDE associated to each method is solved by the Euler-Maruyama method, with the time step $\Delta t = T_{\max}/N$, $N = 100$, and $T_{\max}= \tau_i$. 
The Dupire's model is solved with the calibrated local volatility surface and the Black-Scholes with the implied volatilities. The ground truth prices are then the Heston ones. 

\begin{table}[!ht]
\centering
\begin{tabular}{|c|ccc|ccc|ccc|}
\hline
 &\multicolumn{3}{c|}{$\tau = 0.1$}&\multicolumn{3}{c|}{$\tau = 0.5$}&\multicolumn{3}{c|}{$\tau = 1$}\\
 \hline
$\log(K/S_0)$& 0 & -0.1 & 0.1 &  0 & -0.1 & 0.1 & 0 & -0.1 & 0.1\\
\hline\hline
Heston & 0.0330 & 0.1011 & 0.0046 & 0.0688 & 0.1245 & 0.0313 & 0.0972 & 0.1509 & 0.0560\\
L.Vol. & 0.0317 & 0.0986 & 0.0042 & 0.0709 & 0.1269 & 0.0328 & 0.0962 & 0.1485 & 0.0557\\
B-S & 0.0314 & 0.1004 & 0.0044 & 0.0648 & 0.1236 & 0.0274 & 0.0832 & 0.1394 & 0.0427\\
\hline
\end{tabular}
\caption{European Asian call option prices.}
\label{tab:asian}
\end{table}

\begin{table}[!ht]
\centering
\begin{tabular}{|c|ccc|ccc|}
\hline
 &\multicolumn{3}{c|}{Local Volatility}&\multicolumn{3}{c|}{Black \& Scholes}\\
 \hline
$\log(K/S_0)$ & 0 & -0.1 & 0.1 &  0 & -0.1 & 0.1\\
\hline\hline
$\tau = 0.1$ & 0.0247 & 0.0387 & 0.0985 & 0.0067 & 0.0478 & 0.0519\\
$\tau = 0.5$ & 0.0189 & 0.0317 & 0.0495 & 0.0076 & 0.0576 & 0.1246\\
$\tau = 1.0$ & 0.0157 & 0.0103 & 0.0057 & 0.0757 & 0.1436 & 0.2370\\
$\tau = 1.5$ & 0.0400 & 0.0420 & 0.0426 & 0.1244 & 0.1791 & 0.2592\\
\hline
\end{tabular}
\caption{Relative errors in Asian call option prices.}
\label{tab:asian2}
\end{table}

Table~\ref{tab:asian} presents the Asian option prices at different maturities and strikes;
we denote the prices evaluated with the local volatility model by {\em L.Vol.},
and the ones evaluated with the Black-Scholes model by {\em B-S}. 
Comparing these to the Heston prices, we can see that Dupire prices are more precise than the Black-Scholes ones 
for longer maturities, as expected. 
This is confirmed by the absolute relative errors presented in Table~\ref{tab:asian2}. The normalized residual of  local volatility prices is $0.0213$, whereas that of the Black-Scholes prices is $0.0623$, almost three times larger.  So, we see that
the local volatility model is more reliable for evaluating 
path-dependent option prices than simply using a technique based on implied volatilities, 
the latter having more difficulties to  capture the variance evolution in a time interval.

\section{Concluding Remarks}\label{sec:future}

In this paper we have presented and tested a calibration methodology that
is suitable for treating commodity futures and their derivatives.
This is done by first modeling the futures with a fixed maturity
according to a stochastic differential equation whose volatility
depends on time and on the price of the future contract. Such model
falls within the framework of Dupire's local volatility theory and
thus is amenable, after suitable adaptation, to the online calibration
approach developed in~\citep{vvla2}.
The complete methodology allows to perform the calibration starting
from market quoted prices of calls (or puts) on futures, and from there to
estimate the local volatility surface. Once such volatility surface is
estimated one can compute arbitrage free market compatible prices of
various derivatives, including path dependent ones. In Section~\ref{exotic}
we exemplify this for the case of an Asian option.
We have validated our methodology extensively, using both real and synthetic data.

The local volatility calibration problem is an ill-posed one,
and there are many solutions to the associated data fitting problem.
To determine a suitable one among those, we regularized the problem by incorporating
{\em a priori} information. Importantly, we have also tackled the underlying-price
uncertainty. This is done by judiciously choosing the regularization
parameters in a Tikhonov-type approach, and introducing a splitting optimization
method for determining both the local volatility surface and the underlying price. 
Our approach is amenable to handling scarce data as well as substantial
increase in data volume.

The numerical tests reported herein confirm the robustness and
effectiveness of our methodology.  Indeed,  numerical experiments with
synthetic data show that, as more observations of option
prices are included, reconstructions improve and the error decreases.
Furthermore, as we adjust the underlying prices, the reconstructions
come closer to the ground truth one. In the case of real data,  the
computed prices of vanilla options display adherent implied
volatilities.

\section*{Acknowledgements}
V.A. acknowledges and thanks CNPq through grant 201644/2014-2, Petroleo Brasileiro S.A. and Ag\^encia Nacional do Petr\'oleo for the financial support during the time when this work was developed.
U.A. Acknowledges with thanks a {\em Ciencias Sem Fronteiras} (visiting scientist) grant from CAPES, Brazil.  
J.P.Z. acknowledges and thanks the financial support from CNPq through grants 302161/2003-1 and
474085/2003-1, and from FAPERJ through the programs {\em Cientistas do Nosso Estado} and {\em Pensa Rio}.


\end{document}